\documentclass[twocolumn,10pt]{aastex631}

\usepackage [english]{babel}
\usepackage [autostyle, english = american]{csquotes}
\MakeOuterQuote{"}

 \newcommand{\head}[2]{\multicolumn{1}{>{\centering\let\newline\arraybackslash}p{#1}}{{#2}}}

\graphicspath{{./}{figures/}}

\received{November 24, 2021}
\revised{February 8, 2022}
\accepted{February 23, 2022}
\submitjournal{PSJ}

\shorttitle{Neptune's mid-IR Variability}
\shortauthors{Roman et al.}

\begin{document}

\title{Sub-Seasonal Variation in Neptune's Mid-Infrared Emission from Ground Based Imaging}


\correspondingauthor{Michael T. Roman}
\email{m.t.roman@leicester.ac.uk}

\author[0000-0001-8206-2165]{Michael T. Roman}
\affil{School of Physics and Astronomy 
University of Leicester 
University Road, Leicester, LE1 7RH, UK}

\author[0000-0001-5834-9588]{Leigh N. Fletcher}
\affil{School of Physics and Astronomy 
University of Leicester 
University Road, Leicester, LE1 7RH, UK}

\author[0000-0001-7871-2823]{Glenn S. Orton}
\affil{Jet Propulsion Laboratory/California Institute of Technology, 4800 Oak Grove Dr, Pasadena, CA 91109, USA}

\author[0000-0001-6613-5731]{Thomas K. Greathouse}
\affil{Southwest Research Institute, San Antonio, TX, USA }

\author[0000-0002-8837-0035]{Julianne I. Moses}
\affil{Space Science Institute, Boulder, CO 80301, USA}

\author[0000-0001-8692-5538]{Naomi Rowe-Gurney}
\affil{Department of Physics and Astronomy, Howard University, Washington, DC 20059, USA}
\affil{Astrochemistry Laboratory, NASA/GSFC, Greenbelt, MD 20771}
\affil{Center for Research and Exploration in Space Science and Technology, NASA/GSFC, Greenbelt, MD 20771}

\author[0000-0002-6772-384X]{Patrick G. J. Irwin}
\affil{University of Oxford
Atmospheric, Oceanic, and Planetary Physics, Department of Physics\\ Clarendon Laboratory, Parks Road, Oxford OX1 3PU, United Kingdom}

\author[0000-0001-9206-6960] {Arrate Antuñano}
\affil{UPV/EHU, Escuela Ingernieria de Bilbao, Fisica Aplicada, Spain}

\author[0000-0001-5374-4028]{James Sinclair}
\affil{Jet Propulsion Laboratory/California Institute of Technology, 4800 Oak Grove Dr, Pasadena, CA 91109, USA}

\author[0000-0002-8160-3553]{Yasumasa Kasaba}
\affil{Planetary Plasma and Atmospheric Research Center, Graduate School of Science, Tohoku University,\\ Sendai, Miyagi, 980-8578, Japan}

\author{Takuya Fujiyoshi}
\affil{Subaru Telescope, National Astronomical Observatory of Japan, 650 North A’ohoku Pl., Hilo, HI 96720, USA}

\author[0000-0002-4278-3168]{Imke de Pater}
\affil{Department of Astronomy, 501 Campbell Hall, University of California at Berkeley, Berkeley, CA, USA}

\author[0000-0001-8751-3463]{Heidi B. Hammel}
\affil{Association of Universities for Research in Astronomy,\\ 1212 New York Avenue NW, Suite 450, Washington, DC 20005, USA}



\begin{abstract}

We present an analysis of all currently available ground-based imaging of Neptune in the mid-infrared.  Dating between 2003 and 2020, the images reveal changes in Neptune's mid-infrared ($\sim$8--25 $\mu$m) emission over time in the years surrounding Neptune's 2005 southern summer solstice. Images sensitive to stratospheric ethane ($\sim$12 $\mu$m), methane ($\sim$8 $\mu$m), and CH$_3$D ($\sim$9 $\mu$m) display significant sub-seasonal temporal variation on regional and global scales.  Comparison with H$_2$ S(1) hydrogen-quadrupole ($\sim$17.035-$\mu$m) spectra suggests these changes are primarily related to stratospheric temperature changes. The stratosphere appears to have cooled between 2003 and 2009 across multiple filtered wavelengths, followed by a dramatic warming of the south pole between 2018 and 2020.  Conversely, upper-tropospheric temperatures--inferred from $\sim$17--25-$\mu$m imaging---appear invariant during this period, except for the south pole, which appeared warmest between 2003 and 2006.  We discuss the observed variability in the context of seasonal forcing, tropospheric meteorology, and the solar cycle. Collectively, these data provide the strongest evidence to date that processes produce sub-seasonal variation on both global and regional scales in Neptune's stratosphere.  

\end{abstract}


\section{Introduction} \label{sec:intro}

Despite being the most distant giant planet from the Sun, the ice giant Neptune possesses an extremely dynamic atmosphere, with meteorological phenomena evolving over a surprising range of timescales. With the most powerful zonal winds in the solar system \citep{limaye1991winds}, Neptune's highest clouds evolve so rapidly that the planet's appearance can change dramatically over the course of days \citep[\textit{e.g.},][]{smith1989voyager,sromovsky1993dynamics,sromovsky2001neptune}. Over time, analyses of cloud activity suggest intriguing trends in Neptune’s cloud cover and albedo \citep{lockwood2002photometric,lockwood2019final,karkoschka2011neptune}, occurrences of long-lived cloud features \citep{huesoneptunelonglived2017}, and dark vortices \citep{polvani1990simple,lebeau1998epic,stratman2001epic,wong2018new,hsu2019lifetimes,hadland2020epic}---each varying over periods of years.  All this relatively rapid variability occurs against a backdrop of Neptune's 165-year orbital period, which theory predicts should modulate stratospheric temperatures and chemistry slowly across seasons lasting several decades given Neptune's 28.3$^{\circ}$ axial tilt \citep{conrath1990temperature, moses2018seasonal}. 

While observational evidence of cloud and haze variability have been well documented over the past decades \citep[\textit{e.g.},][]{lockwood2002photometric, lockwood2006photometric,lockwood2019final, smith1989voyager,sromovsky2001neptune,sromovsky2003seasonal_response,hammel2007long,karkoschka2011neptune,irwin2011neptune,Roman2013DPS,irwin2016neptunetimevary,huesoneptunelonglived2017,simon2019darkspot,molter2019neptune}, unequivocal evidence of temporal variation of the lower stratospheric temperatures and chemistry has been scarce \citep{roques1994neptune,hammel2006mid, hammel2007long, greathouse2011spatially,fletcher2014neptune, sinclair2020spatial}. The paucity of these measurements is largely a consequence of the challenges of making reliably accurate mid-infrared observations from the ground, given Neptune's frigid atmospheric temperatures \citep{orton1987spectra,conrath1989neptune, fletcher2014neptune}. This challenge is further compounded by Neptune's small angular diameter ($\sim$2.3'').  As a result, spatially-resolved mid-IR observations only really became feasible within the last two decades---with current temporal coverage amounting to roughly a 1/9 of the full seasonal cycle.  

Although challenging, spatially resolved mid-infrared observations can be used to diagnose the thermal and chemical structure of the atmosphere and reveal changes and processes undetectable by other means. The mid-infrared provides a unique and essential window into the chemistry, dynamics, and radiative processes that define Neptune's lower stratosphere and upper troposphere --- a region that spans the interface between the turbulent weather-layer below and the presumably stably stratified, photochemically rich layer above.  Observations at these wavelengths are dominated by the collision-induced opacity of H$_2$ and emission from various hydrocarbons---primarily stratospheric methane, ethane, and acetylene \citep[\textit{e.g.},][]{moses2020icegiantchem}. Challengingly, the intensity of the hydrocarbon emission is modulated by both temperature and chemical abundances, neither of which is typically independently constrained, often rendering the physical nature of the observed changes ambiguous. 

Comparing five disk-averaged mid-infrared spectra dating between 1985 and 2004, \citet{hammel2006mid} were able to show an apparent trend of increasing emission at $\sim$12 $\mu$m over time, followed by a slight decrease in 2004. Changes at $\sim$8 $\mu$m---sensitive to methane---were present but weaker. \citet{hammel2006mid} argued that these relatively rapid changes were likely caused by chemical changes and not by changing atmospheric temperatures, given previous estimates of relatively long radiative time constants in the stratosphere \citep{conrath1998thermal}. Subsequent radiative modeling has since reduced the expected radiative time constants at stratospheric pressures from decades to years or less \citep{li2018high}, thus reopening the possibility of either a chemical or thermal interpretation. Likewise, \citet{fletcher2014neptune} compared spectra between 2003 and 2007 and found slight variability in the $\sim$8-$\mu$m methane emission, but strong variability in the ethane emission, which they recognized as ambiguous but consistent with a 2$\times$ drop in ethane abundance provided that the stratospheric temperatures remained constant. In either case, significant calibration uncertainties and the limited number of observations left the veracity and nature of these diverging trends somewhat in doubt. 

Models of Neptune's radiative heating and cooling \citep{greathouse2011spatially} and models of seasonally varying photochemistry by \citet{moses2018seasonal} provide expectations for how the temperatures and chemical abundances will change in time.  Following Neptune’s southern summer solstice in 2005, models suggest the south pole would eventually grow warmer in response to the increased insolation, but with a lag indicative of the radiative time constant \citep{greathouse2011spatially}.  At the same time, the increased ultraviolet (UV) flux should result in an increase in the stratospheric photochemical production.  While the methane abundance itself should not vary appreciably because its overall abundance far outweighs its chemical loss, the amount of photochemically derived ethane should increase gradually but significantly (by 30\% at 0.5 mbar from southern summer solstice to autumn equinox), resulting in increased emission at $\sim12$ $\mu$m. However, the amount and timing of the chemical response will depend on the chemical and transport time scales, which vary with season, latitude, and atmospheric pressure.  At higher pressures, photochemical variation is generally less significant with longer lags given longer chemical and radiative time scales.  But all these expectations can be modified by dynamics, which can hypothetically alter temperatures and chemical distributions through large-scale circulation patterns or small-scale processes such as convective storms and inertia-gravity waves \citep{conrath1989neptune,fletcher2014neptune,dePater2014neptune}. 

The first inferences of the three-dimensional temperature structure and implied circulation in Neptune's atmosphere were derived from spatially resolved Infrared Interferometer Spectrometer and Radiometer (IRIS) data from Voyager 2 in 1989 \citep{smith1989voyager,conrath1989neptune}. Acquired during Neptune's southern late autumn (solar longitude\footnote{The apparent planet-centered longitude of the Sun, \textit{L$_s$} is used here to represent the seasonal phase over the planet's orbital period.  Cyclic, with values of 0$^{\circ}\leq$ \textit{L$_s$}$<$360$^{\circ}$, \textit{L$_s$} is defined as 0$^{\circ}$ at the time of the planet's northern spring equinox, 90$^{\circ}$ at the northern summer solstice, 180$^{\circ}$ at the northern autumnal equinox, and 270$^{\circ}$ at the northern winter solstice.} \textit{L$_s$}$\sim$236$^{\circ}$), the observations showed anomalously cool mid-latitudes in the upper troposphere that were attributed to local upwelling and adiabatic cooling, with compensating sinking and warming at the equator and poles \citep{conrath1990temperature}.  Later ground based measurements analyzed by \citet{fletcher2014neptune} and \citet{dePater2014neptune} showed this pattern more or less continued near the time of the 2005 southern summer solstice (\textit{L$_s$}=270$^{\circ}$), with a possible increase in the polar temperatures by 2003, remaining warmer until at least 2006. In the stratosphere, Voyager suggested a distinct equatorial maximum in temperatures or acetylene \citep{bezard1991hydrocarbons}, but later observations suggested a rather more latitudinally uniform distribution in ethane in 2003 and 2007 \citep{greathouse2011spatially,fletcher2014neptune}. 

Whether these observations indicate a discrepancy in hydrocarbon distributions, an apparent seasonal change, or a transient meteorological event is unclear.  Nor do these limited observations tell us which state is more typical of Neptune’s atmospheric temperatures and chemical distributions.  

Now, with a body of ground-based imaging spanning 17 years, we take a fresh look at Neptune’s atmospheric temperatures and chemistry across time.  By compiling and analyzing all currently available mid-IR imaging data, gathered from multiple observatories from 2003 to 2020 (Section \ref{sec:data}), we are able to show that sub-seasonal variation is unequivocally present on both regional and global scales (Section \ref{sec:analysisresults}). By additionally utilizing Spitzer IRS and ground-based spectroscopy at wavelengths sensitive to stratospheric temperatures (i.e. the $\sim$17 $\mu$m H$_2$ S(1) quadrupole), we find that temperature changes are likely primarily responsible for Neptune’s observed variability.  We discuss these changes in context of model predictions and possible causes (Section \ref{sec:discuss}).  And finally, we conclude by summarizing our results (Section \ref{sec:conclude}), which altogether suggest that Neptune’s stratosphere is likely more complicated and temporally variable than previously realized.

\section{DATA} \label{sec:data}

To characterize variability in Neptune's mid-infrared emission, we analyzed mid-infrared images (7--25 $\mu$m) from various ground-based facilities, and supplement these observations with spectral data for further temporal and spectral context. A summary of all data sources is listed in Table \ref{tab:datasource}.

\begin{deluxetable*}{ccccccc}[h]
\centering
\tabletypesize{\footnotesize}
\tablecolumns{7} 
\tablewidth{6.in}
\tablecaption{Data Sources}
\tablehead{
 \head{4 cm}{\vspace{-0.15cm} Observatory, Instrument} & 
 \head{1.6cm}{\vspace{-0.15cm} Type } &
 \head{2.1cm}{\vspace{-0.15cm} Plate Scale (arcsec/pixel)$\dagger$} &
 \head{3.3cm}{\vspace{-0.15cm} Year} &
 \head{2.6cm}{\vspace{-0.15cm} Molecular Emission} &
 \head{2cm}{\vspace{-0.15cm} P.I./Reference} 
 }
 \startdata
 Keck, LWS & imaging & 0.0847 & 2003 & CH$_4$, CH$_3$D, C$_2$H$_6$, H$_2$ & 1, 2 \\
Gemini-North, Michelle  & imaging & $\sim$0.099 & 2005 & CH$_4$, C$_2$H$_6$ & 3 \\
Very Large Telescope, VISIR & imaging & 0.075 & 2006, 2008, 2009 & CH$_4$, CH$_3$D, C$_2$H$_6$, H$_2$ & 4, 5, 6, 7 \\
Gemini-South, T-ReCS & imaging &  0.090 &  2007, 2010 & CH$_4$, CH$_3$D, C$_2$H$_6$, H$_2$ & 8, 9\\
Subaru, COMICS & imaging & 0.13 & 2008, 2011*, 2012*, 2020*  & CH$_4$, CH$_3$D, C$_2$H$_6$, H$_2$ & 10\\
Very Large Telescope, VISIR & imaging & 0.0453 & 2018* & CH$_4$, C$_2$H$_6$, H$_2$ & 11 \\ 
& & & & & \\
Very Large Telescope, VISIR & spectroscopy & 0.127 & 2006* & H$_2$ S(1) quadrupole & 4 \\
Gemini-North, TEXES & spectroscopy & $\sim$0.137 & 2007, 2019* & H$_2$ S(1) quadrupole & 12 \\
Spitzer, IRS & spectroscopy & 1.8--4.5 & 2004, 2005, 2006 & all of the above & 13 \\
\enddata
\label{tab:datasource}
\tablecomments{Principal Investigator (program ID) / References: [1] M. Brown (C14LSN); [2] I. de Pater (U38LS) / \citet{dePater2014neptune}; [3] H. Hammel (GN-2005A-DD-10)/ \citet{hammel2007distribution}; [4] Th. Encrenaz (077.C-0571(A))/ \citet{orton2007evidence}, [5] \citet{fletcher2014neptune}; 077.C-0571(A) [6] T. Encrenaz (081C-0496(A);  [7] G. Orton (083.C-0163(A,B)); [8] G. Orton (GS-2007B-Q-47)/\citet{orton2007evidencehotspot};  [9] G. Orton (GS-2010B-Q-42) / \citet{orton2012recovery}  [10] Y. Kasaba (o07150, o08161, o12143) \citet{orton2012recovery}; [11] L. N. Fletcher (0101.C-0044(A)); [12] T. Greathouse (GN-2007B-C-8)/ \citet{greathouse2011spatially}; [13] J. Houck / 10.5281/zenodo.5254503\\  Instruments References: Long Wavelength Spectrometer (LWS), \citep{jones1993keck}; Michelle, \citep{glasse1997michelle}; The VLT Imager and Spectrometer for mid Infrared (VISIR), \citep{lagage2004visir}; the Thermal-Region Camera Spectrograph (T-ReCS), \citep{deBuiz2005Trecs}; and the Cooled Mid-Infrared Camera and Spectrometer (COMICS), \citep{kataza2000comics}
\\$\dagger$ Neptune's angular diameter is $\sim$2.3 arcseconds
\\The asterisks (*) indicate previously unpublished observations.}
\end{deluxetable*}

\subsection{Imaging Data} 
\subsubsection{Aggregate Properties and Classification}\label{sec:groupings}
Combining unpublished observations with archival data, we collected all currently available imaging data that spatially resolved Neptune's disk in the mid-infrared (see Table \ref{tab:datasource}). In total, this amounts to more than 95 images acquired between 2003 and 2020 at wavelengths ranging from 7.7 to 25.5 $\mu$m. Data were acquired using infrared instruments on 8\textendash10 meter class telescopes at multiple ground-based facilities, listed in Table \ref{tab:datasource}.  Images prior to 2011 were previously analyzed and published, in part, in  \citet{hammel2007distribution,orton2007evidence,orton2012recovery,dePater2014neptune,sinclair2020spatial}; and further summarized by \citet{fletcher2014neptune}, who investigated temporal variability between the Voyager-2 encounter (1989) and Neptune's southern summer solstice (2005). All subsequent imaging data are analyzed here for the first time, including Subaru-COMICS images from 2011, 2012, and 2020, and VLT-VISIR images from 2018. Observational details for all individual images are provided in the Appendix.

\begin{figure}
 \centering
    \includegraphics[clip, trim=.04in .05in 0.in 0.0in,scale=.95]{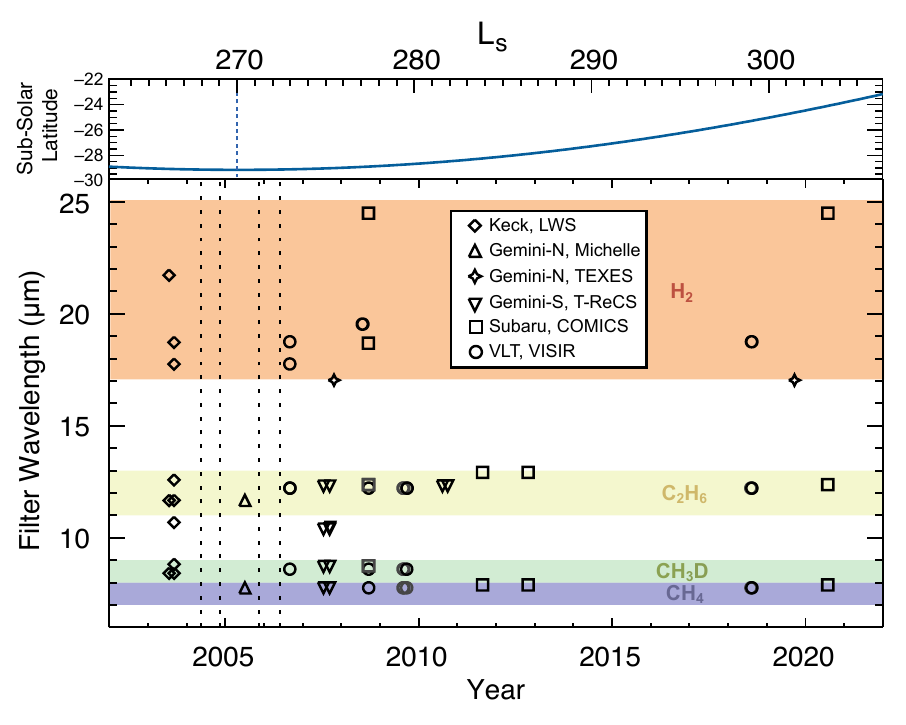}
    \caption{The effective mean wavelength of filtered observations, plotted versus time and solar longitude. The observatory and instrument observations are indicated by the symbols as defined in the key, with colors shading the groupings by sensitivity to H$_2$, C$_2$H$_6$, CH$_3$D, and CH$_4$, as discussed in the text. The vertical dotted lines mark the dates of Spitzer-IRS observations. The top panel displays the corresponding sub-solar latitudes and solar longitude (\textit{L$_s$}), with the southern summer solstice (\textit{L$_s$}=270$^{\circ}$) indicated with the blue dotted-line line.}
        \label{fig:allobslamfig}
\end{figure}

\begin{figure*}
 \centering
    \includegraphics[clip, trim=0in .00in 0.in 0.0in,scale=.78]{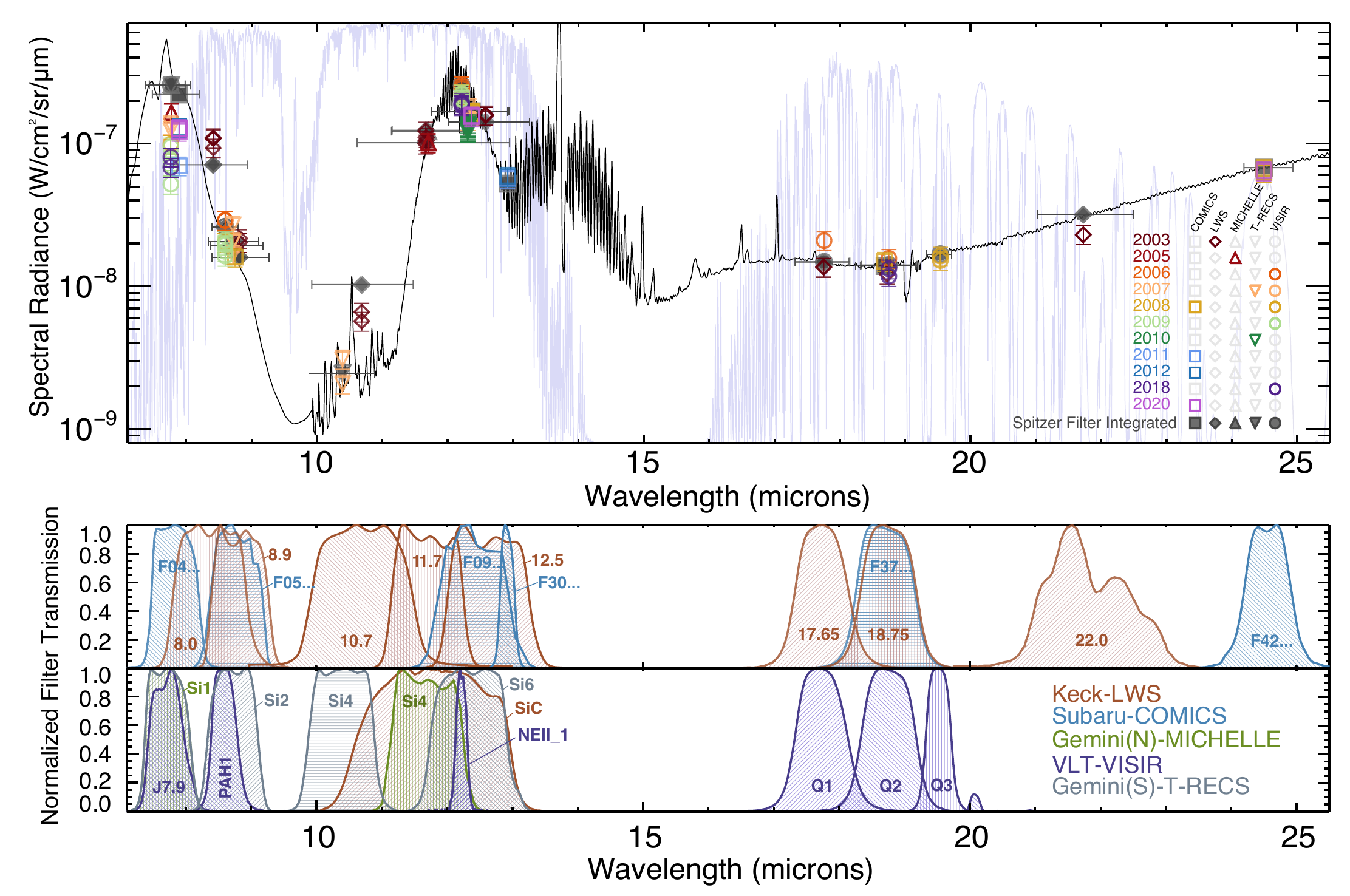}
    \caption{(top) Neptune's mid-infrared spectrum (black) from the Spitzer Infrared Spectrometer (IRS) in 2005 compared to ground-based imaging data. Disk-integrated radiances of each ground-based image are plotted upon the Spitzer spectrum, with the color and symbol indicating the year and instrument. Equivalent filter-integrated radiances derived from Spitzer IRS spectra are shown in dark-gray solid symbols for comparison. The light purple line represents the atmospheric transmission, scaled between 0 and $\sim$1. Note that there is generally good agreement between Spitzer observations and the ground-based observations across most of the spectrum above 8 $\mu$m. Below 8 $\mu$m,  the observations are spread over a greater range of values, but the atmospheric transmission is also relatively lower, which can affect the calibrations. (bottom) Corresponding transmission curves for each filter, as labeled, with colors indicating the instrument.
        \label{fig:filters}}
\end{figure*}

Collectively, images were acquired in 26 different bandpass filters\footnote{Additionally, a broad N-band filter (7.70--12.97 $\mu$m, central wavelength of 10.39 $\mu$m) was used solely for spectroscopy acquisition by Gemini-T-ReCS in 2007. Given its broad spectral range, large calibration uncertainty, and single usage, we do not include the filter and lone image in our analysis, but we display and list it for completeness in the Appendix.} transmitting at wavelengths within the N-band (7--13 $\mu$m) and Q-band (17--25 $\mu$m), as summarized in Table \ref{tab:filters} and Figures \ref{fig:allobslamfig} and \ref{fig:filters}. Clustered at wavelengths of enhanced telluric transparency, these filters sense Neptune's emission spectrum in just a few distinct spectral regions, with sensitivity to different molecular transitions and pressures. 

\begin{figure*}
 \centering
    \includegraphics[clip, trim=0in .00in 0.in 0.0in,scale=.41]{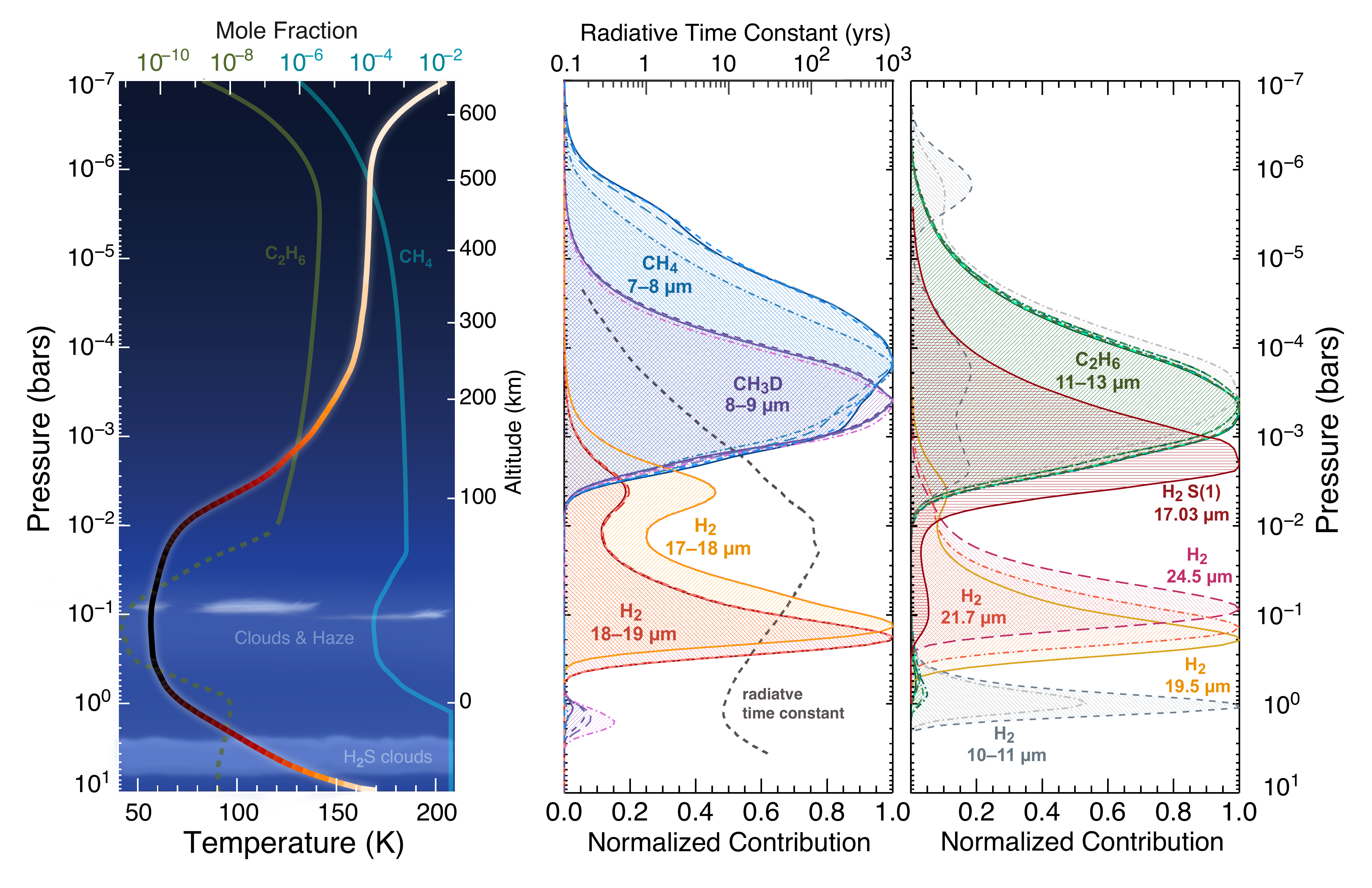}
    \caption{Average vertical profiles and contribution functions for Neptune's atmosphere. (Left) Mean temperature profile from \citet{moses2018seasonal} and \citet{greathouse2011spatially}, with methane (cyan curve) and ethane (olive curve) profiles from \citet{moses2018seasonal}.  From the tropopause minimum, the temperatures increase throughout the lower stratosphere, becoming nearly isothermal by 10$^{-5}$ bar.  Ethane is poorly constrained in the troposphere (dashed curve), but is thought to increase with height above its condensation region in the lower stratosphere (solid curve), as methane decreases in abundance. (Right) Computed from the Jacobian of the temperature, contribution functions for each filter showing the pressures from which the observed emission originates. Similar wavelengths have similar contribution functions, which are labeled and colored by group or filter for clarity, with line styles indicating the instrument as VISIR (solid), COMICS (long dash), Keck (dot-dash), and T-ReCS/Michelle (short dash). The 10--11 $\mu$m and 17--25 $\mu$m H$_2$ filters sense a wide range of pressures, including most deeply, while the discrete H$_2$ S(1) line senses pressures of several millibars.  C$_2$H$_6$, CH$_3$D, and CH$_4$ filters sense progressively lower pressures. Curves are shown for an emission angle of 45$^{\circ}$, but shift upward at larger emission angles (i.e., towards the limb) and downward at smaller angles (i.e., nearer nadir), generally varying by less than $\pm$ 20\% in pressure; likewise, in the bimodal curves for the hydrogen-sensing filters, the relative contributions from the lower pressure maxima ($\sim$5 mbar) increase with the emission angle. The radiative time constant from \citet{li2018high} is also shown for context.}
        \label{fig:temp_conts}
\end{figure*}

For clarity, we broadly sort the majority of the images into the following four distinct groups defined by their passband and, correspondingly, the molecules and pressures they sense: 
\begin{itemize}
  \item 17--25-$\mu$m sensing emission from molecular hydrogen (H$_2$), primarily at pressures near the tropopause ($\sim$80--200 mbar). [15 images]
  \item 11--13-$\mu$m, sensing emission primarily from stratospheric ethane (C$_2$H$_6$), with maximum contributions from $\sim$0.1--2 mbar. [42 images]
  \item 8--9-$\mu$m, sensing emission from stratospheric methane and its isotopologue, deuterated methane (CH$_3$D), primarily from $\sim$0.1--1 mbar. [17 images]
  \item 7--8-$\mu$m, sensing emission from stratospheric methane (CH$_4$) primarily from $\sim$0.01--1 mbar. [15 images]
\end{itemize}

Additionally, there are six images with filtered central wavelengths of 10--11 $\mu$m that are weakly sensitive to both stratospheric hydrocarbons (mainly ethane and ethylene (C$_2$H$_4$)) and H$_2$ emission from deeper in the upper troposphere (at $\sim$1 bar).  However, given their broad but weak sensitivity, low signal-to-noise ratio (SNR), and limited temporal coverage, these images provide few constraints and do not factor prominently into our analysis. 


The radiances at mid-infrared wavelengths are fundamentally determined by the combination of the kinetic temperature of the atmosphere and the abundance of the emitting molecules at the pressures sensed. The abundance of hydrogen is assumed to be well-mixed and known in the atmosphere, and thus emission from H$_2$ at 17--25 $\mu$m provides an unambiguous indication of the atmospheric temperatures in the upper troposphere.  However, since the abundance of hydrocarbons are likely spatially and temporally variable given chemical sources, sinks, and transport \citep{moses2018seasonal,moses2020icegiantchem}, variation in the observed emission from methane (7--8 $\mu$m), deuterated methane (8--9 $\mu$m), and ethane (11--13 $\mu$m) alone cannot differentiate between variation in stratospheric temperature or composition.  To help resolve this ambiguity, we include additional spectroscopic data sensitive to the stratospheric temperatures, as discussed in Section \ref{sec:specdata}.

\begin{deluxetable*}{cccccc}
\centering
\tabletypesize{\footnotesize}
\tablecolumns{6} 
\tablewidth{6.in}
\tablecaption{Filter Properties}
\tablehead{
 \head{3.3cm}{\vspace{-0.15cm} Observatory-Instrument} & 
 \head{2.2cm}{\vspace{-0.15cm} Filter} &
 \head{1.5cm}{\vspace{-0.15cm} Passband ($\mu$m)} &
 \head{1.6cm}{\vspace{-0.15cm} Mean Wavelength ($\mu$m)}& 
 \head{1.9cm}{\vspace{-0.15cm} Mean Atmospheric Transmission} & 
 \head{4.9cm}{\vspace{-0.15cm} Year of Neptune Observations     [Number of Images]}}
 \startdata 
Keck-LWS   & 8.0   & 7.81--8.93 & 8.41 &  82.4$\%$  & 2003 [3] \\
Keck-LWS   & 8.9   & 8.39--9.26  & 8.82 & 92.5$\%$ &  2003 [2] \\
Keck-LWS   & 10.7  & 9.92--11.5  & 10.68 & 94.1$\%$ &  2003 [2] \\
Keck-LWS   & 11.7  & 11.1--12.2  & 11.67 & 96.6$\%$ &  2003 [2] \\
Keck-LWS   & SiC   & 10.6--12.9  & 11.66 & 95.3$\%$ &  2003 [2] \\
Keck-LWS   & 12.5  & 12.0--13.2  & 12.58 & 89.9$\%$ &  2003 [2] \\
Keck-LWS   & 17.65 & 17.3--18.2  & 17.75 & 66.1$\%$ &  2003 [2] \\
Keck-LWS   & 18.75 & 18.3--19.2  & 18.72 & 67.3$\%$ &  2003 [1]\\
Keck-LWS   & 22.0  & 21.0--22.5  & 21.73 & 42.6$\%$ &  2003 [1] \\
Gemini(N)-MICHELLE & Si1-7.9  & 7.37--8.07 & 7.76 & 45.0$\%$ &  2005 [2] \\
Gemini(N)-MICHELLE  & Si5-11.6 & 11.1--12.2 &  11.66 & 96.2$\%$ &  2005 [3]  \\
Gemini(S)-T-ReCS & Si1-7.9  & 7.37--8.07 & 7.76 & 40.1$\%$ & 2007 [2]  \\
Gemini(S)-T-ReCS & Si2-8.8     & 8.34--9.11 & 8.72 & 93.6$\%$ & 2007 [3]  \\
Gemini(S)-T-ReCS & Si4-10.4    & 9.87--10.9 & 10.39 & 93.7$\%$ & 2007 [3] \\
Gemini(S)-T-ReCS & Si6-12.3    & 11.7--12.9 & 12.31 & 94.1$\%$ & 2007 [2], 2010 [10] \\
VLT-VISIR & J7.9      & 7.43--7.98  & 7.76 & 39.3$\%$ & 2008 [1], 2009 [4], 2018 [2]\\
VLT-VISIR & PAH1      & 8.39--8.79  & 8.60 & 92.0$\%$ & 2006 [2], 2008 [1], 2009 [5]\\
VLT-VISIR & NEII$\_$1 & 12.1--12.3  & 12.22 & 97.7$\%$ & 2006 [2], 2008 [1], 2009 [6], 2018 [4]\\
VLT-VISIR & Q1      & 17.3--18.1  & 17.76 & 61.6$\%$ & 2006 [1]\\
VLT-VISIR & Q2        & 18.3--19.2  & 18.75 & 63.8$\%$ & 2006 [1], 2018 [2] \\
VLT-VISIR & Q3        & 19.3--19.7  & 19.54 & 46.7$\%$ &  2008 [2] \\
Subaru-COMICS  & F04C07.80W0.70 & 7.48--8.20 &  7.89 & 53.0$\%$ & 2011 [1] ,2012 [1], 2020 [2]\\
Subaru-COMICS  & F05C08.70W0.80 & 8.38--9.17 & 8.75 & 92.6$\%$ &  2008 [2]\\
Subaru-COMICS  & F09C12.50W1.15 & 11.9--12.9 & 12.37 & 92.9$\%$ &  2008 [4], 2020 [2]\\
Subaru-COMICS  & F30C12.81W0.20 & 12.8--13.0 & 12.92 & 89.5$\%$ &  2011 [1], 2012 [1]\\
Subaru-COMICS  & F37C18.75W0.85 & 18.2--19.2 & 18.68 & 66.9$\%$ &  2008 [1]\\
Subaru-COMICS  & F42C24.50W0.80 & 24.2--24.9 & 24.49 & 52.7$\%$ &  2008 [2], 2020 [2]\\
\\
\enddata
\label{table:filters}
\tablecomments{The Si1-7.9 used with Michelle and T-ReCS are identical.  Passbands are defined by the minimum and maximum wavelengths for which the filter's laboratory transmission (i.e. neglecting atmospheric transparency) is 50$\%$ or greater. Mean wavelength calculations include atmospheric transmission and assume average observing conditions.  Likewise, mean atmospheric transmissions assume roughly climatological averages of precipitable water vapor (PWV) and average airmass of the observations.}
\label{tab:filters}
\end{deluxetable*}

The atmospheric pressures sensed by each filter are estimated using radiative-transfer modelling (NEMESIS, \citet{irwin2008nemesis}), assuming vertical temperature and chemical profiles based on \citet{greathouse2011spatially} and \citet{moses2018seasonal}. As illustrated in Figure \ref{fig:temp_conts}, these contribution functions also group by filter passband, with some overlap among the defined groups. The precise shapes and locations of these contribution functions are dependent on the assumed atmospheric model and observational emission angle, but broadly show sensitivity to gases at a range of pressures extending from the upper troposphere to the stratosphere. The hydrogen filters sense the deepest, but also have additional minor contributions from a wide range of pressures. As noted by \citet{fletcher2014neptune}, the contributions functions for the H$_2$ filters are somewhat bimodal, with contributions from lower pressures increasing with increasing emission angle. The ethane, deuterated methane, and methane filters sense progressively lower pressures in the stratosphere, with significant contribution from the strongest methane lines continuing at pressures less than 10$^{-5}$ bar.

\begin{figure}
 \centering
    \includegraphics[clip, trim=0in .00in 0.in 0.0in,scale=.8]{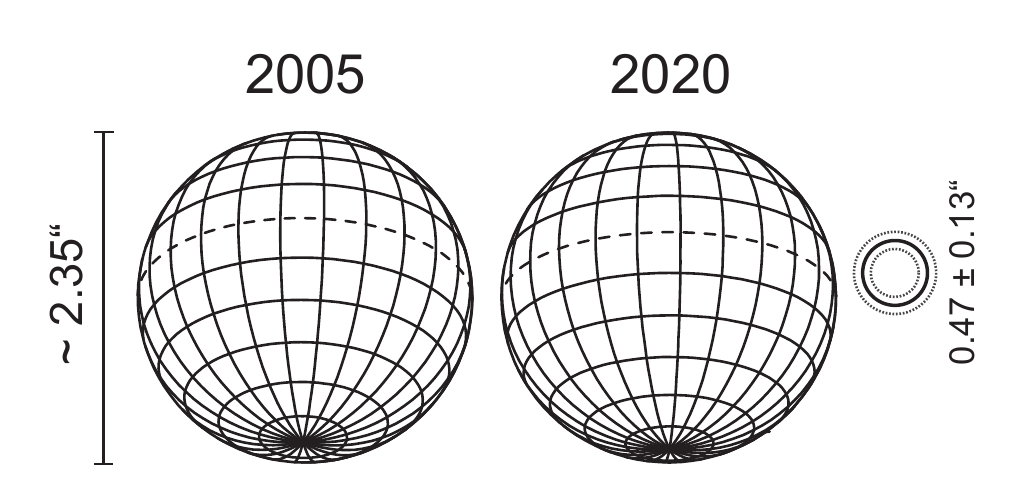}
    \caption{Disk geometry for Neptune in 2005, at southern summer solstice (\textit{L$_s$}=270$^{\circ}$), and 2020 (\textit{L$_s$}=302$^{\circ}$), showing the maximum difference in observing geometries across the data set. South is down and contours represent 15$^{\circ}$ intervals, with the equator marked by the dashed contour. The vertical bar (far left) depicts the $\sim$2.35-arcsecond angular resolution of Neptune's disk, and the concentric disks (far right) illustrate the FWHM of calibration stars (average $\pm$ standard deviation), indicative of the typical image resolution.}
        \label{fig:geom}
\end{figure}

The effective spatial resolution of the images vary owing to differences in the telescope apertures, focal lengths, detector plate scales, filtered wavelengths, and atmospheric seeing. The mean spatial resolution amounts to Neptune's disk subtending $\sim$28 pixels at 4.4$^{\circ}$ of latitude per pixel, although the highest resolution images---acquired in 2018 with VLT-VISIR at a plate scale of 0.0453 arcsecs/pixel---offer twice this resolution. Subaru-COMICS images have the poorest resolution at only 0.13 arcsecs/pixel, subtending closer to 7$^{\circ}$ of latitude per pixel at the disk center. The effective seeing disks varied between approximately 0.3'' and 0.75'', with an average of 0.47'' based on the full-width-half-max of the calibration stars (see Figure \ref{fig:geom}). This limited resolution has the effect of blurring the edges of the disk with the sky, artificially suppressing the emission for the outer 10--30$\%$ of the disk. This must be kept in mind when comparing and interpreting observations. 

Over the seventeen years covered by ground-based thermal imaging, the sub-observer latitude changed by less than 4.4$^{\circ}$, from 29.1$^{\circ}$ S to 24.7$^{\circ}$ S (see Figure \ref{fig:geom}). This small change in sub-observer angle amounts to a change of two pixels or fewer across all images, which implies that most observed changes cannot be attributed to changes in the observing geometry. Changes in emission angle can potentially modify the perceived radiances along the extreme northern and southern edges of the disk, but this is unlikely to be significant over the span of our data (although it will become significant as the south pole appears closer to the limb in the coming years). Likewise, changes in the sub-solar latitude are modest, with a minimum value of 29.2$^{\circ}$S in 2005 images (near southern summer solstice, \textit{L$_s$} = 270$^{\circ}$) and maximum of 24.1$^{\circ}$ S in 2020 images (\textit{L$_s$} $\approx$ 303$^{\circ}$).

\subsubsection{Acquisition, Calibration, and Uncertainties} \label{sec:calibration}
As standard for mid-infrared observations, images were acquired using the chopping and nodding technique to effectively remove the contribution of the sky and telescope's thermal emission from the image.  By this process, hundreds of $\sim$100 milliseconds exposures--short enough to preclude detector saturation--were combined to build up a sufficient SNR for the target.  Total on-target integration times ranged from several minutes to roughly 30 minutes, depending on the filter and observatory.  In most, but not all cases, accompanying observations of standard stars were acquired for flux calibration. 

Radiances for each pixel were calibrated using radiance conversion factors derived from observations of standard stars.  These conversion factors are derived by comparing the observed stellar signal in digital units to the expected flux densities for standard stars.  Stellar flux densities are often provided by the observatories, but we calculated stellar flux densities from Cohen stellar spectra \citep[\textit{e.g.},][]{cohen1995spectral} for the filter transmissions assuming typical atmospheric conditions for the observatories \citep{noll2012atmospheric}. These flux densities were in very close agreement---within 3$\%$ or less---of values currently maintained by Gemini\footnote{\url{http://www.gemini.edu/sciops/instruments/mir/Cohen_list.html}} and the European Southern Observatory\footnote{\url{https://www.eso.org/sci/facilities/paranal/instruments/visir/tools/zerop_cohen_Jy.txt}}. The standard stars used are listed along with each Neptune observation in the Appendix.  To account for differences in airmass between Neptune and the calibration star, airmass extinction corrections were calculated from observations of stars repeated at different airmasses, when available, or otherwise estimated from values of \citet{krisciunas1987atmospheric}.  Extinction corrections typically amounted to less than 5$\%$. In the few cases when stellar observations were absent, calibrations were estimated from observations on proximate nights.

Resulting uncertainties in the image radiances are dominated by uncertainties in the radiance conversion factors, which generally vary by star, wavelength, and time. Statistical analysis of mid-infrared conversion factors at VLT by \citet{dobrzycka2008calibrating} has shown typical temporal variability in these conversion factors of up to 10\% in filters at 8.6 $\mu$m and 11.2 $\mu$m--where atmospheric transparency is greatest--and less than 20\% for the Q2 filter--where telluric water vapor decreases transparency.  With minutes to hours typically lapsing between Neptune and stellar observations, errors can become significant, particularly at wavelengths at which the emission from the sky is greatest (see Figure \ref{fig:filters} and atmospheric transmissions in Table \ref{tab:filters}).  

Similar wavelength dependence in the uncertainties can be inferred from variability in repeated observations of Neptune on relatively short timescales. For example, six VISIR observations acquired at 12.2 $\mu$m (NeII$\_$1) on five nights in August and September 2009 varied by less than $\pm$ 10\%, with a standard deviation of only 6\%. During the same period, five images at 8.9 $\mu$m varied by less than 14$\%$ with a standard deviation of 10\%, while the methane images at 7.8 $\mu$m---where the atmospheric transparency is less than 50\%---varied by $\pm$ 30\% of their mean, with a standard deviation of about 25$\%$.  

Differences in the spatial resolution and filter passbands can also lead to relative differences in the inferred trends across the disk, even at similar wavelengths. For example, Neptune was observed by both VISIR (12.2 $\mu$m, NeII$\_$1) and COMICS (12.4 $\mu$m, F09C12.50W1.15) on the same night in September 2008, but the VISIR image shows significantly greater limb brightening (see Appendix Figure \ref{fig:ethane_imgs}).  Given the contemporaneity, the difference is observational, likely owing to the relatively finer plate scale resolution, lower airmass, and narrower passband (centered on the ethane emission) of the VISIR observation.  As such, even images at very similar wavelengths should be compared with some caution. 

Aside from the conversion factors, repeated attempts to calibrate data with plausible alternative choices for parameters regarding the atmospheric model, sky subtraction, area of aperture photometry, and airmass correction resulted in changes in radiance of less than 5\%.  As a whole, the data do not exhibit any obvious trends in relative radiances versus airmass or precipitable water vapor. Altogether, we estimate uncertainties of up to 10\% in radiance for ethane images (11--13 $\mu$m), 20\% for the CH$_3$D images (8--9 $\mu$m), 25\% for methane images (7--8 $\mu$m), and, conservatively, 30\% for hydrogen images (17--25 $\mu$m). 

\subsection{Spectroscopic Data}\label{sec:specdata}

To help interpret variations in stratospheric emission, we additionally analyzed spectroscopic observations of the H$_2$ S(1) quadrupole at $\sim$17.03 $\mu$m \citep{fletcher2018hydrogen}, as summarized in Table \ref{tab:specdat}. In contrast to the stratospheric imaging data, which are sensitive to both temperature and composition, measurement of this hydrogen emission line serves as a nearly unequivocal indicator of the lower stratospheric temperatures (see Figure \ref{fig:temp_conts}), assuming the para-hydrogen fraction is in local equilibrium. 

Spectra measuring Neptune's H$_2$ S(1) quadrupole emission as a function of latitude were acquired with VLT-VISIR in September 2006.  With the 1''-wide slit aligned with the central meridian, observations spanned over 8 hrs and airmasses from 1.01 to 2.7, with a cross-dispersion spatial resolution of 0.127''/pixel and at a spectral resolution of R$\sim$14000.  Spectra were calibrated via comparison to observations of HD211416.  Random errors at these wavelengths are estimated to be less than 20\%, but we find a larger systematic error in the calibration, as discussed in Section \ref{sec:quadrupole}. 

Subsequent observations of the quadrupole emission were made in 2007 and 2019 from Gemini North using TEXES, the Texas Echelon cross-dispersed Echelle Spectrograph \citep{lacy2002texes} as part of larger spectroscopic studies. The 2007 observations were previously analyzed and published by \citet{greathouse2011spatially}, while the 2019 quadrupole data are presented here for the first time.  Both TEXES observations used a scanning technique in which the 0.8'' slit was sequentially positioned over different portions of the disk, gathering spectra at each 0.25'' step that can then be combined to effectively provide a spatially resolved spectral-image of the H$_2$ S(1) quadrupole emission at R$\sim$80000.  The detector has a plate scale of $\sim$0.137''/pixel, but the spatial resolution is effectively diffraction-limited ($\sim$0.53'') in the direction along the slit. In the perpendicular scanning direction, the spatial resolution of the composite image is effectively limited by the slit width, as all the disk falling within the 0.8'' slit (perpendicular to the spatial dimension) will contribute to the signal. This means the observed signal can potentially be more diluted towards the edges of Neptune's disk more in the scanning direction if the slit width is only partly filled. The scanning direction was parallel to Neptune's rotational axis in 2007, but perpendicular to it in 2019, and this difference is considered when interpreting the data.  In each case, radiance calibration was achieved by accompanying observations of the thermal blackbody emission from a black, metal plate at ambient temperature within the observatory. Following \citet{greathouse2011spatially}, uncertainties are taken to be 15\% or less. 


Finally, for further temporal and spectral context, we relate our ground-based observations to Spitzer-IRS spectra of Neptune observed between 2004 and 2006 \citep{rowe2021neptune}. The Spitzer-IRS observations are essentially disk-integrated, but offer more precisely calibrated radiances over the full spectral range of our filtered images at four different times---first in May 2004 and then again 6, 12, and 24 months later. The Long-High (LH) module data provides measurements of the disk-integrated H$_2$ S(1) quadrupole emission at R$\sim$600, to which we compare our ground-based spectra.  For comparison to filtered images, we computed filter-integrated fluxes of the Spitzer-IRS spectrum for each imaging filter passband with mean observing conditions. Errors in the spectral radiances alone are taken to be less than 6\% \citep{rowe2021neptune}. Tables listing the disk-integrated radiance for each image, along with the ratio of each relative to filter-integrated Spitzer radiances, are provided in the Appendix.

\begin{deluxetable}{cccc}[th!]
\centering
\tabletypesize{\footnotesize}
\tablecolumns{4} 
\tablewidth{3.4in}
\tablecaption{H$_2$ S(1) Spectroscopy}
\tablehead{
\head{1.9cm}{\vspace{-0.1cm} Date  (yyyy-mm-dd)} & \head{2.4cm}{\vspace{-0.0cm}Observatory-Instrument} &
\head{1.5cm}{\vspace{-0.0cm}Spectral Range        ($\mu$m)} &
\head{1.6cm}{\vspace{-0.0cm}  Spectral Resolution ($\lambda/\delta\lambda$)}}
 \startdata 
 2004-05-15  &  Spitzer-IRS  & 5.2--36.8  &   600*  \\ 
 2004-11-15  &  Spitzer-IRS &  5.2--36.8 & 600*    \\
 2005-11-19     &  Spitzer-IRS  &  5.2--36.8    &   600*     \\
 2006-05-31     &  Spitzer-IRS  &  5.2--36.8    &   600*     \\
 2006-09-04     &  VLT-VISIR  &  17.004--17.056      &  14000   \\   
 2007-10-24     &  Gemini(N)-TEXES &  17.043--17.036 &  80000 \\    
 2019-09-19     &  Gemini(N)-TEXES  &  16.986--17.134&  80000   \\ 
\enddata
\label{tab:specdat}
\tablecomments{Spitzer-IRS data are effectively disk-integrated; VLT-VISIR data are latitudinally resolved; and the Gemini-TEXES data are effectively disk-resolved.\\
$*$For wavelengths $<$9.89 $\mu$m, the Spitzer low-resolution modules provided a resolution R$\sim$60--127.}
\end{deluxetable}

\begin{figure*}
 \centering
    \includegraphics[clip, trim=0in .00in 0.in 0.0in,scale=.8]{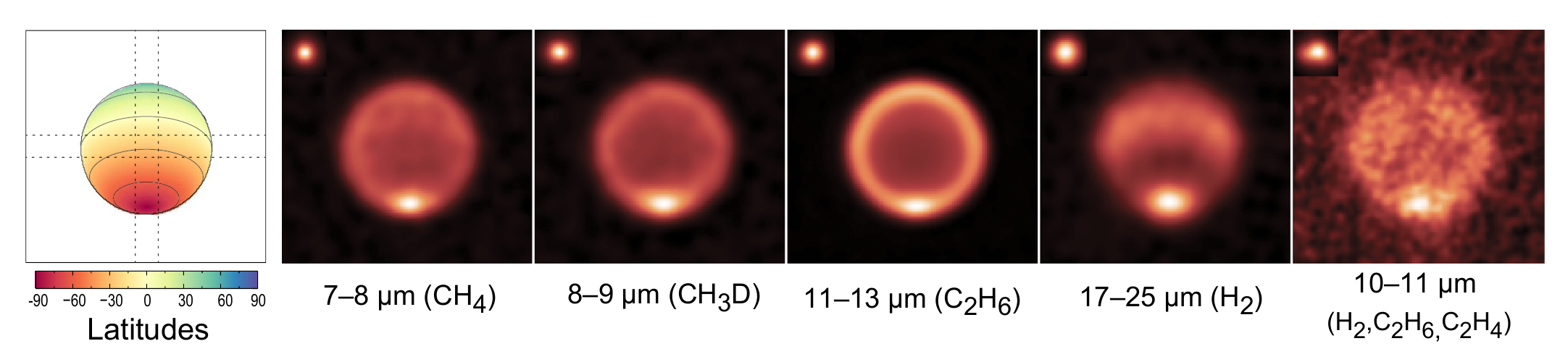}
    \caption{Averaged images for each distinct spectral group---methane, deuterated methane, ethane, and hydrogen, plus the limited 10--11 $\mu$m images, as noted in the text. Insets of averaged stellar calibrators, indicative of the average spatial resolution of each group, are included to the upper left of each disk. (Far left) Corresponding averaged disk geometry. The color scale indicates the latitudes, with 0$^{\circ}$, 30$^{\circ}$ S, and 60$^{\circ}$ S marked with solid lines. The dotted lines mark the boundaries of the cross-sectional areas taken to define the meridional- and approximately zonal-radiance profiles, as discussed in the text of Section \ref{sec:spatialvstime}.}
        \label{fig:ave_imgs}
\end{figure*}

\section{Analysis and Results}\label{sec:analysisresults}
Images were sorted into distinct spectral groups, defined by wavelengths and based on their sensitivities and typical image characteristics (see Section \ref{sec:groupings}). For each group, images were then evaluated to display mean characteristics and temporal behavior, including variation in spatially distributed and disk-integrated radiances. We compared calibrated imaging repeated with the same instrument and filter over time to reveal absolute changes in Neptune's mid-infrared emission at different wavelengths.  In addition, we evaluated and compared relative changes within each group of images by expressing their radiances as ratios relative to the equivalent 2005 radiances inferred from the Spitzer-IRS spectrum.  By examining these relative changes, we were able to evaluate trends in the aggregate data and mitigate the sparse sampling in time.  

\subsection{Image Mean Characteristics}\label{sec:meanimges}
Average images displaying the distinct characteristics of these groups are shown in Figure \ref{fig:ave_imgs}.  For these averages, images were normalized in size to that of the finest resolution data (i.e., to a disk with an equatorial width of 51.8 pixels, as imaged by VLT-VISIR on August 13, 2018) and normalized in relative radiance to the equivalent disk-integrated values inferred from the Spitzer spectrum. By combining and averaging the images as such, we greatly improve the SNR to reveal the temporally averaged structure across the disk.  While all images show enhanced emission at the south pole---interpreted as an indicator of vertically extensive downwelling \citep{dePater2014neptune}-- each group exhibits different spatial variation in radiance across the disk that can be attributed to differences in the pressures and molecules sensed within the structure of Neptune's stratosphere and upper troposphere \citep[\textit{e.g.},][]{hammel2006mid, hammel2007distribution,orton2007evidence, fletcher2014neptune, dePater2014neptune, sinclair2020spatial}. 

The Q-band images (17-25 $\mu$m) show the enhanced H$_2$ emission from the equator and pole relative to mid-latitudes, mostly from pressures between 30 and 300 mbar. This latitudinal pattern is consistent with the equator and pole being regions of descending and warming while mid-latitudes being regions of upwelling and adiabatic cooling, as first inferred from Voyager IRIS data \citep{conrath1990temperature,fletcher2014neptune}. Suggestions of a similar pattern appear in the averaged 10--11-$\mu$m image, but these images suffer from very low SNR given the extremely weak radiances at these wavelengths (see Figure \ref{fig:filters}) and sense a broad range of pressures, making their physical interpretation less clear.

Broadly sensing emission from ethane at pressures of 10$^{-4}$--10$^{-3}$ bars, the average ethane images are dominated by the strongest limb-brightening, attesting to the combination of a weakly negative temperature lapse rate (i.e. increasing temperature with height) and increasing ethane mixing fraction with height at these pressures (see Figure \ref{fig:temp_conts}). The strongest limb brightening is seen in the VISIR imaging, likely owing to the high spatial resolution and narrow passband.  A subtle latitudinal gradient is also evident, with generally lower radiances at southern mid-latitudes, but variable in time as discussed in Section \ref{sec:spatialvstime}.  Given the excellent sky transparency and strong signal at these wavelengths, the 11--13 $\mu$m ethane images have the greatest SNR. 

Despite sensing similar pressures as the ethane images, the average CH$_3$D images exhibit a lower SNR and comparatively less limb brightening than the ethane emission. This is consistent with a lower abundance of CH$_3$D---inferred to be only 0.0264$\%$ as abundant as methane \citep[with uncertainties exceeding 50\%; \textit{e.g.}, ][]{lecluse1996deuterium,feuchtgruber2013d,fletcher2010neptune}---and expectations that CH$_3$D, like methane, decreases in abundance with height at these pressures, \citep{moses2018seasonal}, partly countering the effects of the negative lapse rate.  

Likewise, the 7--8 $\mu$m images sensing methane ($^{12}$CH$_4$ and $^{13}$CH$_4$) show slightly less limb brightening, as these images sense even higher altitudes where the lapse rate approaches isothermal.   Despite the relatively strong emission from abundant methane, these images suffer from poor atmospheric transparency, which reduces the SNR and increases the calibration uncertainty compared to the ethane images.  Careful inspection of the images nonetheless reveals latitudinal variation with subtle signs of possible zonal banding as previously noted by \citet{sinclair2020spatial}, suggesting that dynamical processes shaping zonal variations continue to be important up to at least 10$^{-5}$ bars.

\subsection{Spatial Variation in Time}\label{sec:spatialvstime}
We evaluate spatial-temporal variation by comparing images and profiles of the radiances across the disks at different times. By comparing profiles along the polar and perpendicular axes of Neptune's disk (see Figure \ref{fig:ave_imgs}), we separate trends in latitude from the trends in the emission angle (which are indicative of vertical gradients), while displaying differences due to seeing and instrumental diffraction that can significantly alter the spatial brightness distribution. To improve SNR and better reveal trends in time, annually averaged images were also calculated by averaging similar or appropriately normalized images (as discussed below) acquired over different nights within a calendar year. Details on the individual images can be found in the Appendix Tables \ref{table:ethanetab}, \ref{tab:images_h2_ch4_ch3d_imgs}, and \ref{table:methanetab}.

\begin{figure*}
    \centering
    \includegraphics[clip, trim=.0in 0.0in 0.in 0.0in,scale=0.88]{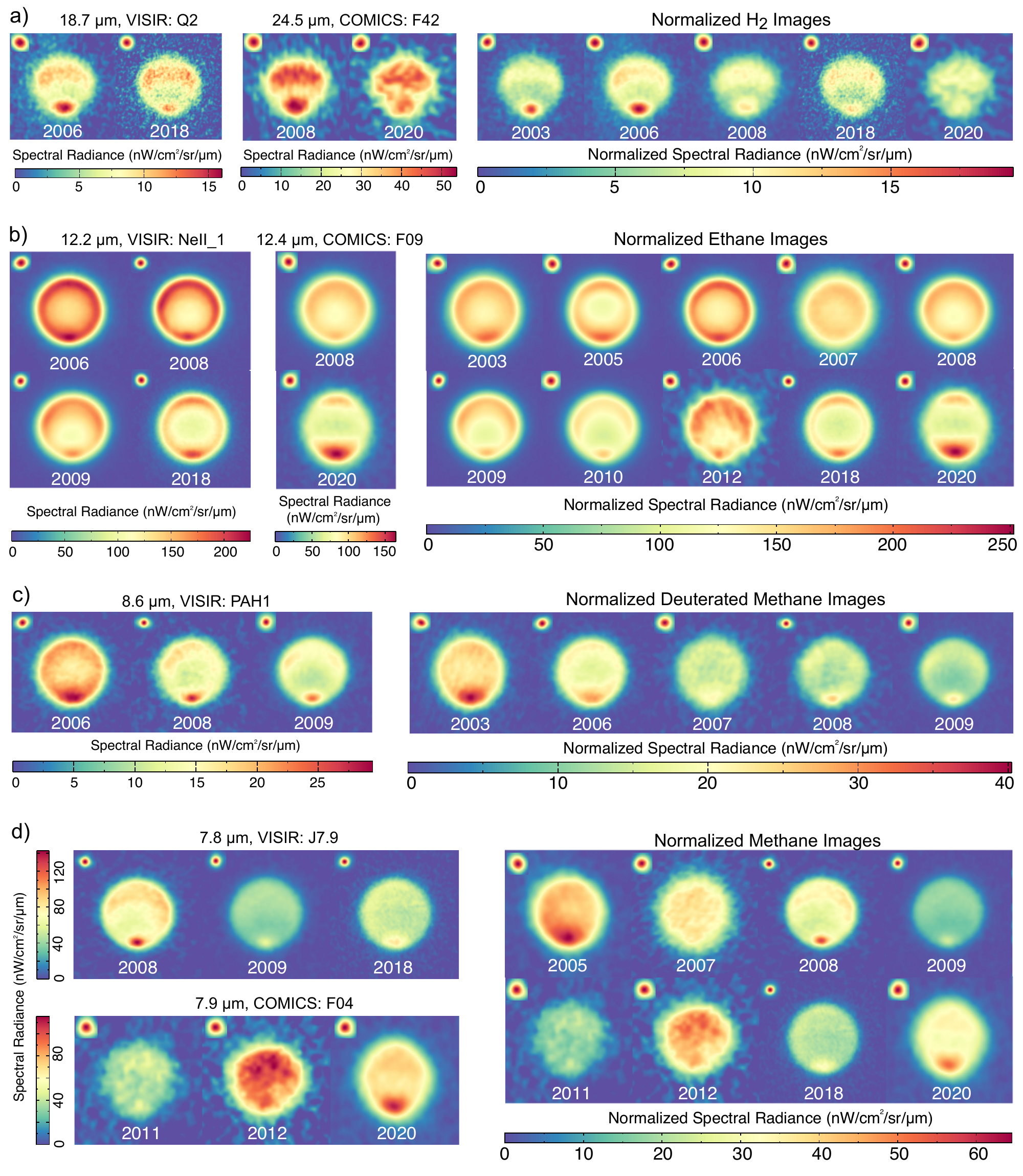}
    \caption{Comparison of observed radiances in images from different years. On the left, images with identical filters are compared over time in the different spectral regions as described in the text for images sensitive to a) H$_2$, b) C$_2$H$_6$, c) CH$_3$D, and d) CH$_4$.  On the right, image sequences including all filters, normalized relative to Spitzer disk-filter-integrated radiances and scaled by a single fiducial value for each spectral group---specifically the Spitzer filter-integrated values for Q2 (H$_2$), NeII$\_$1 (C$_2$H$_6$),  PAH1 (CH$_3$D), and J7.9 (CH$_4$). Variation in methane images d) are largest, and indicate significant global changes in radiances or larger calibration uncertainties than expected. Uncertainties are assumed to be up to 10\% for ethane images (11--13 $\mu$m), 20\% for the CH$_3$D images (8--9 $\mu$m), 25\% for methane images (7--8 $\mu$m), and no more than 30\% for hydrogen images (17--25 $\mu$m). }
    \label{fig:img_comparison}
\end{figure*}

Beginning with the Q-band images, Figure \ref{fig:img_comparison}a shows a direct comparison between pairs of annually averaged images acquired 12 years apart using the same filters:  first by VISIR (Q2, 18.8 $\mu$m) in 2006 and 2018, and then by COMICS (F42, 24.5 $\mu$m) in 2008 and 2020. These images sense thermal emission from well-mixed hydrogen near the tropopause. Though the image quality differs, we see that the basic structure of thermal emission from the 100-mbar level remains unchanged over the intervening 12 years.  The south pole, however, appears significantly brighter in both the 2006 and 2008 images compared to 2018 and 2020.  

\begin{figure*}
    \centering
    \includegraphics[clip, trim=.0in 0.0in 0.in 0.0in,scale=0.84]{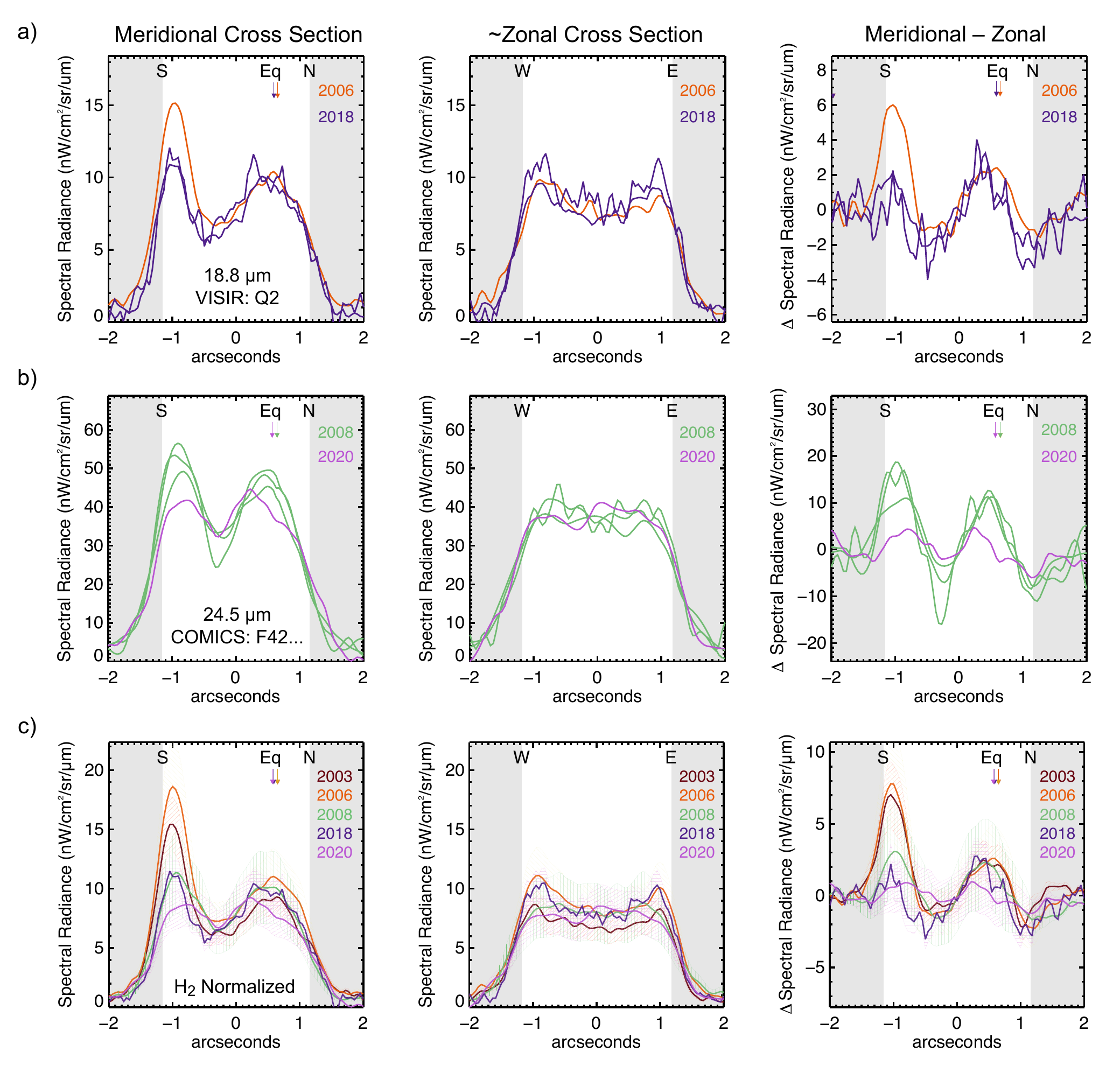}
    \caption{Profiles of radiances across the disk for 17--25 $\mu$m hydrogen-sensing images that form the average images shown in Figure \ref{fig:img_comparison}a. a) Profiles from the VISIR Q2 (18.8 $\mu$m) images showing the meridional cross-section (vertically bisecting the image, north-south), the perpendicular cross-sections ($\sim$zonal, horizontally bisecting the image, roughly west--east), and the difference between the profiles (meridional--zonal) to help separate latitudinal variation from center-to-limb behavior. Plots average 9 central lines for each profile, amounting to 0.4'', or roughly just over 17$\%$ of the disk diameter in our normalized resolution images.  Observation years are defined by color, and the location of the disk edges and changing equator are also indicated. For clarity, error bars are omitted but taken to be 30$\%$ or less. Plots show relatively enhanced emission at the pole in 2006. b) Corresponding curves for COMICS F42C24.50W0.80 (24.5 $\mu$m) images showing greater amplitude in images from 2008 than in 2020. c) Annually averaged curves for all Q-band images (17--25$\mu$m, eight separate filters) normalized by their passband-integrated Spitzer radiances and scaled by the corresponding Q2 filter (18.8 $\mu$m) radiance. Fainter crosshatching envelopes represent the uncertainty. The resulting curves show remarkable consistency in radiance in time for all latitudes except for the south pole, which appeared relatively brighter in 2003 and 2006.}
    \label{fig:diskh2}
\end{figure*}

These same absolute differences between annually averaged image pairs are readily seen in the profiles of radiance across the disk shown in Figure \ref{fig:diskh2}a and \ref{fig:diskh2}b, which show a remarkable consistency in the 18.8 $\mu$m radiances at all latitudes with the exception of the south polar edge. Although the 2018 images have lower SNR, profiles slicing perpendicularly across the center of the images show little difference in the center-to-limb behavior, suggesting that the differences at the pole cannot be easily attributed to observational differences in the seeing and image quality. Differences between the meridional and $\sim$zonal\footnote{Only approximately zonal, not precisely, given that Neptune's south pole was tilted $\sim$27$^{\circ}$ towards the observer and the image was not reprojected (see Figure \ref{fig:ave_imgs})}. cross-sections clearly show the pole and equator are latitudes of enhanced emission.  

While a direct comparison between identical filters provides the strongest evidence of temporal changes, with only two pairs of Q-band images acquired with the same filters spanning multiple years (Figure \ref{fig:img_comparison}a, left and center; Figure \ref{fig:diskh2}a,b), the opportunity for direct comparisons of images are limited.  In order to provide greater temporal coverage, we also compared images at similar but different Q-band wavelengths (17.5--24.5 $\mu$m) by attempting to normalize their intrinsically different radiances. This was done by dividing the images by the filter-integrated radiances derived from the 2005 Spitzer spectrum (to effectively remove the wavelength-dependent variation) and then scaling by a single fiducial radiance---in this case, the filter-integrated radiance at 18.8 $\mu$m. Although the fiducial scaling is somewhat arbitrary, the relative differences that emerge are meaningful. Cautiously, we note that not all changes observed in these sequences are necessarily temporal, as differences in filter passbands, contribution functions, and normalization relative to Spitzer can potentially introduce differences, in addition to inescapable variation owing to image quality and calibration.  Nevertheless, by including additional images, the resulting normalized image sequence provides a fuller account of the potential temporal behavior, as seen in Figure \ref{fig:img_comparison}a (right). 

The corresponding plots of annually averaged, normalized radiance profiles are shown in Figure \ref{fig:diskh2}c; similar plots showing profiles from all images, before and after normalization, are provided in the Appendix Figure \ref{fig:diskh2_c2h6all}. The profile-plots show how temporally uniform the radiances are at most latitudes after normalization.  The exception is the south pole, which appears significantly brighter in 2003 and 2006.  While observational differences may still contribute to some of the variation given the limited number of images averaged (\textit{e.g.}, note the range in south polar radiances observed in 2006 alone in Appendix Figure \ref{fig:diskh2_c2h6all}), these data indicate that the upper-tropospheric south polar temperatures were intrinsically warmer in 2003 and 2006.

\begin{figure*}
    \centering
    \includegraphics[clip, trim=.0in 0.0in 0.in 0.0in,scale=0.86]{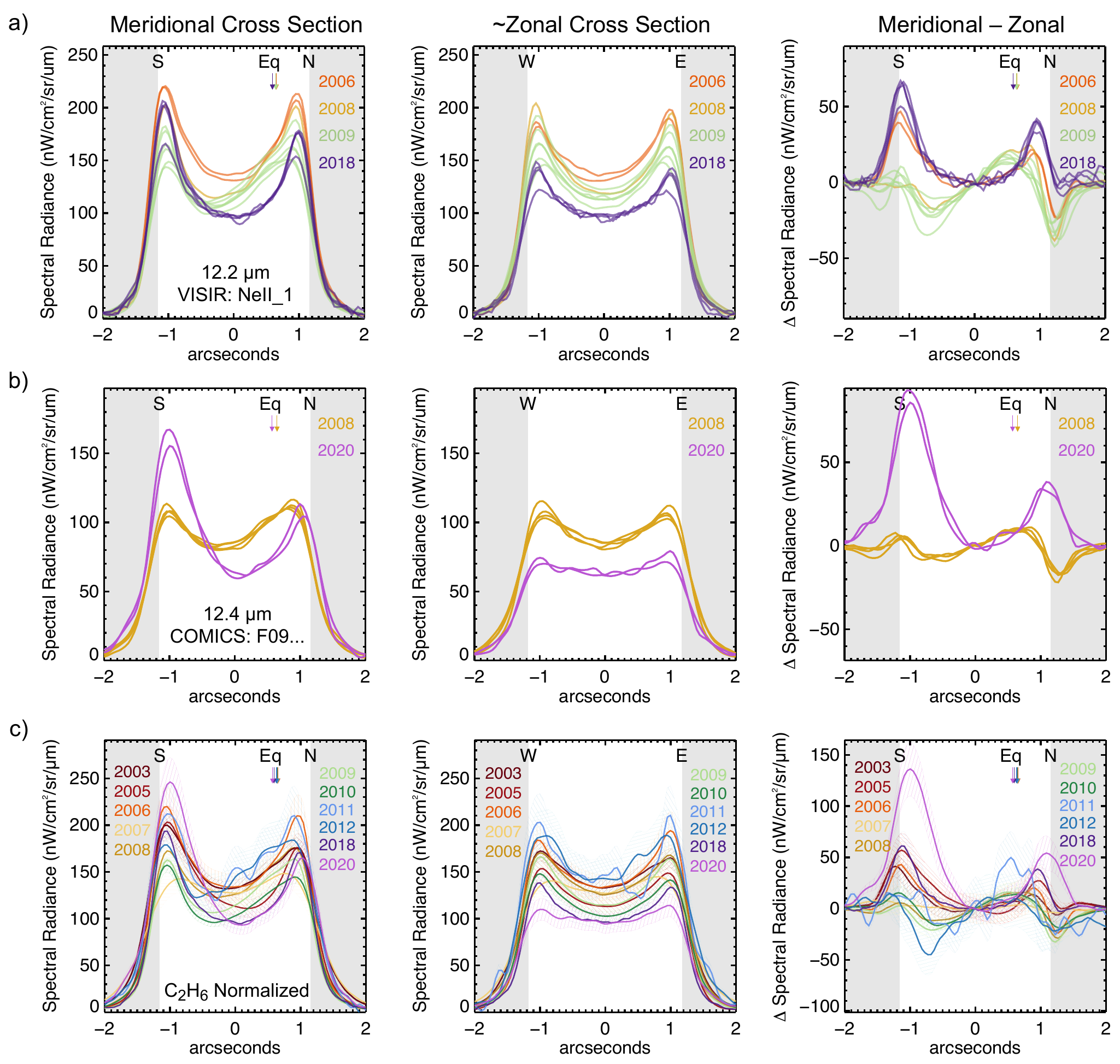}
    \caption{As in Fig. \ref{fig:diskh2}, profiles of radiances across the disk, now for 11--13 $\mu$m ethane-sensing images that form the average images shown in Figure \ref{fig:img_comparison}b. a) Profiles from the VISIR NeII$\_$1 (12.2 $\mu$m) images showing the meridional cross-section, the perpendicular cross-sections, and the difference between the profiles (meridional$-$zonal) to help separate latitudinal variation from center-to-limb behavior. For clarity, error bars are omitted but taken to be less than 10\%.  A roughly symmetric meridional profile with enhancement at both southern and northern limbs in 2006 gives way to an asymmetric distribution in 2008--2010 as radiances drop in the southern hemisphere first, before similarly falling at the equatorial latitudes in 2018 to become once again roughly symmetric.  b) Corresponding curves for COMICS F09C12.50W1.15 (12.4 $\mu$m) in 2008 and 2020 showing the most dramatic difference in relative profile among image pairs, with significant brightening at the south pole accompanied by darkening at all other latitudes. c)  Annually averaged curves for all ethane images (11--13$\mu$m, seven separate filters) normalized by their passband-integrated Spitzer radiances and scaled by the corresponding VISIR NeII$\_$1 filter (12.2 $\mu$m) radiance, with uncertainty represented by the fainter envelopes.}
    \label{fig:diskc2h6}
\end{figure*}

We applied the same approach to the 11--13-$\mu$m ethane images shown in Figure \ref{fig:img_comparison}b. Neptune was observed most frequently at these wavelengths---in 11 separate years between 2003 and 2020---allowing for the fuller assessment of the stratospheric temporal behavior. A direct comparison using the VISIR 12.2-$\mu$m (NeII$\_$1) filter alone shows significant changes in the brightness distribution on scales of just two years (Figure \ref{fig:img_comparison}b, left, and Figure \ref{fig:diskc2h6}a). The 2006 images appear globally brightest, dominated by limb brightening with no additional strong latitudinal variation except for enhanced emission at the south pole and, to a lesser extent, along the northern limb. However, the radiances appear to decrease in the following years, and by 2008, the radiance in the southern hemisphere drops by roughly 20\% in radiance, and the resulting asymmetry across the disk, from the northern to southern limbs, continues throughout 2009, with further reduction along the southern limb and pole.  By 2018, the equatorial region also drops in brightness, resulting in a now globally dimmer but once again more symmetric pattern with relatively enhanced emission at the northern and southern limbs. This is clearly seen in the meridional cross-sections shown in Figure \ref{fig:diskc2h6}a. This surprising asymmetric change in radiance cannot be explained by errors in calibration ($\leq$10\%). 


Even larger changes are seen between COMICS 12.4 $\mu$m images taken in 2008 and 2020 (Figure \ref{fig:img_comparison}b, center, and Figure \ref{fig:diskc2h6}b). The 2008 image shows an asymmetric pattern consistent with what was seen in the contemporaneous VISIR images, but the 2020 image shows a dramatic $\sim$50\% increase in the polar radiance accompanied by a $\sim$25\% decrease at all other latitudes (see Figure \ref{fig:diskc2h6}b). While the small changes in the polar emission angle (from 59$^{\circ}$ in 2008 to 64$^{\circ}$ in 2020) can potentially lead to differences in observed radiances, the changes observed here far exceed any observational biases. The comparison of zonal cross-sections also shows that the degree of disk limb-brightening slightly lessened in 2020 (as evidenced by the slightly flatter curve), which suggests a greater reduction in emission from relatively lower pressures.

Data from other years are not as reliable owing to fewer observations and poorer weather conditions in some cases (particularly the 2007 T-ReCS image), but normalizing (wavelength-correcting using Spitzer and scaled to the average NeII$\_$1 radiance) and combining all the data in the 11--13 $\mu$m spectral region reveals a fuller timeline of these stratospheric changes, as shown in Figures \ref{fig:img_comparison}b (right) and \ref{fig:diskc2h6}. In the normalized ethane data, Neptune's brightness appears to fluctuate, appearing brighter in 2003, 2006, 2011, and 2012 than in other years.  Starting with a somewhat symmetric distribution of disk radiances in 2003, significant meridional asymmetry in the disk profiles develop sometime between 2006 (or 2007 if the T-ReCS image is to be trusted) and mid-2008 and lasts until at least 2011 and 2012, when images again become brighter.  By 2018, the equatorial radiance drops by a quarter or more in radiance and remains at roughly this value in 2020, just as the south pole begins to dramatically brighten.

\begin{figure*}
    \centering
    \includegraphics[clip, trim=.0in 0.0in 0.in 0.0in,scale=0.86]{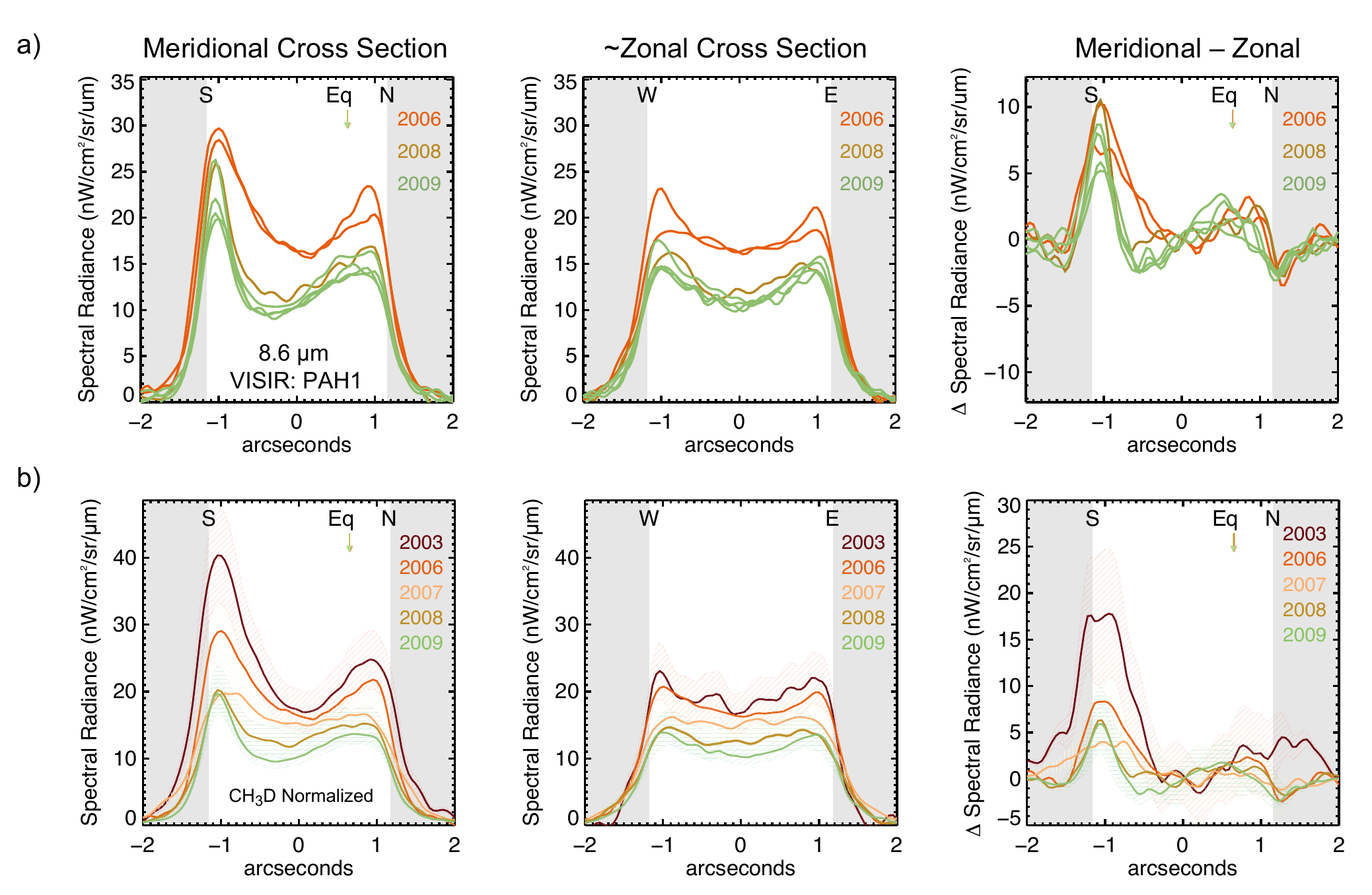}
    \caption{As in Figures \ref{fig:diskh2} and \ref{fig:diskc2h6}, but for 8--9 $\mu$m CH$_3$D-sensing images: a) direct comparison of 8.6 $\mu$m VISIR images. Error bars are omitted, but taken to be less than 20\%. b) Annually averaged profiles from all the CH$_3$D-sensing images (from five different filters) normalized by Spitzer values and scaled by the corresponding Spitzer filter-integrated value for the VISIR PAH1 filter (8.6 $\mu$m).}
    \label{fig:diskch3d}
\end{figure*}

Images sensing deuterated methane emission are fewer, but show substantial trends in time between 2003 and 2009.  VISIR 8.6-$\mu$m (PAH1) images are globally brighter in 2006 than in 2008 and 2009, with increasing contrast between equatorial and southern mid-latitudes in 2008 and 2009 (see Figure \ref{fig:img_comparison}c and \ref{fig:diskch3d}). Normalizing the set to include and compare 2003 images at 8.9 $\mu$m from Keck-LWS, we find 2003 images are the brightest, with radiances steadily decreasing with time. 

\begin{figure*}
    \centering
    \includegraphics[clip, trim=.0in 0.0in 0.in 0.0in,scale=0.85]{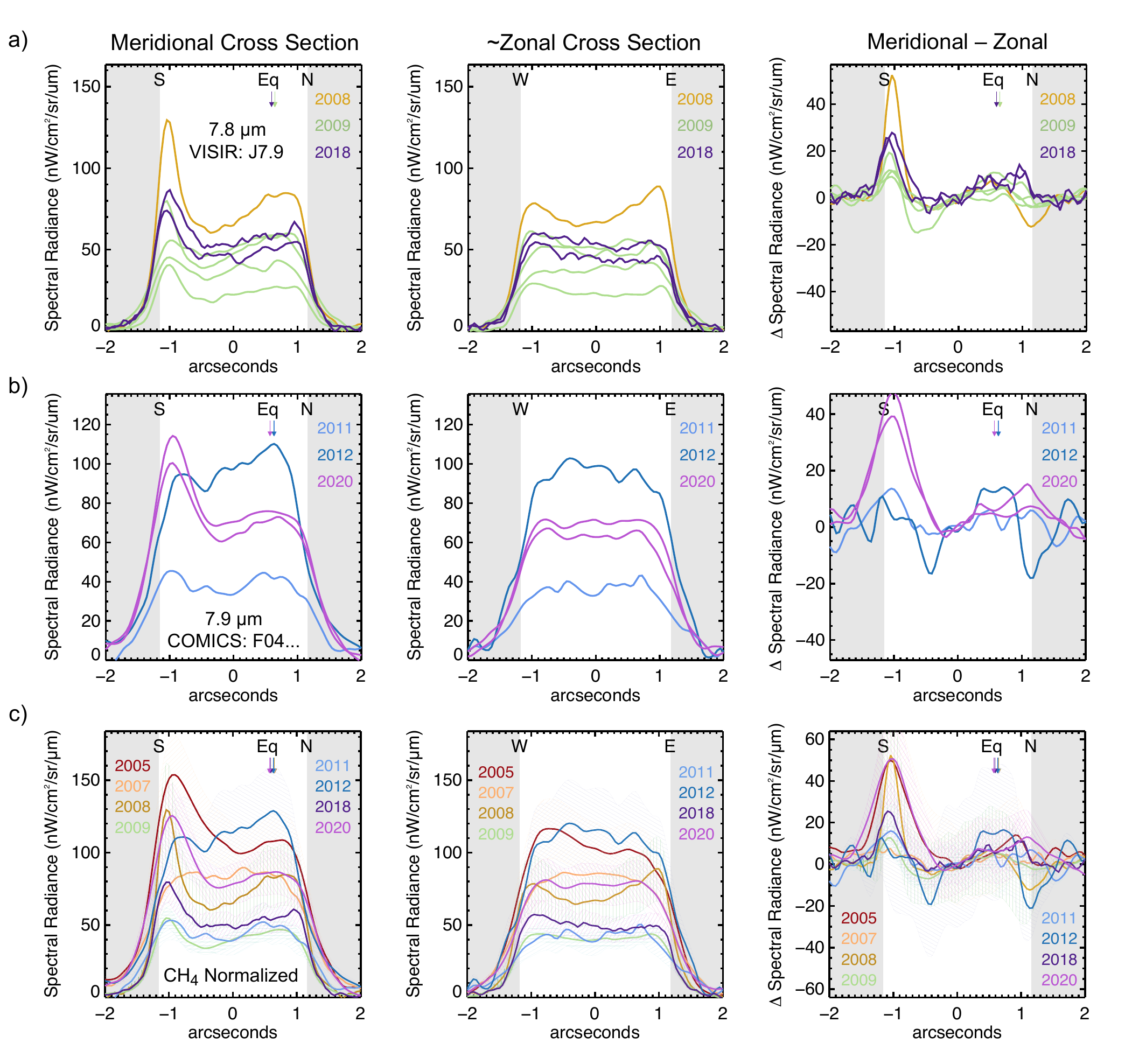}
    \caption{As in Figures \ref{fig:diskh2}, \ref{fig:diskc2h6}, and \ref{fig:diskch3d} but for 7--8 $\mu$m methane images: a) profiles from VISIR 7.8-$\mu$m filter and b) COMICS 7.9-$\mu$m filter, as calibrated, with considerable spread owing to larger calibration uncertainties (estimated at 25$\%$).  c) Annually averaged profiles from all methane-sensing images (four filters), normalized by Spitzer values and scaled by the corresponding Spitzer filter-integrated value for the VISIR J7.9 filter (7.8 $\mu$m).  In this case, the normalization fails to bring radiances into close agreement (\textit{e.g.}, compared to the ethane images). This may indicate greater intrinsic variability in the methane emission, but it may also be partly attributed to calibration uncertainties exceeding the 25$\%$ estimate.}
    \label{fig:diskch4}
\end{figure*}

Applying the same analysis to methane images (7-8 $\mu$m) reveals far greater variability with time. Much of this variability may potentially be owing to calibration errors, as the images show a wide range in quality and large uncertainties.  Variation in 2009 alone may suggest even larger errors in calibration than expected, with disk-integrated radiances ranging from 52 $\pm$ 13 to 95 $\pm$ 24 nW/cm$^2$/sr/$\mu$m (see Appendix Figure \ref{fig:diskch3dplusch4all}). Latitudinal and temporal trends are also noisier with a greater spread in values compared to the C$_2$H$_6$ and CH$_3$D images, exceeding the estimated 25\% calibration uncertainty in both direct and normalized comparisons (Figures \ref{fig:img_comparison}d and \ref{fig:diskch4}). Unfortunately, with some years limited to a single observation, it is impossible to know for certain whether the observed variation can be attributed to errors exceeding the best estimates of the typical uncertainty or true physical changes in Neptune's atmosphere. Interestingly, the general trends in time and latitude are roughly similar in behavior to those seen in ethane and CH$_3$D images, as discussed in Section \ref{sec:diskintvstime}, so at least some variation is likely physical. From a peak in 2005, radiances fall across the disk, with a minimum at southern mid-latitudes in 2009.  We also see a relative increase in the polar radiance between 2018 and 2020, as seen in the other stratospheric-sensing images.  

\begin{figure*}
   \centering
   \includegraphics[clip, trim=.0in 0.0in 0.in 0.0in,scale=0.85]{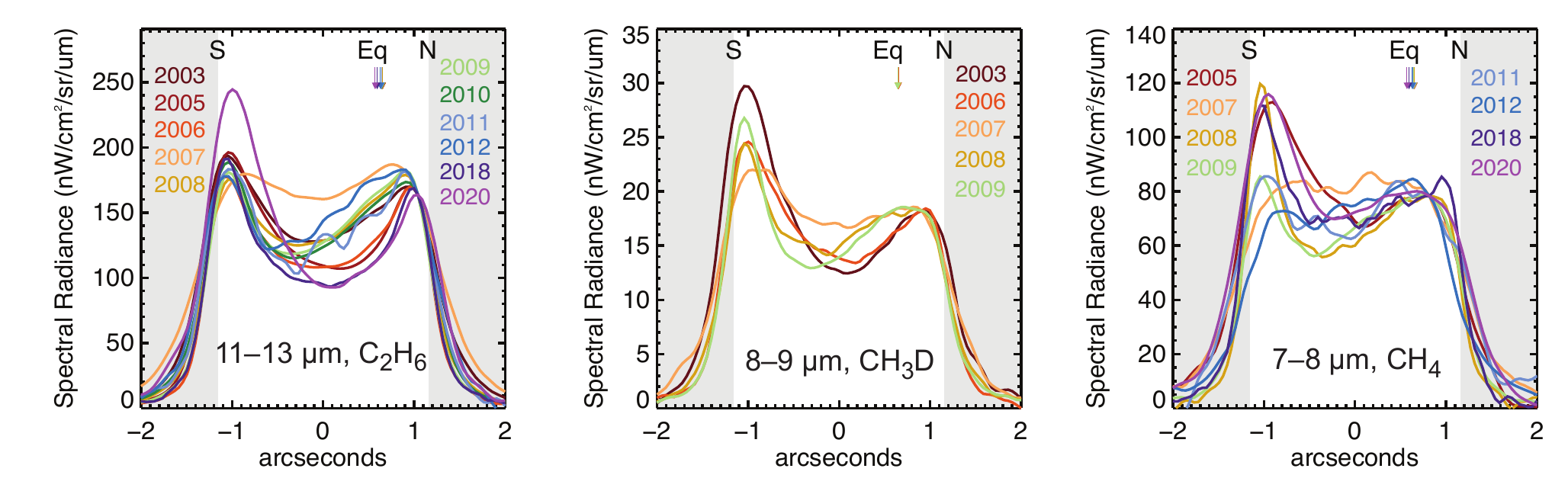}
    \caption{Annually averaged meridional variation in radiance, normalized and arbitrarily adjusted in magnitude to reduce offsets and emphasize meridional trends for C$_2$H$_6$ (left), CH$_3$D (center), and CH$_4$ (right). For comparison, curves are shifted so that the northern limbs---the least variable region in the relatively reliable ethane images---are nearly invariant in radiance.}
    \label{fig:diskfixed}
\end{figure*}

\begin{figure}
    \centering
    \includegraphics[clip, trim=.0in 0.0in 0.in 0.0in,scale=0.95]{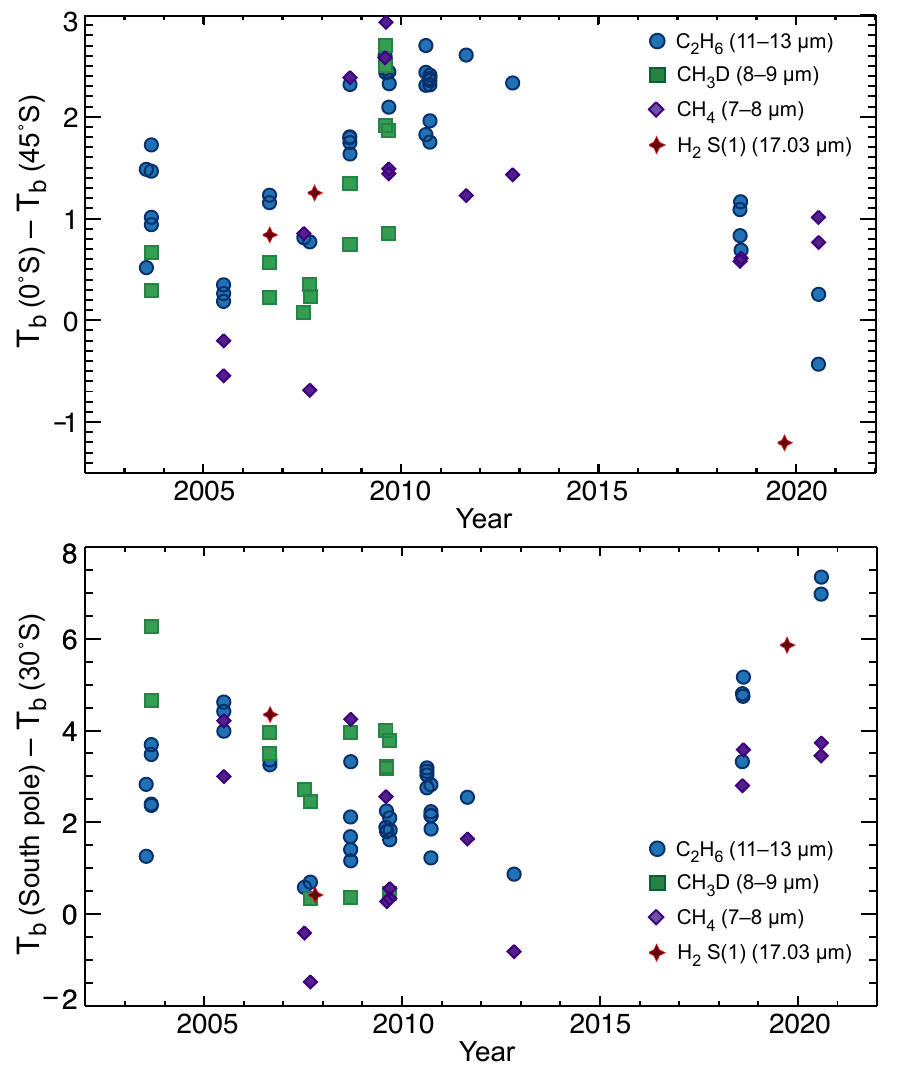}
    \caption{Latitudinal brightness temperature contrasts as a function of time for all stratosphere-sensing images.  (Top panel) The contrast is calculated as the equatorial $T_b$ minus the $T_b$ at 45$^{\circ}$ S, for C$_2$H$_6$ (blue circle), CH$_3$D (green square), and CH$_4$ (purple diamond) images, as indicated by the key.  Contrasts determined from hydrogen quadrupole measurements (red star) are also shown, as discussed in Section \ref{sec:quadrupole}. Nearly all data appear to follow a similar trend with increasing contrast between 2005 and 2011, as radiances dropped at southern latitudes first.  The contrast weakens in 2018--2020 as the equatorial radiances diminished while radiances at southern mid-to-high latitudes increased. (Bottom panel) Corresponding $T_b$ differences between the south pole and the center of the disk, at roughly 30 $^{\circ}$ S latitude, show a slight decrease trend in time between 2003 and 2012. Uncertainties in brightness temperature differences are roughly 0.5 K.}
    \label{fig:asymm}
\end{figure}

If we disregard global variations as purely calibration errors, then we are left with just the aforementioned latitudinal behavior, as shown in Figure \ref{fig:diskfixed}.  In the ethane images, curves generally fall into two groups---those with relatively elevated radiances near the equator and those without. The 2007 data curve is anomalously flat, but may be dismissed due to the poor observing conditions.  Latitudinal trends in the methane and CH$_3$D images are somewhat less variable considering the range in image quality, but they do show a relative reduction in radiance at southern mid-latitudes in 2008 and 2009.  This latitudinal contrast can be quantified as the difference in brightness temperature ($T_b$) between the equator, mid-latitudes, and south pole as a function of time as shown in Figure \ref{fig:asymm}.

The contrast in brightness temperature between the south pole and other latitudes generally ranges between 2 and 7 K but shows no obvious long-term trends (Figure \ref{fig:asymm}). Some, but certainly not all, of the polar variability is likely observational, owing to differences in image resolution and astronomical seeing. The observed contrast also appears to be wavelength dependent---while the extreme brightening seen at the pole in 2020 is clearly exceptional in the ethane images, it appears somewhat less remarkable in the methane images, with values comparable to some previous years (i.e. 2008, 2005, 2018). Likewise, the Q-band images sensing temperatures near the tropopause from 2018 and 2020 show only a relatively weak polar vortex, albeit with poorer image conditions. If the enhanced polar emission is produced by a temperature enhancement, this would suggest it is vertically localized near the peak of the ethane contribution functions.  Alternatively, the wavelength dependence suggests that ethane enhancement is contributing to the rising polar brightness.   

\subsection{Disk-Integrated Variation in Time}\label{sec:diskintvstime}
\begin{figure*}
    \centering
    \includegraphics[clip, trim=.0in 0.0in 0.in 0.0in,scale=.88]{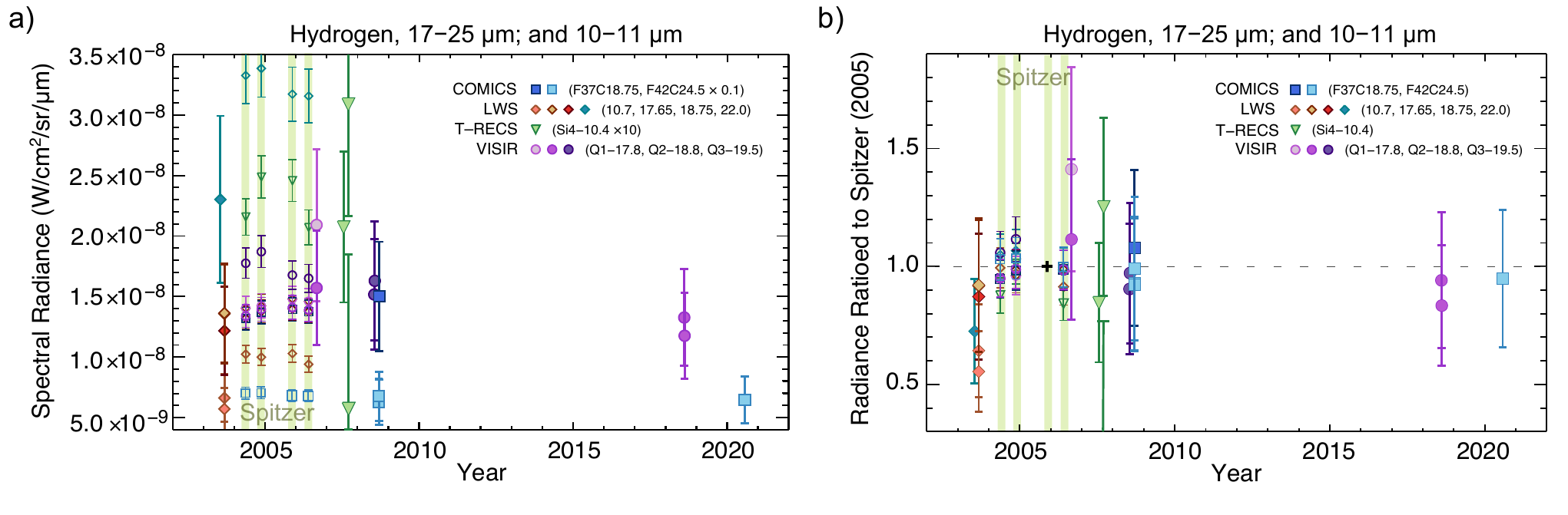}
    \caption{Disk-integrated radiances versus time for all hydrogen-sensing images (17--25 $\mu$m) along with the 10--11 $\mu$m images. Symbols correspond to different instruments and filters as indicated by the key, with error bars of 30\%. Spitzer filter-integrated equivalent values are also shown as smaller, open symbols in the green shaded regions of 2004, 2005, and 2006, with error bars of 7\%. a) The absolute radiances for each observation versus time, which span a wide range of values given differences in filter wavelengths.  Note that COMICS 24.5 $\mu$m and the T-ReCS 10.4 $\mu$m are shown at 1/10 and 10 $\times$ their actual values, respectively, for clarity.  b) The relative radiance as a ratio to Spitzer observations in 2005 (indicated by the + sign).  As a ratio, the wavelength-dependent differences are essentially removed, and the relative differences of the group over time become evident. The images show no significant disk-integrated trends in time beyond the uncertainties.}
    \label{fig:h2vstime}
\end{figure*}

\begin{figure*}
    \centering
    \includegraphics[clip, trim=.0in 0.0in 0.in 0.0in,scale=.9]{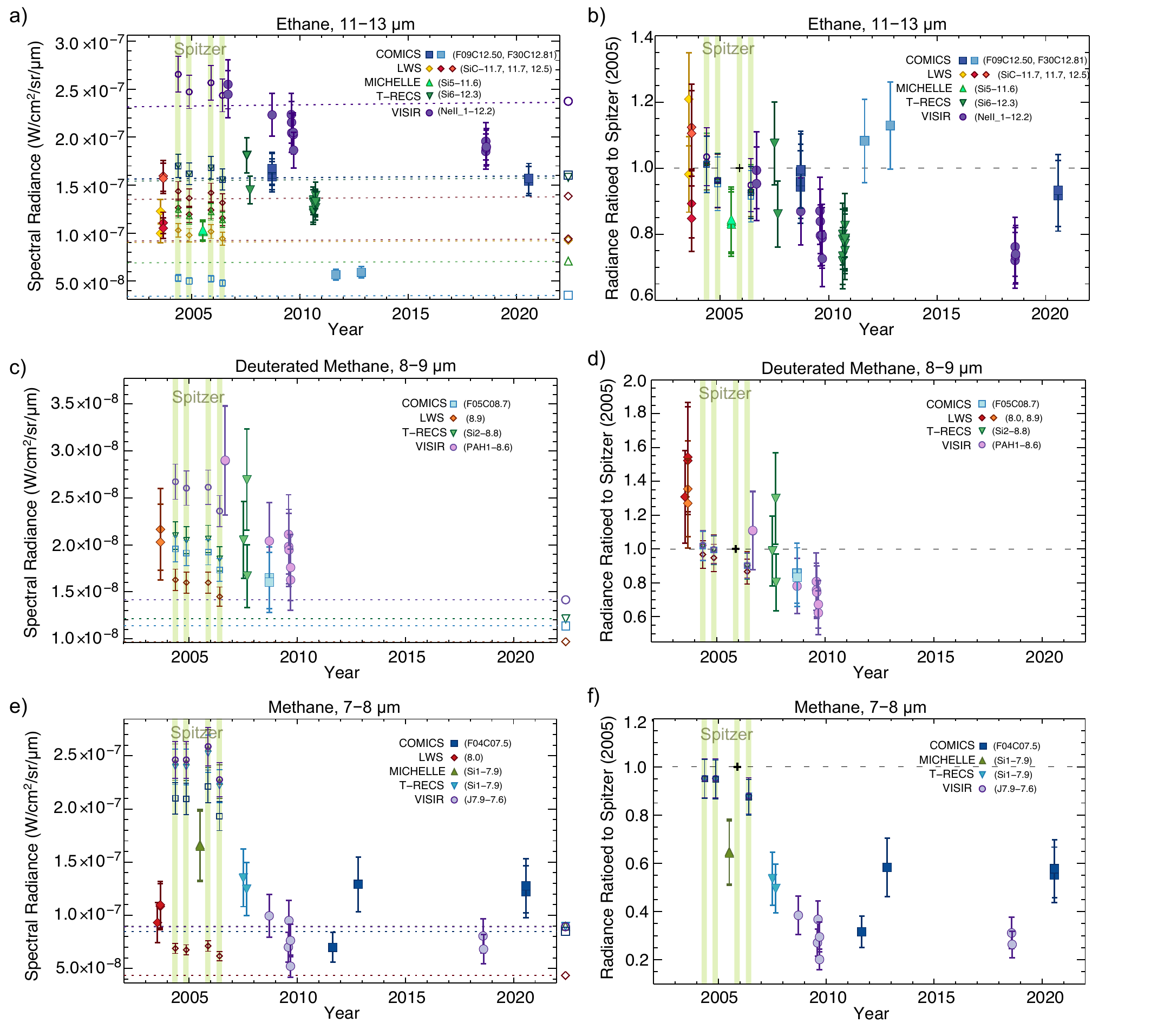}
    \caption{As in Figure \ref{fig:h2vstime}, disk-integrated radiances versus time for images sensing  stratospheric C$_2$H$_6$ (panels a and b), CH$_3$D (panels c and d), and CH$_4$ (panels e and f).  Dotted lines in the panels on the left show the disk-integrated radiances predicted from our atmospheric temperature/chemical model for each filter, as indicated by the corresponding open symbols to the right of the panels.  Panels on the right show radiances ratioed to Spitzer as described in the text. Deuterated methane (8--9 $\mu$m) images (panel d) show nearly linear relative drop between 2003 and 2009.  Methane (7--8 $\mu$m) images (panel f) show a similar trend prior to 2010, but with a possible systematic offset compared to our filter-integrated values for Spitzer.}
    \label{fig:stratvstime}
\end{figure*}

We investigated the potential global-scale variability suggested by the images in Section \ref{sec:spatialvstime} by evaluating the disk-integrated radiances versus time.  For these calculations, we performed aperture photometry on calibrated, sky-subtracted images by summing Neptune's observed signal and dividing by the true solid angle of Neptune's disk.  Results are plotted as a function of time for each of the image groups in Figures \ref{fig:h2vstime} and \ref{fig:stratvstime}. The left panels show the disk-integrated radiances for each observation, with a different icon for each instrument. Filter-integrated Spitzer values are also shown (in green shaded regions), representing the radiances we would have observed from the ground at our filtered wavelengths given the Spitzer spectral radiances in 2004 (May and November), 2005, and 2006. For example, in Figure \ref{fig:stratvstime}a, we see that the VISIR 12.2 $\mu$m measurements (purple closed circles) appear to decrease in radiance from 2006 to 2018. Integrating the Spitzer spectrum over the VISIR 12.2-$\mu$m filter passband shows that values in 2004 and 2005 (purple open circles) were similar, within uncertainties, to the ground-based image from 2006, extending the trend back two years further in time. 

For further comparison and temporal context, we also plot the radiances relative to the equivalent Spitzer filter-integrated values, as a ratio, to display relative changes seen across all filters within the group (Figures \ref{fig:h2vstime} and \ref{fig:stratvstime}, right panels).  By expressing these radiances as a ratio to Spitzer, we once again effectively normalize the observations by removing the wavelength dependence.  The results reveal trends across multiple filters that are indicative of changes in the image group in general.  

As indicated with the disk profiles (Figure \ref{fig:diskh2}), the hydrogen-sensing filters span a wide range of intrinsic radiances given their range of wavelengths (17-25 $\mu$m) (Figure \ref{fig:h2vstime}a), but their global variations when ratioed to Spitzer suggest no significant change beyond the expected uncertainties (Figure \ref{fig:h2vstime}b).  

Images sensing stratospheric pressures, however, show more variability with coherent trends (Figure \ref{fig:stratvstime}).
The VISIR ethane imaging shows a decline of approximately 30$\%$ between 2003 and 2018, while other instruments were used too sparsely to establish trends (Figure \ref{fig:stratvstime}a). When normalized (Figure \ref{fig:stratvstime}b), we see a rough decline of approximately 30$\%$ between 2003 and 2010, followed by a secondary peak in 2012, and a 25$\%$ rise in brightness from 2018 to 2020. CH$_3$D images using the same filter are very limited (Figure \ref{fig:stratvstime}c), but altogether, the normalized data display an almost linear decline of $\sim$50$\%$ in relative radiance from 2003 to 2010 (Figure \ref{fig:stratvstime}d). Finally, CH$_4$ images show a similar decline for the years measured within the same period (Figure \ref{fig:stratvstime}, panels e and f).  The methane decline is even greater ($\sim$70$\%$) if the filter-integrated values from Spitzer-IRS are accurate.  Radiances remain low in subsequent methane imaging, but like the ethane images, we see signs of a secondary peak in 2012 and a rise by more than 80$\%$, between 2018 and 2020.


Although the uncertainties on all these measurements are considerable, the similar temporal behavior seen across different nights, observatories, instrument filters, and spectral groups lends credibility to these variations.  The shared behavior also indicates that a single mechanism may be primarily responsible for altering the emission. As the dotted lines in Figures \ref{fig:stratvstime}a, \ref{fig:stratvstime}c, and \ref{fig:stratvstime}e show, the expected variation from seasonal photochemistry alone produces an almost negligible effect on disk-integrated radiances over this timescale. Far larger variation coordinated across different hydrocarbons would be needed.  Alternatively, variation in the stratospheric temperatures is the simplest explanation, as investigated in the next section.   


The plots in Figure \ref{fig:stratvstime} also reveal a possible systematic error in radiances at wavelengths less than 10 $\mu$m. The near linear trend in the radiance ratio plots for CH$_3$D images (Figure \ref{fig:stratvstime}d) would be better fit if all the ground-based observations were systematically reduced by 10\%, or the Spitzer values were systematically raised by 10\%. Likewise, the trend in methane images (Figure \ref{fig:stratvstime}f) would fall into better agreement with the Spitzer values if the Spitzer radiances were systematically decreased by 50\%.  Additionally, our radiative-transfer modeling of the radiances, based purely on previously published temperature and chemical profiles, also shows a discrepancy with respect to our Spitzer integrated values.  Assuming the average temperature profile (Figure \ref{fig:temp_conts}) based on \citet{greathouse2011spatially} and \citet{moses2005photochemistry} \citep[the latter originally based on][]{orton1992thermal,yelle1993distribution,roques1994neptune}, combined with the seasonally varying photochemical hydrocarbon model of \citet{moses2018seasonal}, we computed synthetic radiances using NEMESIS \citep{irwin2008nemesis} to model the radiative transfer. Disk-integrated values were calculated by creating synthetic spatially-resolved images, from which modeled disk-integrated radiances were extracted using aperture photometry, just as we had done with true data. The modeled disk-integrated radiances are plotted alongside observations in the left panels of Figure \ref{fig:stratvstime} and are nearly invariant in time. Although, these modeled radiances are derived from ground-based measurements of the temperatures and hydrocarbon distributions---the latter partly constrained by numerous measurements of hydrocarbon mole fractions from various ground- and space-based observations made between 1981 and 2009 \citep[as described in][]{moses2018seasonal}---they are independent of the observed radiances extracted from our calibrated imaging data.  Yet, as Figure \ref{fig:stratvstime}e shows, these modelled radiances appear consistent with most of the ground-based methane data (dotted lines versus icons), but they appear systematically lower than Spitzer values.

Systematic errors in the calibrations across different filters, observatories, and times are difficult to explain, as it would indicate nearly all ground based imaging is underestimating the 7--8-$\mu$m emission by roughly 50\% while overestimating those at 8--9-$\mu$m.  Combined with the modeling results, this apparent offset suggests that our filter-integrated Spitzer radiances at for 7--8-$\mu$m are systematically too bright.  If the Spitzer calibrations at these wavelengths are accurate, one possible explanation for this error may be the lower spectral resolution (R$<$127) of the Spitzer-IRS spectra at wavelengths of 7--10 $\mu$m, which may conceivably lead to larger errors when interpolated and integrated over the filter passbands. As a test, we interpolated the entire Spitzer N-band spectra to a resolution R$\sim$127 (from R$\sim$600) and recomputed filter-integrated radiances for the $\sim$12-$\mu$m ethane filters; we found values only changed by no more than 5$\%$. However, the radiances and atmospheric transmission vary more strongly with wavelength below 9 $\mu$m, which may accentuate these errors. Furthermore, ground-based observations are subject to absorption by telluric CH$_4$, which can selectively suppress the observed emission from CH$_4$ in Neptune's atmosphere at the wavelengths of the most intense emission lines. While the correlation between emission and telluric absorption is imperfect because the stratospheric emission lines in Neptune’s spectrum are generally Doppler-shifted and narrower than the pressure-broadened lines of telluric methane, the combined effect can nonetheless significantly alter the expected radiances observed from the ground. When calculating the equivalent Spitzer filter-integrated radiances, we attempt to account for this telluric absorption by convolving the Spitzer spectrum with telluric transmission; however, the lower spectral resolution of Spitzer 7--10-$\mu$m spectrum means this correlation between emission and absorption may not be adequately captured, potentially explaining the discrepancy between ground-based and filter-integrated Spitzer observations. 




Interestingly, at 8--9-$\mu$m, nearly all observations---both Spitzer and ground-based---appear too bright compared to the model predictions (dotted lines versus icons, Figure \ref{fig:stratvstime}c). These differences suggest that our assumed ratio of CH$_3$D to CH$_4$ of 2.64$\times10^{-4}$  \citep[from Herschel-PACS analysis,][]{feuchtgruber2013d}, is likely too low. Increasing the D/H in CH$_4$ to 4$\times10^{-4}$---the upper limit suggested by \citet{fletcher2010neptune} from 2007 AKARI satellite measurements and \citet{irwin2014line} from 2010 VLT/CRIRES spectra---brings the models into better agreement with the ground based observations in 2009, but they still fall short of the Spitzer equivalent values.  Considering that the modeled CH$_4$ emission (which, we note, we model using temperature profiles constrained by ground-based observations) is also too dim compared to Spitzer, these data suggest a need to increase the D/H ratio and methane abundance or temperatures in stratosphere to match the Spitzer values between 2004 and 2006.  Alternatively, as mentioned, it is conceivable that, at $\sim$7--9 $\mu$m, our filter-integrated radiances for Spitzer are either systematically too great, or our ground-based radiances are too small, owing to an unaccounted error. In 2022, James Webb Space Telescope MIRI spectra will provide an opportunity for potentially comparing coordinated ground- and space-based observations in order to assess ground-based calibrations and further investigate the source of any possible discrepancy in future work.  For the present investigation, the Spitzer radiances primarily serve as a fiducial point for comparison, and the apparent offsets do not alter the inferred trends and conclusions.

\subsection{Hydrogen Quadrupole Emission}\label{sec:quadrupole}

\begin{figure*}[hb]
    \centering
    \includegraphics[clip, trim=.0in 0.0in 0.in 0.0in,scale=.88]{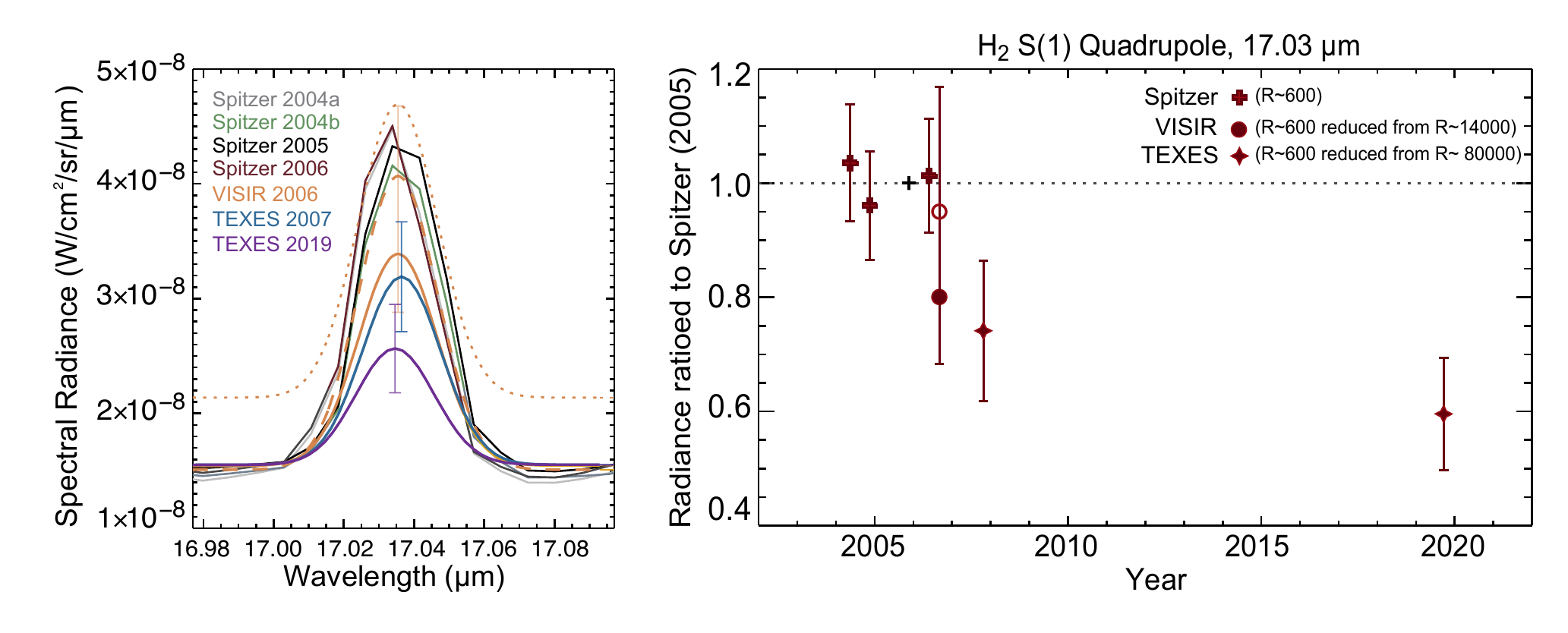}
    \caption{Comparison of the H$_2$ S(1) emission line from different instruments and times, as listed in Table \ref{tab:specdat}. (Left panel) The 2006 VISIR data (gold lines) are shown for three different assumptions regarding calibration: first, as flux calibrated with no correction (dotted gold line), and then corrected so that the continuum radiances match Spitzer, equally accomplished by subtracting $6.3\times 10^{-9}$ W/cm$^2$/sr/$\mu$m (dashed gold line) or dividing by a factor of 1.385 (solid gold line).  VISIR and TEXES spectra were reduced to a resolution of R$\sim$600 from intrinsic resolutions of R$\sim$14000 and R$\sim$80000, respectively, to match Spitzer-IRS, and all were shifted to an averaged central wavelength of 17.03 $\mu$m.  Error bars express an uncertainty of 15$\%$ for TEXES and VISIR data. Spitzer measurements have an uncertainty of $6\%$, with error bars omitted for clarity. (Right panel) Corresponding disk radiances ratioed to the Spitzer 2005 value, including the two plausible values of VISIR (filled and open circles corresponding to the solid and dashed curves, respectively), show a clear decrease over time. }
    \label{fig:quadvsspitz}
\end{figure*}

The H$_2$ S(1) quadrupole emission at $\sim$17.035 $\mu$m provides an unambiguous indicator of stratospheric temperatures at millibar pressures.  However, given the differences in spectral and image resolutions among data (see Table \ref{tab:specdat}), some manipulation is first necessary in order to make a valid comparison. The Spitzer spectra are disk averaged by virtue of their low spatial resolution, while the VISIR and TEXES data sample subsets of the disk as seen through their 1'' and 0.8'' slits, respectively. The lateral scanning by TEXES allows us to construct and average an effective image of brightness across the disk, but the VISIR data are limited to the average of the slit area projected along the meridian. The Spitzer spectral resolution of R$\sim$600 is also considerably lower than that of VISIR (R$\sim$14000) and TEXES (R$\sim$80000), which directly affects the measured intensity of the line. As the spectral resolution decreases, the discrete emission becomes more strongly convolved with the continuum emission, effectively weakening the line radiance.  To account for this, we convolved the VISIR and TEXES spectra with a Gaussian to reduce all spectra to an equivalent  resolution of R$\sim$600.  Comparisons between the resulting disk-integrated, resolution-normalized radiances as shown in Figure \ref{fig:quadvsspitz}.  

We find that the VISIR quadrupole spectra appear systematically offset from the other observations in continuum radiance, which originates from deeper in the atmosphere, near the tropopause. The Q-band imaging that senses the continuum emission from these deeper pressures shows no significant variation over the years observed, and we therefore conclude that the discrepancy in the continuum likely indicates a systematic offset in the VISIR spectral calibration. The source of the error is unknown, and since the calibration process involves both division and subtraction, either a multiplicative or additive factor may be appropriate for correcting the offset. The choice of correction is significant since it directly affects the ratio of the line emission relative to the continuum.  Given that both sources of error are equally plausible, we compare results assuming multiplicative (solid line) and additive(dashed line) corrections separately in Figure \ref{fig:quadvsspitz}.  

As Figure \ref{fig:quadvsspitz} shows, Spitzer values from 2004 to 2006 are in strong agreement with each other, varying in disk-integrated radiances by less 5\%---within expected uncertainties.  Assuming an additive correction brings the VISIR quadrupole emission in very close agreement with the Spitzer values, while assuming a multiplicative correction reduces the line radiance, bringing it closer to that observed for TEXES in 2007.  Given the uncertainties, it is safest to assume that the true VISIR value is somewhere between the two values, as would be expected for a smooth trend in time. The TEXES 2007 quadrupole emission appears significantly less than the Spitzer values, and the 2019 TEXES value is smaller still, amounting to a roughly 40\% decline between 2006 and 2019. 

\begin{figure*}[h]
    \centering
    \includegraphics[clip, trim=.0in 0.0in 0.in 0.0in,scale=0.7]{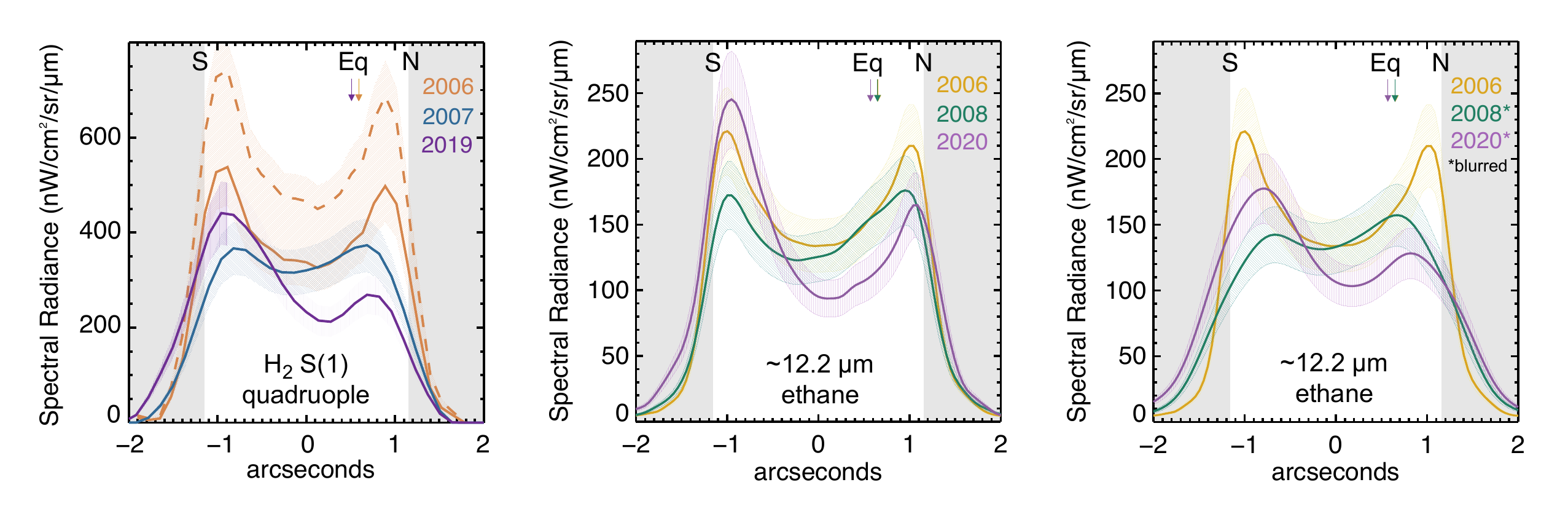}
    \caption{Meridional variation in the H$_2$ S(1) quadrupole emission compared to emission measured in ethane images. (Left panel) Plots show profiles of meridional radiance across the disk for the 2006 VISIR (orange), 2007 TEXES (blue), and 2019 TEXES (purple) observations. The solid lines represent the nominal values, with absolute uncertainties represented by the hatching.  For VISIR, the dashed line represents the corrected value assuming an additive offset, while the solid line assumes the multiplicative correction.  For the TEXES observations, spectral resolutions were reduced from 80000 to 14000 to match that of VISIR, although differences in acquisition may still lead to systematic differences. (Center panel) Meridional profiles from roughly contemporaneous ethane imaging (12.2-$\mu$m VISIR, 2006 and 2008; 12.5-$\mu$m COMICS, 2008 and 2020), averaged and color-coded by year as indicated.  (Right panel) The same ethane imaging data, but now with the observations in 2008 and 2020 smoothed by a boxcar average with a width of 0.08'' to mimic the effect of blurring likely seen in the TEXES data.  The similarity between the quadrupole (left) and blurred ethane images (right) suggests that the observed radiance variations are primarily owing to variation in temperatures.}
    \label{fig:quadvlat}
\end{figure*}

\begin{figure}[h]
    \centering
    \includegraphics[clip, trim=.0in 0.0in 0.in 0.0in,scale=0.80]{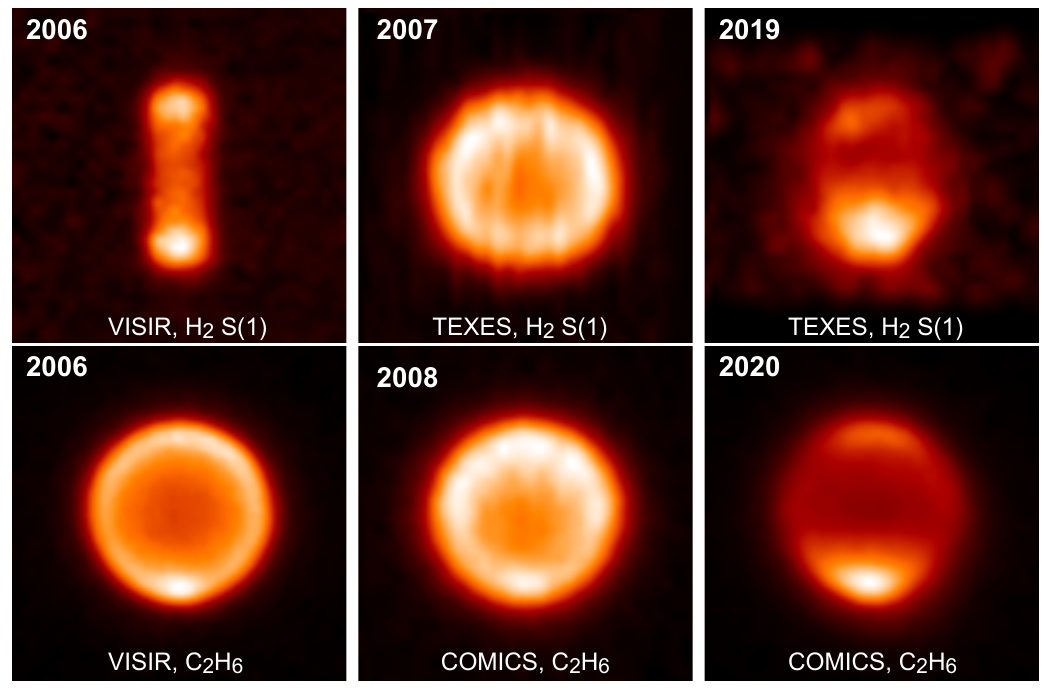}
    \caption{Relative spatial variation in the H$_2$ S(1) quadrupole emission (top row) compared to ethane emission (bottom row). The 2006 VISIR spectra show radiances within the slit projected over the central meridian, while the sequential scanning acquisition of the 2007 and 2019 TEXES spectra allow for the full disk to be reconstructed, roughly revealing the spatial distribution of emission across the disk. The spatial brightness in quadrupole emission appears remarkably similar to that seen in roughly contemporaneous imaging of ethane emission at $\sim$12 $\mu$m, suggesting that the appearance and variation of ethane images is largely owing to the temperature structure.}
    \label{fig:quadethanesame}
\end{figure}

Meridional profiles of the peak quadrupole emission extracted from VISIR and TEXES spectra reveal latitudinal trends, as shown in Figure \ref{fig:quadvlat}.  These trends are similar to those seen in the ethane images around the same time, with a drop in mid-disk radiance and an increase in meridional asymmetry between 2006 and 2019. Indeed, a qualitative comparison of the quadrupole emission images constructed from the spectra show a remarkable similarity to nearly contemporaneous imaging at 12--13 $\mu$m (see Figure \ref{fig:quadethanesame}).  However, the TEXES profiles appear significantly smoother with softer limbs compared to the VISIR data.  This difference is likely observational and attributable to the difference in image resolution, seeing, and acquisition method. The scanning approach applied by TEXES would have produced more blurring had the shifting slit position had included significant background sky, particularly in 2007 when the scanning direction was perpendicular to the polar axis.  As Figure \ref{fig:quadvlat} shows, the TEXES 2007 and 2019 quadrupole profiles also appear remarkably similar to the VISIR ethane (12.2 $\mu$m) image profiles from 2008 and 2020 when the VISIR ethane image profiles are smoothed by a boxcar average equivalent to the TEXES slit width (0.8'') to mimic the effect of blurring. The relative difference in radiance between the 2006 ethane profiles and the other years is also similar to the relative differences among the quadrupole profiles, assuming the multiplicative correction to the 2006 VISIR spectra. The overall similarities suggest that the spatial and temporal variations seen in both ethane images and hydrogen quadrupole spectra are associated with changes in the Neptune's stratospheric temperatures.    

\subsection{Inferred Temperature Changes}
To relate the observed changes in radiances to changes in atmospheric temperatures, we used a combination of radiative transfer inverse and forward modelling to derive consistent atmospheric temperature models. 

\subsubsection{Retrieval Preparation and Process} \textit{Correcting for blurring}:  Given limited spatial resolution and Neptune's $\sim$2.3'' angular diameter, a simple direct inversion of the images will provide unreliable temperatures nearer the edges of the disk, where blurring with space surrounding the disk artificially reduces the observed radiances. In theory, a deconvolution of the images with the image point spread function (estimated from the stellar seeing disk) in Fourier space can correct for this degradation; however, direct deconvolutions amplify the noise, which we find can become dominant in most images, requiring yet more smoothing to overcome.  We instead adopted an alternative approach to effectively deconvolve the images by modeling idealized data and the effects of blurring, following \citet{roman2020uranus}. 

We began by extracting radiances from the central meridian of the disk in the unaltered images, but limited to points on the disk with modest emission angles (i.e. $\mu\geq0.5$). Selected ethane (11--13 $\mu$m), hydrogen (17--25$\mu$m), and methane (7--9 $\mu$m) images were used, with multiple images combined to give averages for each filter group representative of each observing epoch. These extracted radiances were then inverted using an optimal estimation retrieval algorithm, NEMESIS \citep{irwin2008nemesis}, to create a rough model of the atmospheric temperature structure versus pressure and latitude. A vertical temperature profile based on \citet{moses2005photochemistry} and  \citet{greathouse2011spatially}, with chemical abundances of \citet{moses2018seasonal}, was used as the prior (as shown in Figure \ref{fig:temp_conts}). Inferred temperatures at modest emission angles were extrapolated to latitudes viewed at higher emission angles (i.e. $\mu<0.5$) to complete the disk. Assuming zonal uniformity, we then forward-modelled radiances for all locations on the disk from this initial model of temperatures (as a function of pressure and latitude), using NEMESIS to solve the radiative transfer at the relevant observing geometries for each pixel.  Forward models were calculated at the resolution of the spatially normalized data (i.e., for a disk with an equatorial width of 51.8 pixels), with care taken to adequately represent the contributions from the limb. The resulting modeled radiances yielded an initial simulated image of the idealized disk as a first approximation.

\begin{figure}[h]
    \centering
    \includegraphics[clip, trim=.0in 0.0in 0.in 0.0in,scale=0.45]{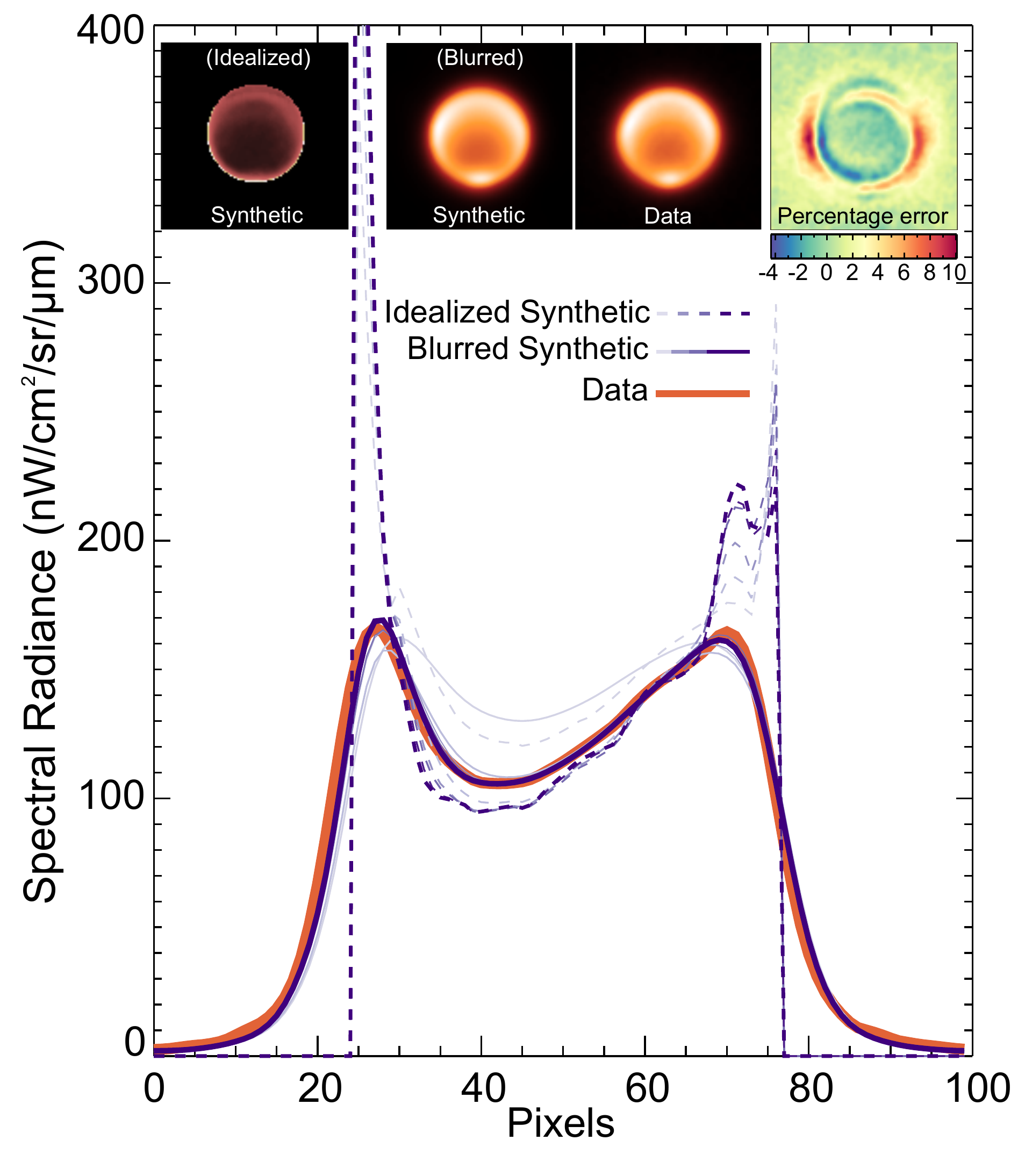}
    \caption{Example illustrating our technique to mitigate the effects of blurring in the images. The orange curve plots the radiance across the central meridian of the real image---a 12.2-$\mu$m VISIR image from August 8, 2009, labeled as data in the inset. Initial temperature retrievals from the image are used to create an idealized model atmosphere from which radiances are plotted (dashed, faint blue line). The idealized model image is then convolved with the stellar image to simulate the effects of atmospheric blurring, yielding the solid blue line. Differences between the blurred model and data are used to iteratively correct the idealized model, and subsequent iterations are shown in darker blue. The resulting final idealized and blurred synthetic images are shown in the insets. Differences between the data and synthetic image are shown as percentage errors. }
    \label{fig:imgfix}
\end{figure}

We then used this simulated image to estimate the effects of blurring in the data.  The simulated image was convolved with its corresponding data's calibration star image (which we assume approximates the effective PSF of the data) to mimic the blurring suffered by the real disk of Neptune in the observations. We then directly compared the idealized and blurred disks to define a flux-conserving multiplicative correction (for each pixel) that converts between the two. These correction factors, determined from the simulated images, therefore represent the transformation of Neptune's disk owing to atmospheric blurring. We applied this factor inversely to the real data, with the goal of approximating how the true disk would appear prior to blurring. Then, in order to evaluate whether this corrected disk was truly consistent with the data, the corrected image was then artificially blurred (via the same stellar convolution) for direct comparison with the actual observations. Fractional differences between observed and artificially blurred disks were applied to the correction factors, and the process was repeated, iteratively adjusting the factors as needed until the data and model agreed to within 10$\%$ or less (see Figure \ref{fig:imgfix}). The process is imperfect, as it can introduce subtle spurious structure in the idealized model, but this fine structure can be removed by carefully smoothing the solutions prior to retrieval while preserving the larger latitudinal trend.

\textit{Retrievals from corrected images}:  The resulting idealized images---now effectively our best approximation of the data prior to being blurred by the atmosphere---were then used as the inputs into the retrieval algorithm.  We extracted strips of radiances along the central meridians spanning all visible latitudes (averaged over nine pixels in longitude, amounting to $\sim$0.4'', or 20$^{\circ}$ of longitude at the equator), but now extended to emission angles of $\sim$81$^{\circ}$ ($\mu = 0.15$).  These radiances were inverted using the NEMESIS code \citep{irwin2008nemesis}, again with the same \textit{a priori} temperature and chemical profiles and vertical correlation length scale of 1.5 pressure scale heights. For clearer interpretation of results, we allowed only temperature to vary in our retrievals, as the dearth of contemporaneous quadrupole measurements makes it impossible to constrain both temperature and chemical abundances independently across all the years.  While both temperature and composition likely vary in Neptune's atmosphere, the analysis of Section \ref{sec:quadrupole} suggests it is reasonable to attribute the observed changes primarily to variation in temperatures. The seasonal photochemical models of \citet{moses2018seasonal} do predict slight changes in ethane mixing ratios over the observed period ($<$19$\%$ at 0.5 mbar between \textit{L$_s$}$\sim$265$^{\circ}$ and \textit{L$_s$}$\sim$304$^{\circ}$), but we use the 2009 chemical model (\textit{L$_s$}$\sim$278$^{\circ}$, roughly corresponding to the average date of our data) for all retrievals and forward models for consistency.  We tested the sensitivity of our results to this choice, and we found that retrieved 0.1-mbar temperatures differed by less than 2 K when neglecting time variability in the chemical model over this period, with maximum differences at the south pole. A more thorough study of simultaneous temperature and compositional gradients across the disk using contemporaneous quadrupole and ethane emission \citep[\textit{e.g.}, following][]{greathouse2011spatially} is left to future work.

Temperature profiles were modeled from $\sim$10 to 1$\times 10^{-8}$ bars of pressure, over 180 vertical layers. However, the limited sampling and poor vertical resolution of the filter contribution functions means only pressures near the tropopause ($10^{-1}$ bars) and lower stratosphere ($10^{-3}$--$10^{-5}$) bars were strongly constrained; temperatures at other pressures tended towards the \textit{a priori} profile and the implied lapse rates are only approximate. Clouds and aerosols were neglected, given their presumed negligible opacity at these infrared wavelengths.  Absorption coefficient ($k$) distributions from line-by-line calculations were used to model the gaseous opacity, calculated from the GEISA 2003 database \citep{jacquinet_geisa}. Collision-induced opacities were taken from \citet{fletcher2018hydrogen} for H$_2$--H$_2$ and  \citet{orton2007evidence,borysow1988collison,borysow1989collision} for H$_2$--He and H$_2$--CH$_4$. 
Images were sorted into six epochs for retrieval---those dating from 2003, 2005, 2006, 2008/2009, 2018, and 2020.  All retrievals included selected ethane (11--13 $\mu$m) and hydrogen (17--25 $\mu$m) images, with multiple images averaged over each epoch to improve SNR. In 2005, since no hydrogen images were available, images from 2006 were substituted. Likewise, stratospheric images from 2008 and 2009 were combined, owing to their limited number and similar radiances, and paired with the hydrogen images from 2008 (no hydrogen images were acquired in 2009).  Given the questionable photometric calibrations of the methane images, we perform two sets of retrievals---those with and without representative methane or CH$_3$D (7--9 $\mu$m) images included---to isolate their effect to the inferred temperature structure.  

We assume local thermodynamic equilibrium (LTE) in our retrievals, neglecting the effect of non-LTE on the emission. Non-LTE emission becomes increasing significant as the pressure and wavelength decrease. \citet{appleby1990ch4} estimated that at 7.7 $\mu$m, neglecting non-LTE can result of errors up to 2 K at 0.1 mbar, increasing to 20 K at 0.1 $\mu$bar. While the 12-$\mu$m ethane images have a peak contribution around 0.5 mbar, the $\sim$8-$\mu$m methane images do have peak contributions near 0.1 mbar, with lesser contribution from as little as 1 $\mu$bar, particularly at low emission angles near the limb. This leads to an additional error of roughly 2 K in temperatures retrieved from $\sim$8-$\mu$m data, growing slightly at extreme high and low latitudes. 

Finally, the retrieved meridional temperature structures were used in a forward model, assuming zonal uniformity, to once again produce synthetic images to validate by comparison with the data. The resulting idealized images were convolved with corresponding stellar PSFs and degraded with synthetic noise to allow a direct comparison with the observations. Random noise was modeled as a normal distribution of randomly generated values with a mean of zero and a standard deviation equivalent to that measured in the corresponding images (off of the disk). 

As is typically the case with such retrievals, our derived solutions are non-unique, especially considering that potential compositional changes will alter the retrieved temperatures. For example, if methane was relatively depleted in the stratosphere at high latitudes---as it is thought to be in the troposphere \citep{karkoschka2011haze,irwin2019latitudinal}---we would be underestimating the polar temperature in retrievals from the same radiance. Nonetheless, given the limits of our data, we attempted to represent the simplest solutions to the temperature field consistent with the observations. We assessed the accuracy of the solutions by evaluating the reduced $\chi^2$ of the retrievals and percentage differences between the modeled and true images. For the reduced $\chi^2$, we assume \begin{equation}
    \chi^2_\nu=\sum_{i=0}^{\nu}\frac{(data_i-model_i)^2}{\sigma_i^2}\frac{1}{\nu}
\end{equation}where we sum over the number of observed radiances $\nu$, and $\sigma$ is taken to be the calibration uncertainty. 

\subsubsection{Retrieval Results}

\begin{figure*}[h]
    \centering
    \includegraphics[clip, trim=.0in 0.0in 0.in 0.0in,scale=0.85]{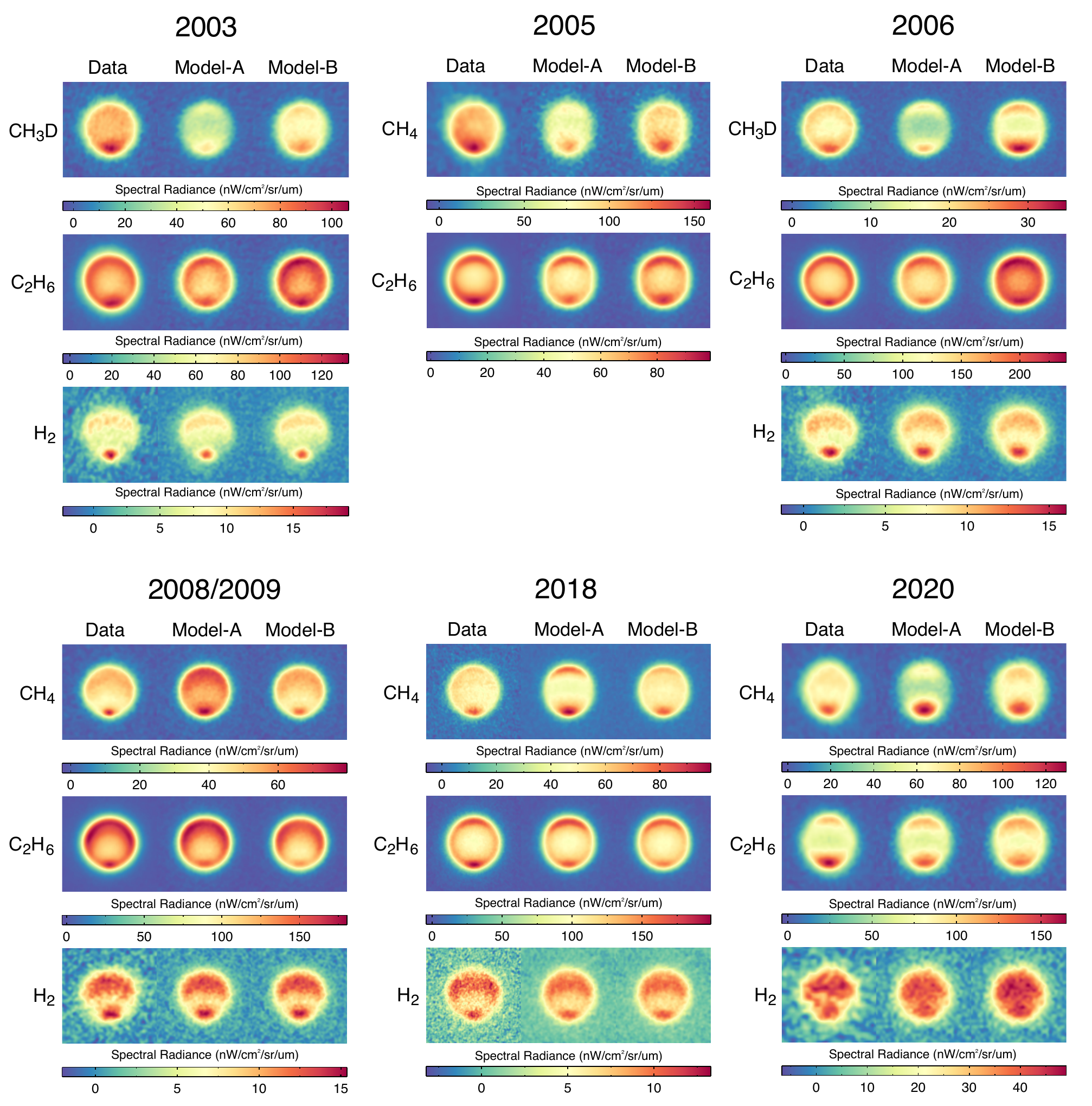}
        \caption{Comparison between observed and modeled radiances for methane CH$_3$D, ethane, and hydrogen images, grouped by epoch. For each trio of images, real data is shown on the left, while synthetic images derived from retrievals are to the right. Model-A represents models derived from temperature retrievals in which the methane observations were omitted, while Model-B included methane observations. Radiance values are indicated by the color bars. Images and models for the following filters are shown: 2003 --- 8.0, 12.5, 18.75 $\mu$m; 2005 --- Si1-7.9, Si5-11.6; 2006 --- PAH1, NEII$\_$1, Q2; 2008/2009 --- J7.9, NEII$\_$1, Q3; 2018 --- J7.9, NEII$\_$1, Q2; 2020 --- F04C07.80W0.70, F09C12.50W1.15, F42C24.50W0.80.}
    \label{fig:datvsmod} 
\end{figure*}

\begin{figure*}[h]
    \centering
    \includegraphics[clip, trim=.0in 0.0in 0.in 0.0in,scale=0.70]{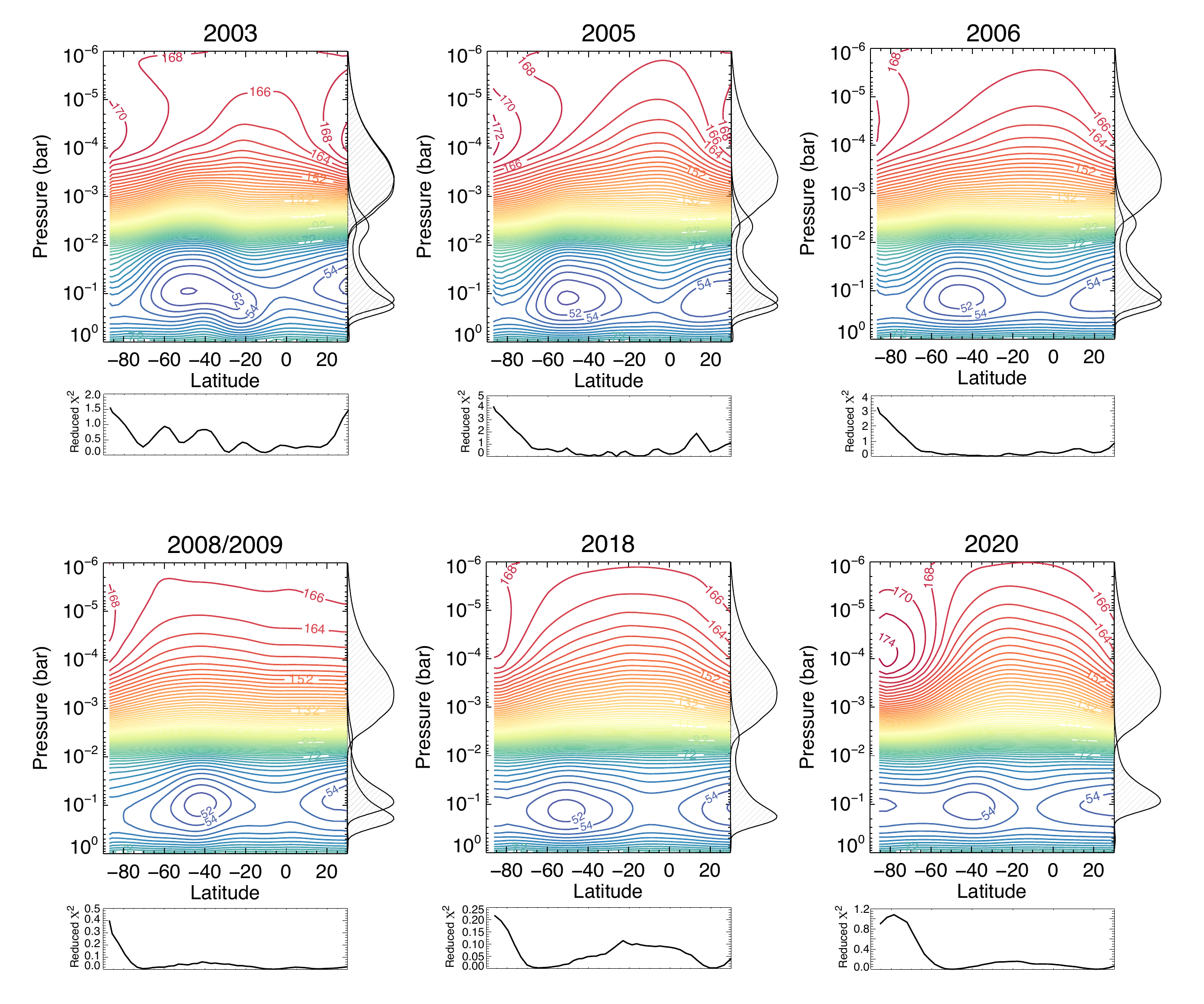}
    \caption{Contours of retrieved temperatures (K) from limited ethane and hydrogen sensing images for each epoch (averaged in time) corresponding to the models A in Figure \ref{fig:datvsmod}. Contours are drawn at 2 K intervals and color coded for clarity. The normalized contribution functions of the filters used for the retrievals are shown on the right of each panel; representative contributions are shown for moderate emission angles (45$^{\circ}$). Temperatures near the peaks of the contributions functions are well constrained, while temperatures at other pressures are not and tend towards the assumed initial profile. The reduced $\chi^2$ of retrievals for each latitude are also shown, indicating the goodness of the fits for each case.}
    \label{fig:tcountours} 
\end{figure*}

\begin{figure*}[h]
    \centering
    \includegraphics[clip, trim=.0in 0.0in 0.in 0.0in,scale=0.7]{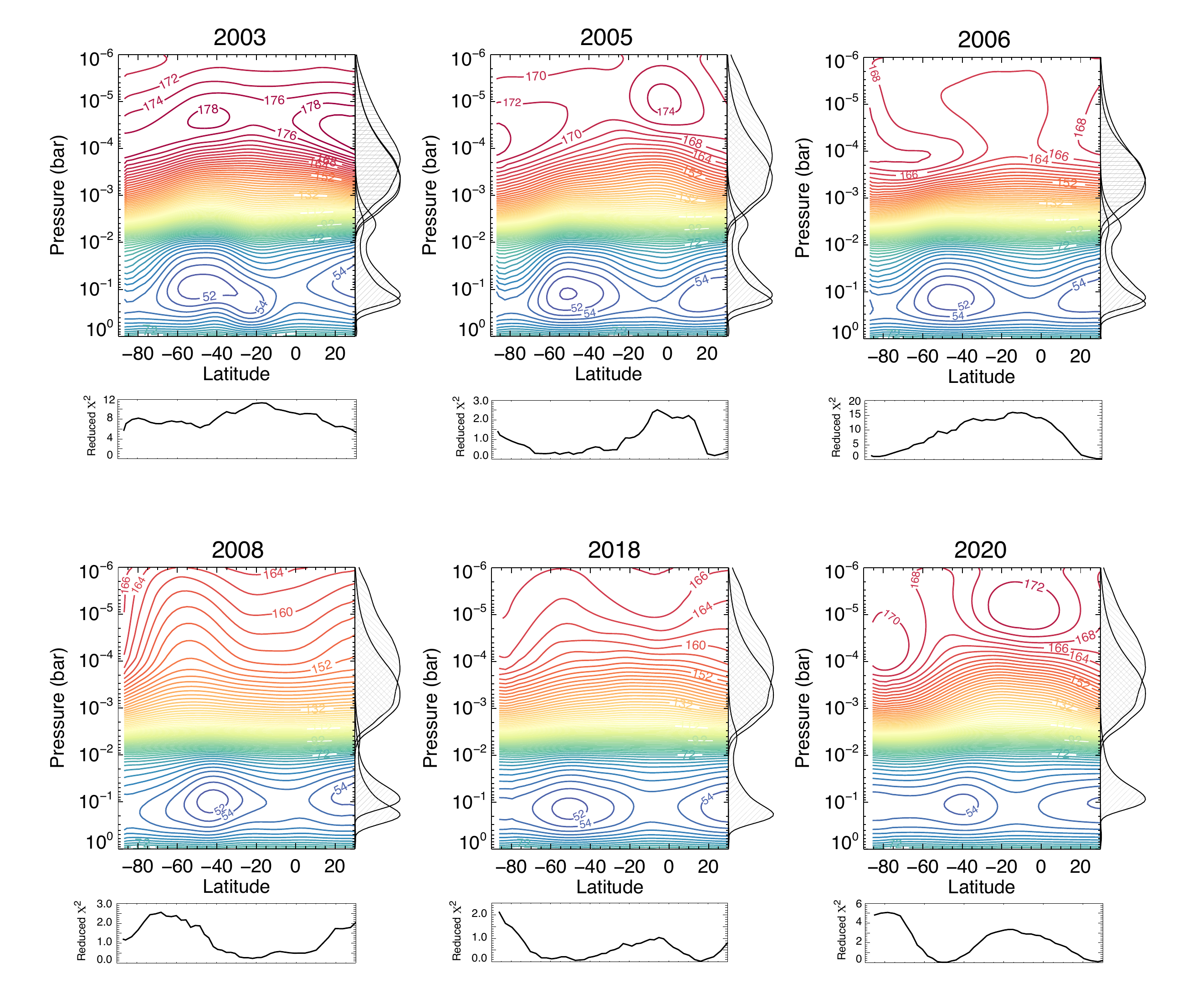}
    \caption{Contours of retrieved temperatures (K), as in Figure \ref{fig:tcountours}, but now with the methane and/or CH$_3$D images included in the retrievals. The methane/CH$_3$D images result in more complicated structures at pressures between $10^{-4}$--$10^{-6}$ bars and warmer stratospheric temperatures in most years, but with significantly poorer fits.}
    \label{fig:tcontoursch4} 
\end{figure*}

Comparisons between data and synthetic observations constructed from retrievals are shown in Figure \ref{fig:datvsmod}. In the figure, we refer to the models constructed from retrievals using only hydrogen and ethane images as \textit{model-A}, and those that additionally include methane images---or, if methane images are absent, CH$_3$D images---as \textit{model-B}. The corresponding temperature structures inferred from the retrievals are shown in Figures \ref{fig:tcountours} and \ref{fig:tcontoursch4}, along with reduced $\chi^2$s of each retrieval versus latitude.

The goodness of the fits and the temperature structures implied by the images vary significantly by epoch. The agreement between data and model is strongest in the hydrogen-sensing filters, where disk-averaged differences are just a few percent or less in nearly all cases. Models of the stratospheric sensing images have greater errors, with values depending on whether methane images are included in the retrieval.  When methane and CH$_3$D images are ignored (i.e. model-A), the ethane images are reproduced to within an average error of up to 10$\%$, with the largest errors seen at the northern and southern limbs. The $\chi^2$s are generally very small, suggesting our assumed uncertainties are too large, but $\chi^2$s increase significantly at latitudes south of -70$^{\circ}$ and north of 20$^{\circ}$ (see Figure \ref{fig:tcountours}).  These errors reflect the challenges of modeling the radiances at high emission angles, even for a single stratospheric filter, and likely indicate residual errors in our attempts to effectively deconvolve the images, as well as possibly errors owing to our plane-parallel approximation of the radiative transfer. The inferred temperatures at these poorer-fit locations should be considered somewhat less accurate, but nonetheless yield a reasonable match between the data and modeled images.  

\begin{figure}[h]
    \centering
    \includegraphics[clip, trim=.0in .20in 0.in 0.0in,scale=0.85]{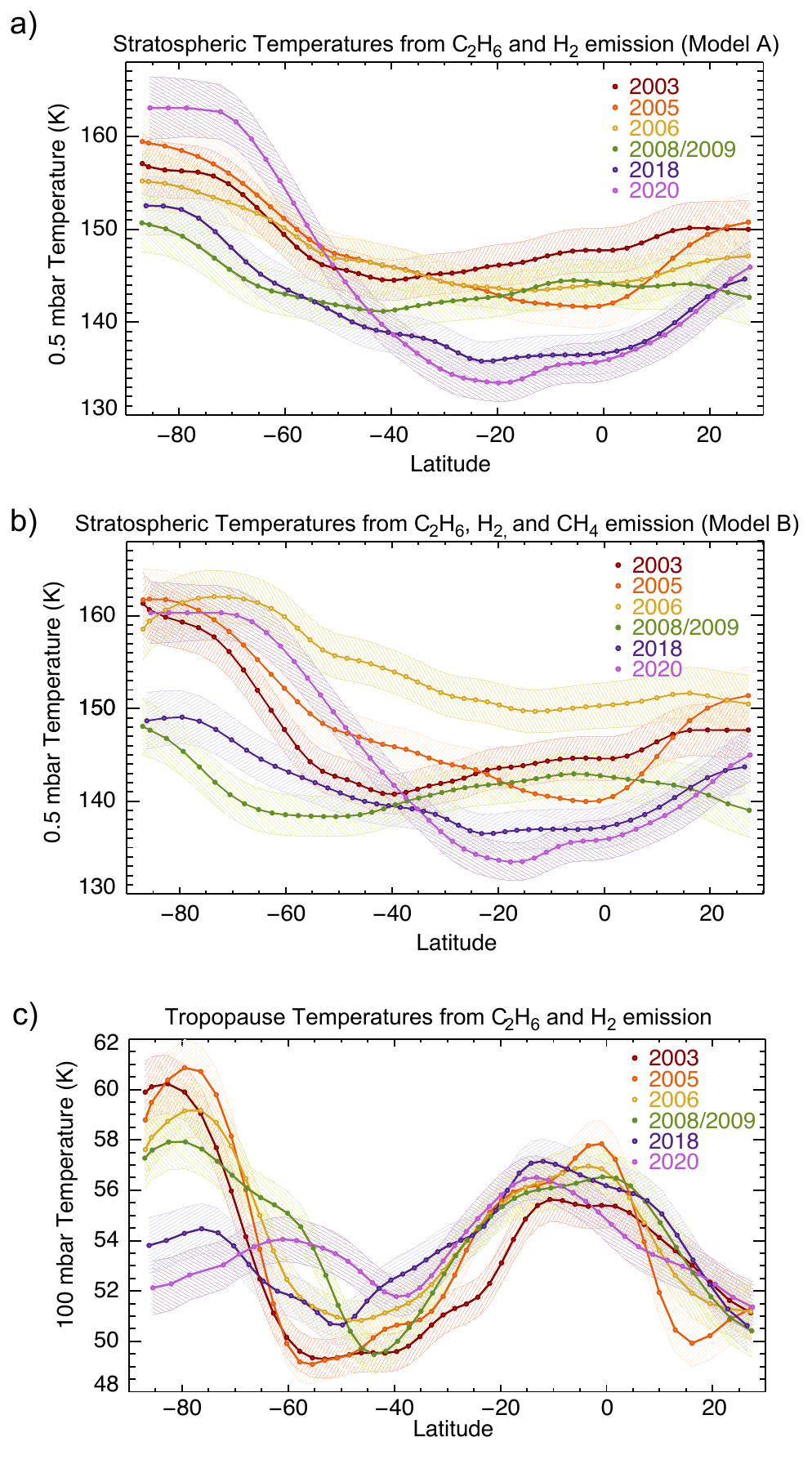}
    \caption{Retrieved temperatures versus latitude. The top two panels show temperatures at the 0.5-mbar peak of the ethane contribution, from retrievals including ethane ($\sim$12 $\mu$m) and hydrogen ($\sim$18-25 $\mu$m) radiances (models A, panel a), as well as those including less reliable methane/CH$_3$D ($\sim$8 $\mu$m) radiances (models B, panel b). Each epoch is indicated by color, and uncertainties from calibration are represented by the corresponding hatched envelopes. The 0.5-mbar temperatures are subject to an additional systematic uncertainty of roughly 10 K owing to uncertainty in the lapse rate.  The bright 2006 methane images result in higher temperatures, but poorer fits. Both models suggest latitudinal temperature contrasts between the warm south pole and cooler equator increased. Corresponding temperatures at 100 mbar (panel c) show no significant changes at low latitudes, but a possible drop in temperature at the poles, insensitive to whether methane radiances are considered or not.}
    \label{fig:tts_vs_lat} 
\end{figure}

The contour plots of Figures \ref{fig:tcountours} show retrieved temperatures for Model-A. Accompanying contribution functions (to the right of each panel) indicate the pressures at which temperatures are actually constrained, and Figure \ref{fig:tts_vs_lat} shows the retrieved temperatures versus latitude near the contribution peaks. In the upper troposphere, observations are consistent with the same basic temperature structure inferred from Voyager-IRIS. The tropopause temperatures in our data reach as low as 49--52 K at mid-latitudes and rise roughly 4--8 K warmer at the equator and south polar regions (Figure \ref{fig:tts_vs_lat}c).  In comparison, the Voyager-IRIS 100-mbar temperatures ranged from  $\sim$51 K at 45$^{\circ}$S to $\sim$57 K at the equator \citep{conrath1998thermal,fletcher2014neptune}. In our measurements, systematic uncertainties in radiances translate to a roughly 3 K systematic uncertainty in temperatures at the $\sim$100-mbar tropopause, with random noise and retrieval error adding uncertainties of $\sim$2 K.  The precise temperatures retrieved are also dependent on the wavelength of the Q-band filter used, since different filters sense slightly different pressures, as indicated by the contribution functions to the right of each panel in Figures \ref{fig:tcountours}. Differences of less than 2 K between epochs are likely not significant.

In the stratosphere, the temperature structures are more variable. When the methane/CH$_3$D emission is ignored, we see a pattern with cooler temperatures at equatorial and southern low-latitudes compared to warmer temperatures at the south pole and, in most cases, northern mid-latitudes.  Accompanying plots in Figure \ref{fig:tts_vs_lat}a show the range in retrieved temperatures versus latitudes at just the 0.5-mbar peak of the ethane contribution function. Latitudinal temperature contrasts between the warmer south pole and cooler equator increased from roughly 8 K to 28 K between 2003 and 2020; polar temperatures rose from 152 K to 163 K between 2018 and 2020 alone.  Uncertainties are roughly $\pm$ 2 K, increasing to 4 K near the poles and edges of the disk, where fits are poorer.  Models in 2008/2009, 2018, and 2020 best match the data, while 2003 and 2005 are poorest but still good at central latitudes. 

The retrieved stratospheric temperatures of Figure \ref{fig:tcountours} were constrained by 11--13 $\mu$m radiances; they did not consider the radiances from methane or CH$_3$D images, and their modeled emission poorly reproduce the observed radiances at 7--9 $\mu$m. At these shorter wavelengths, the forward modeled emission is either too dim (2003, 2005, 2006, 2020) or too bright (2008/2009, 2018), with average errors up to 30$\%$. This discrepancy suggests that the simple temperature and chemical structures inferred from the ethane images and the \textit{a priori} alone are not complex enough in all years.  

When the 7--9 $\mu$m radiances were additionally included in the retrievals (Models-B), the retrievals responded by increasing the 0.01 mbar temperatures in 2003, 2005, 2006, 2020, particularly at central latitudes, while slightly cooling the same heights in 2008/2009 and 2018 (see Figures \ref{fig:tcontoursch4} and \ref{fig:tts_vs_lat}b). The modeled methane/CH$_3$D images are improved, with errors reduced to 15$\%$ or less, but the ethane fits are in most cases worsened from errors of 10$\%$ to 15$\%$, as the retrievals apparently struggle to fit both channels simultaneously. In 2003, in particular, the ethane model appears too bright while the methane model is still too dim. The corresponding reduced $\chi^2$s greatly increase compared to those without methane or CH$_3$D, with values approaching or exceeding 10 in 2003 and 2006.  In contrast, both filter groups can be simultaneously fit to within a few percent in 2018, with reduced $\chi^2$s generally remaining $\le 1$. The reduced $\chi^2$s still increase towards the edges of the disk, but now also show an increase near the equator in most cases, corresponding to warmer equatorial temperatures. The brightness of the south polar feature also tends to be slightly too dim in our modeled ethane images, while simultaneously too bright in our modeled methane images. This inability to fit both simultaneously may indicate additional compositional variation may be present, potentially owing to the interplay between photochemistry and dynamical transport. For example, polar downwelling could potentially increase in the mid and lower stratospheric mixing ratio of ethane while simultaneously reducing that of methane, given the different vertical gradients in the mixing ratios of each \citep[i.e. ethane increases with height, methane does not;][]{moses2005photochemistry, moses2018seasonal}. Alternatively, it may also simply reveal errors in the calibration or the consequence of neglecting non-LTE emission, which makes it impossible to simultaneously reproduce radiances in multiple filters with similar contribution functions. 

The larger calibration uncertainties and lower SNR associated with the 7--9 $\mu$m imaging means retrieved temperatures are only certain to within $\pm$ 3 K, but errors may increase to $\pm$ 8 K near the pole and edges of the disk. 

Previously published values of latitudinal temperature and/or chemical abundance gradients in the stratosphere are limited and appear to differ from our results. Voyager mid-IR measurements of C$_2$H$_2$ emission in 1989 (\textit{L$_s$}$\sim$236$^{\circ}$) showed a maximum range of 20 K (or a factor of two in C$_2$H$_2$ abundance) at the 0.03--2 mbar level, with minimum at 55$^{\circ}$S and maximum between 0-20$^{\circ}$S \citep{bezard1991hydrocarbons}. We find similarly large contrasts at 0.5 mbar in the 2018 and 2020 data (\textit{L$_s$}$\sim$299$^{\circ}$ and \textit{L$_s$}$\sim$303$^{\circ}$, Figure \ref{fig:tts_vs_lat}), but with a minimum at 20$^{\circ}$S and a maximum at the pole. A minimum at mid-latitudes is found in 2003 (\textit{L$_s$}$\sim$266$^{\circ}$) and 2008/2009 (\textit{L$_s$}$\sim$278$^{\circ}$), but with only modest temperature contrast (3.5--5 K) between mid-latitudes and the equator. Given that 15 or more years separate these observations, it is possible that these considerable differences may be explained by genuine temporal variability and/or differences between the ethane and acetylene emission.

In contrast, two subsequent studies of Neptune spectra contemporary to the imaging data show comparatively little variation with latitude. \citet{fletcher2014neptune} analyzed Keck-LWS spectra from 2003---companion spectroscopy to the imaging analyzed here---and determined temperatures only varied by 3 K (155--158 $\pm$2 K) between 20$^{\circ}$N and 80$^{\circ}$S at 0.5 mbar.  Likewise, \citet{greathouse2011spatially} analyzed multiple 2007 Gemini-TEXES spectra, including methane and ethane emission in addition to the quadrupole emission analyzed here, to determine temperatures at 2.1 mbar, 0.12 mbar, and 0.007 mbar. To within measurement uncertainties, their retrieved values were consistent with latitudinally uniform temperatures, with variation of no more than 2 K between $\sim$70$^{\circ}$S and 12$^{\circ}$N (153.5--155.5 $\pm$ 2K at 0.12 mbar, 122--124 $\pm$ 2 K at 2.1 mbar). The variation increased to 5 K if the assumed stratospheric methane volume mixing ratio (VMR) was doubled to 1.5$\times10^{-3}$. Note that our assumed CH$_4$ profile from \citet{moses2018seasonal} has a mixing ratio of $1.15\times10^{-3}$ at 5 mbar, as derived from the Herschel observations of \citet{lellouch2015new}, decreasing to $0.9\times10^{-3}$ at 0.5 mbar as a result of vertical diffusion and chemical loss near the lower-pressure homopause. In both studies, the marginally greater temperatures appeared nearer the equator, with a slight minimum at southern mid-latitudes (i.e.,\ 55$^{\circ}$S in the TEXES analysis). 

%
These published values of 2003 and 2007 temperatures do appear relatively uniform in latitude compared to most of our inferred stratospheric temperature gradients at 0.5 mbar, shown in Figure $\ref{fig:tts_vs_lat}$a, but note that our 2003 and 2008/2009 curves have the weakest gradients. The 2003 curve only varies by 3.5 K between the equator and 60$^{\circ}$S (144.5--148 $\pm$2.5 K), while the 2008/2009 curves vary similarly between 20$^{\circ}$N and 70$^{\circ}$S (141--144.5 $\pm$2.5 K). Both curves also show a minimum at 40-50$^{\circ}$S and an increase towards the equator over this latitude range. Over this limited range, our results are largely consistent in trends with the previous analyses of separate but nearly contemporaneous data. It is only at latitudes observed nearer the edges of the disk that we detect significantly larger gradients in these same years. This discrepancy is perhaps reasonable to expect, given that the spectra were acquired at lower spatial resolution and with no attempted correction for beam dilution near the edges.  As suggested by the comparison between quadrupole spectra and ethane images (Figure \ref{fig:quadvlat}), beam dilution can suppress the radiances and consequent temperature gradients across the disk. 



\begin{figure}[h]
    \centering
    \includegraphics[clip, trim=.0in 0.0in 0.in 0.0in,scale=0.80]{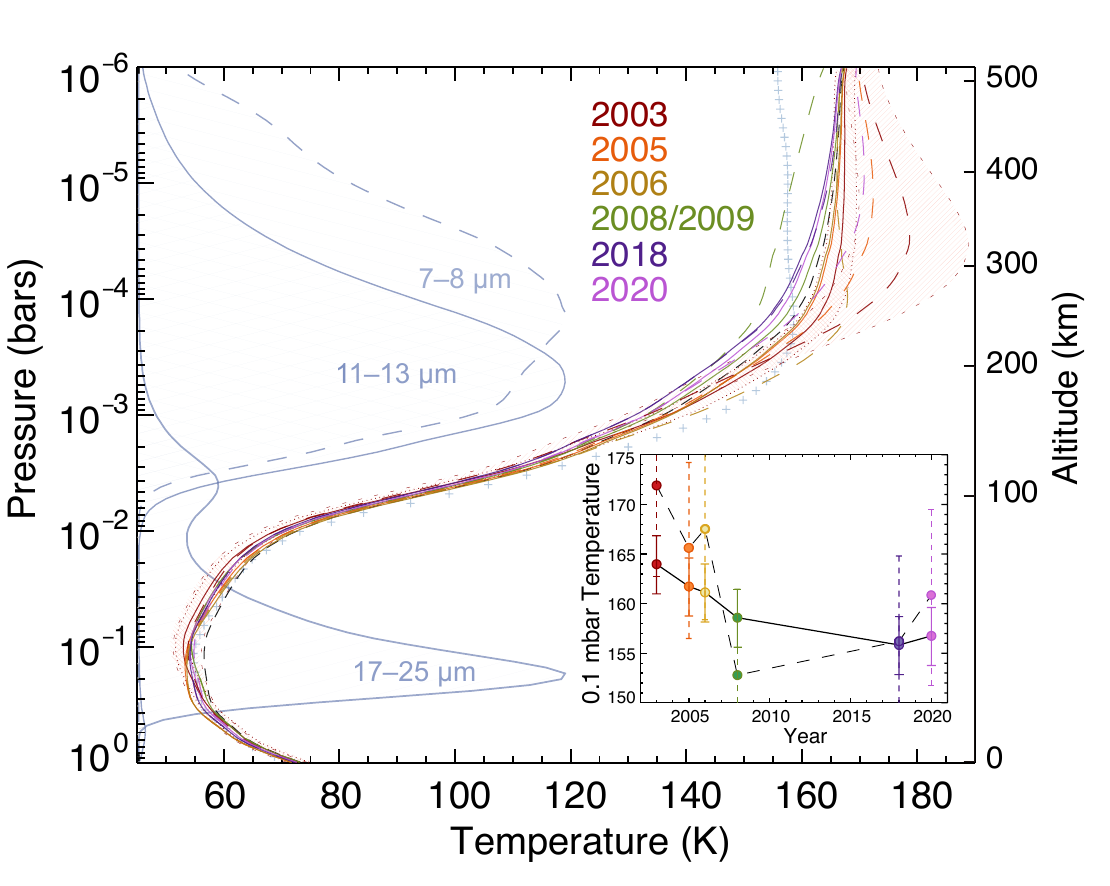}
    \caption{Disk-averaged vertical temperature profiles retrieved from the observations, corresponding Figures \ref{fig:datvsmod}-\ref{fig:tcontoursch4}, assuming all changes in radiance are attributed to temperature changes. Profiles are color-coded by the years as indicated, with multiple observations within each year averaged. Solid lines are temperature profiles retrieved from a combination of 11-13 $\mu$m (ethane) and 17-25 $\mu$m (hydrogen) images (i.e., model-A), while dashed lines additionally include 7-9 $\mu$m (methane or CH$_3$D, i.e., model-B). Typical normalized vertical contribution functions for each image group are suggested on the left. Pressures with weak contributions are not well constrained by the data and simply tend towards the initial \textit{a prior} profile, depicted by the black dashed line (from Figure \ref{fig:temp_conts}).  Hatching, enveloped by short-dashed and dotted-lines, represent the range of possible solutions for the 2003 retrievals, with and without methane radiances, given the assumed calibration uncertainties. Assumed uncertainties for other years are similar or less, but omitted for clarity. The temperature profile of \citet{fletcher2014neptune}, retrieved from 2007 AKARI/IRC spectra, is represented by the blue plus signs. (Inset) Temporal variation can easily be seen in the corresponding  0.1-mbar temperatures, plotted versus year.  Solid lines connect model-A results, while dashed lines apply to model-B results (i.e., additionally including methane/CH$_3$D), with $\sim\pm$3 K (solid) and $\sim\pm$9  (dashed) error bars, respectively; systematic uncertainties may be up to 10 K, as discussed in the text.}
    \label{fig:trets} 
\end{figure}

We also note that our retrieved 2003 temperatures (neglecting methane emission) are also about 10 $\pm$ 3 K colder at 0.5 mbar than \citet{fletcher2014neptune} retrieved from the contemporaneous spectra. Yet, the 0.1-mbar disk-integrated temperatures between 2003 and 2008 (as plotted in the inset of Figure \ref{fig:trets}) are remarkably consistent with contemporaneous 0.1-mbar temperatures reported by \citet{fletcher2014neptune}---well within the $\sim\pm$ 3 K uncertainties for retrievals from ethane images.  The apparent discrepancy at 0.5 mbar may therefore be explained by differences in the vertical resolution of the two datasets. The relatively broad contribution functions of the imaging data creates uncertainty in the retrieved temperatures at precise pressures, particular when the vertical temperature gradient is large.





To better assess our vertical temperature gradients, we computed disk-averaged temperature profiles for each epoch, as shown in Figure \ref{fig:trets}. Compared to the profile of \citet{fletcher2014neptune}, derived from 2007 AKARI/IRC spectra \citep{fletcher2010neptune}, our temperature profiles are slightly colder and increase more slowly with height at 10--1 mbar, but more rapidly with height at $<$0.5 mbar, before becoming nearly isothermal at relatively lower pressures.  As a result, our retrieved temperatures are comparatively cooler at 0.5 mbar (consistent with the discrepancy noted above), but warmer at 10$^{-5}$ bars, such that the integrated thermal emission is comparable. This difference in lower stratospheric lapse rates appears to follow directly from equivalent differences in the chosen \textit{a priori} profiles, indicating that the precise lapse rate at these pressures is not strongly constrained by these imaging data. Hence, considering the discrepancy, we may ascribe an uncertainty of roughly 10 K when defining temperatures at precise pressures in this region of strong vertical gradient.  This weakness is unsurprising, given the relatively broad contribution functions of the imaging filters. The greater vertical resolution provided by spectroscopic data is better suited for vertical structure analysis, and so the profiles and precise pressures reported here should not be interpreted as contradictory to the previous work. Rather, we emphasize that the primary strength of the images is their ability to clearly reveal variation in radiance with latitude and time, and this variation will significantly affect the disk-averaged temperature retrievals. 

We find that our retrieved temperature profiles show variation across the years at pressures less than 1 mbar. If the methane radiances are discounted, the profiles mostly fall between the profiles of \citet{moses2005photochemistry} (based on spectral observations from the early 1990s) and \citet{greathouse2011spatially} (based on 2007 TEXES data), from which studies our prior is derived. Our profiles differ by a maximum of only about 6 $\pm$ 4 K at 1 mbar and 8 $\pm$ 4 K at 0.1 mbar, from a maximum in 2003 to a minimum in 2018.  If the methane radiances are included, the range increases to 20 $\pm$ 14 K at 0.1 mbar, owing mostly to much warmer temperatures in 2003-2006 and cooler temperatures in 2008/2009. These results are roughly consistent with differences of less than 10 K at 1 mbar and 6 K at 0.1 mbar among the multiple retrievals from ground-based spectroscopy (2003-2007) presented in \citet{fletcher2014neptune}. However, our 2003 result based on methane radiances appears as a questionable outlier---$15^{+10}_{-8}$ K warmer at 0.04 mbar than was retrieved from the contemporaneous Keck-LWS spectra. This discrepancy possibly indicates a calibration error in the methane radiances, which may be exaggerating measured temperatures in a year that already appears exceptionally warm based on ethane images alone. 

\section{Discussion}\label{sec:discuss}

\subsection{Temporal Variation in Mid-Infrared Emission}

Although glimpses of mid-infrared temporal variability on Neptune have been suggested by previous studies \citep{hammel2006mid,greathouse2011spatially,fletcher2014neptune}, the sporadic sampling and considerable observational uncertainties prevented a more conclusive evaluation of the extent and nature of these changes. By applying consistent calibration and analysis approaches to the entire mid-infrared imaging dataset, starting with the raw data and accounting for systematic differences, we have revealed a clearer and fuller picture of Neptune's variability over much of the past two decades. The consistency of the observed trends---seen across multiple filters and observatories---provide compelling evidence of both global and latitudinal changes on Neptune. 

The data reveal that Neptune's global stratospheric radiances fell between 2003 and 2010. The reduction in radiances appears unequivocal; the precise temperature changes, however, are sensitive to the assumed \textit{a priori} temperature profile, chemical model, and whether observations in the 7-9 $\mu$m range are considered reliable.  With our adopted chemical model \citep[following][]{moses2018seasonal}, reliable ethane calibrations suggest a global drop of at least 6--8 $\pm$ 4 K, but possibly twice that amount if the anomalously large 2003 methane radiances are trusted (Figure \ref{fig:trets}).  In contrast, the deeper, Q-band hydrogen images show little change in inferred global temperatures. 

If we assume the radiances in 11--13 $\mu$m images are indicative of stratospheric temperature gradients (and not compositional gradients), we see in Figure \ref{fig:tts_vs_lat}a that the relative uniform, warm temperatures at $\sim$0.5 mbar in 2003 became cooler at the equator by 2005 (by $\sim$6 $\pm$ 3.5 K). The equatorial radiances and temperatures rebounded by at least 3 K in 2006, but then temperatures also began to fall at southern mid and high latitudes in the following years. Temperatures were nearly latitudinally uniform in 2007 (as suggested by TEXES and, albeit poor quality, T-ReCS data), but by 2008, the southern mid-latitudes were $\sim$3 K colder than the equator. This trend persisted into 2010 at least, and likely into 2012, when lone COMICS observations showed possibly the largest normalized equatorial radiances in both ethane and methane imaging.  There is a gap in observational records at all wavelengths in the following six years, until VISIR N-band images in 2018 revealed a colder, relatively uniform stratosphere at low latitudes. These low latitude temperatures remained steady into 2020, while the south polar regions warmed dramatically, increasing by 11 K in just two years.

Interestingly, the two methane images from 2020 (Figures \ref{fig:img_comparison}d; Figures \ref{fig:diskch4}c), while brighter at the south pole, do not show as large an increase in polar temperatures and latitudinal contrast as the ethane images, despite sensing similar pressures (see contribution functions in Figure \ref{fig:temp_conts}).  Moreover, while the hydrogen quadrupole spectra show an increase in stratospheric equator-to-pole temperature contrast since 2006, the hydrogen images from 2018 and 2020 show upper tropospheric radiances at the pole actually decreased since 2006. The tropospheric hydrogen data may be misleading given their low SNR, but if accurate, these data suggest that the polar temperatures increased over a relatively limited vertical range in the lower stratosphere while simultaneously decreasing in the upper troposphere.  Such vertical discontinuity in temperature would be difficult to explain with simple adiabatic warming alone within the single down-welling branch of an extended circulation cell \citep[\textit{e.g.},][]{dePater2014neptune} and would suggest the role of additional radiative or chemical processes. The particularly strong $\sim$12-$\mu$m emission may indicate that the south pole has not only grown warmer but also richer in ethane, as theory would predict given increased seasonal insolation \citep{moses2005photochemistry,moses2018seasonal}. This would be similar to what has been observed on Saturn during the formation of its north polar stratospheric vortex \citep{fletcher2018saturnhexagon}. As noted previously, ethane mixing ratios in the lower stratosphere could also be preferentially increased (relative to methane) by stronger polar downwelling simply because the environmental ethane mixing ratio increases with height, while the methane ratio does not \citep[\textit{e.g.},][]{moses2005photochemistry,moses2018seasonal}. 

The decline in stratospheric radiances between 2003 and 2009 appears beyond doubt, and is consistent, in part, with trends hinted at with previous spectral observations.  These include: 
\begin{itemize}
    \item{A roughly 40\% decline in disk-integrated radiances between 2003 and 2004,} reported by \citet{hammel2006mid} using N-band spectra from NASA’s Infrared Telescope Facility (IRTF) and Broadband Array Spectrograph System \citep[BASS;][]{hackwell1990low}.

    \item{A drop in radiances between the 2003 Keck-LWS and 2007 AKARI/IRC spectra, interpreted by \citet{fletcher2014neptune} as a marginal $\sim$4 K decrease in the 0.1-mbar temperatures, or, alternatively, a $\sim$30$\%$ decrease in the ethane mole fractions.}
\end{itemize}   

To add to our assessment of temporal variability, we make use of previous spectral observations by computing equivalent filter-integrated disk radiances from the following N-band spectra: 2003 Keck-LWS; 2005 Gemini-Michelle; 2007 AKARI/IRC; and 2007 Gemini-S-TReCS, all previously analyzed by \citet{fletcher2014neptune}; along with 2002, 2003, and 2004 IRTF BASS spectra \citep{hammel2006mid}.  We chose the passbands of VISIR's J7.9, PAH1, NEII$\_$1 filters to represent of the methane, CH$_3$D, and ethane radiances.  Although \citet{fletcher2014neptune} reported no consistent trend with time for inferred 0.1-mbar temperatures between 2003 and 2007, we find that the filter-integrated radiances are overall roughly consistent with the larger trend seen in the imaging data (as plotted in the top panel of Figure \ref{fig:time_trends}, with the above, archival filter-integrated values in gray). 

A longer trend can be inferred by also including older measurements: the 1985 and 1991 IRTF spectra \citep{orton1987spectra,orton1992thermal}, as previously noted by \citet{hammel2006mid}, along with the 1997 ISO-PHT-S N-band spectra \citep{schulz1999detection} and ISO-SWS H$_2$ S(1) quadrupole spectra \citep{feuchtgruber1999detection}. When similarly compared (Figure \ref{fig:time_trends}; top plot, also in gray), these few measurements suggest an increase in radiances between 1985 and 2003, followed by the decline ever since. However, the sporadic sampling makes it impossible to establish any meaningful trend in mid-infrared radiances with confidence prior to 2003. Likewise, the few measurements between 2010 and 2017 leave the temporal trend unfortunately poorly resolved over the past decade. 

\subsection{Correlations and Possible Causes of Variation}
\begin{figure*}[h]
    \centering
    \includegraphics[clip, trim=.0in 0.0in 0.in 0.0in,scale=0.90]{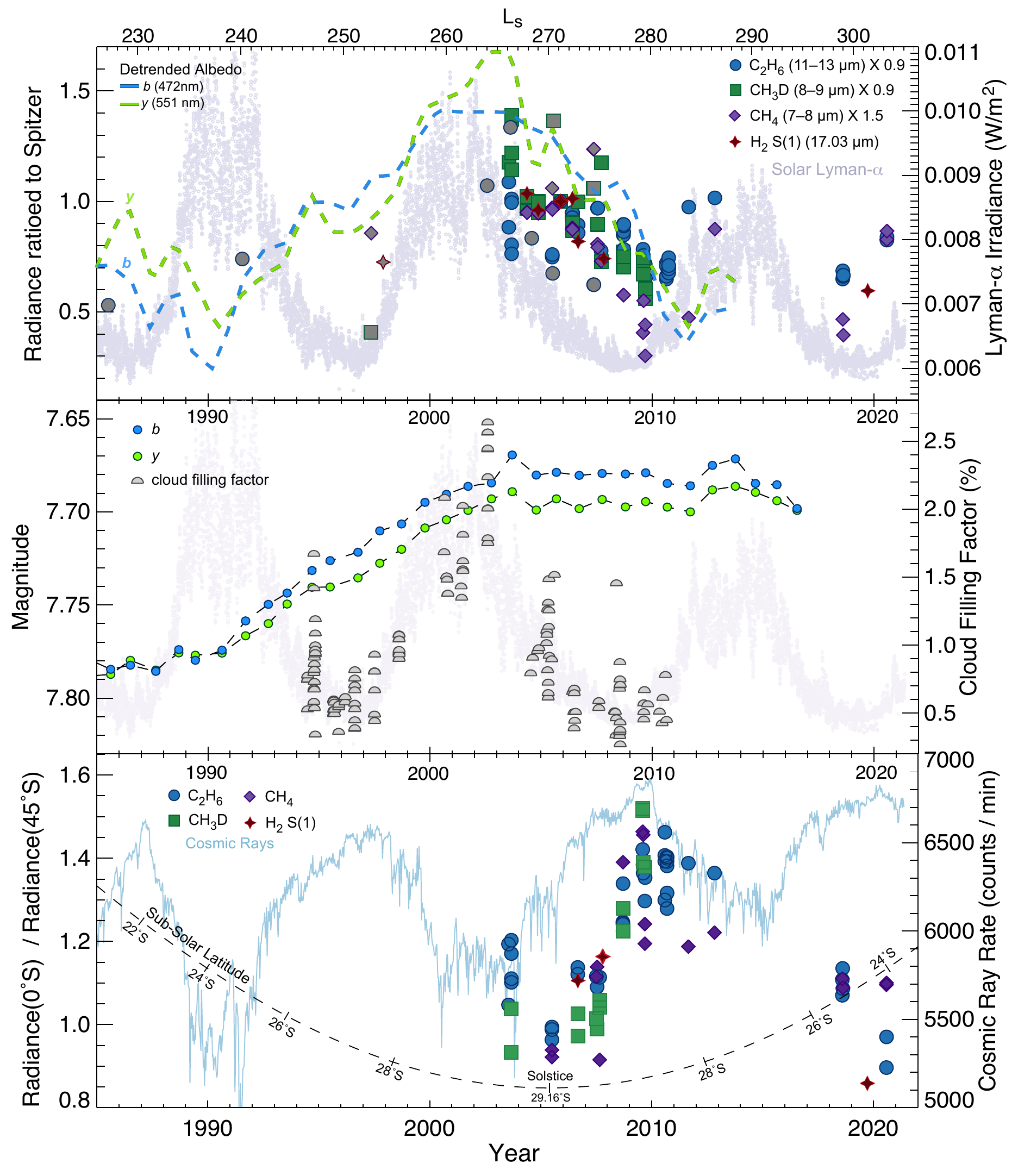}
    \caption{Temporal behavior for our thermal measurements compared to other observables. (Top panel) Disk-integrated radiance ratios versus time for all stratospheric observations, scaled to reduce offset and separated by group as indicated by the symbol key, compared to the composite solar Lyman-$\alpha$ irradiance (light purple) at Earth from \citet{Machol_lymanalpha}.  Gray-filled symbols represent published measurements as described in the text. y- and b- filter magnitude fluctuations  (detrended to remove greater seasonal variability) from \citet{aplin2016determining} (following \citet{lockwood2006photometric}), are also plotted, scaled from 0.02 to -0.02 in magnitude (i.e. brightness increasing upwards). (Middle panel) Photometric magnitudes of Neptune in the \textit{b} (472 nm) and \textit{y} (551 nm) filters from \citet{lockwood2019final} show an increase in scattered light. Also shown are the cloud filling factors of \citet{karkoschka2011neptune}, which represent the abundance of discrete cloud features, along with the solar Lyman-$\alpha$ irradiance again for comparison. (Bottom panel) The asymmetry in our images plotted as the fractional radiance at the equator compared to 45$^{\circ}$ S latitude, with symbols as indicated in the top panel.  Also shown are the sub-solar latitude (dashed line), reaching a minimum at the 2005 southern summer solstice, and the cosmic ray rate measured from the Sodankyla Geophysical Observatory.}
    \label{fig:time_trends}
\end{figure*}

The cause of the variations revealed here are unknown, but we can begin to speculate and evaluate mechanisms based on the characteristics and time scales of these changes. The temporal variation of multiple observed phenomena are summarized in Figure \ref{fig:time_trends} and discussed in the following sections. \newline

\subsubsection{Seasonal Effects}
Seasonal variation owing to changes in the sub-solar latitude are expected to produce variability in Neptune's temperatures \citep[\textit{e.g.},][]{conrath1990temperature, greathouse2008general} and stratospheric photochemistry \citep{moses2018seasonal}, both of which can potentially alter observed mid-IR radiances. 

In radiative modeling, seasonal temperature variation is typically characterized by the length of the radiative time constant relative to the orbital period. Radiative time constants near the tropopause are relatively long (approaching a century, Figure \ref{fig:temp_conts}), and so the observed lack of variation in Q-band images is consistent with expectations.  However, given shorter radiative time constants in the stratosphere \citep{li2018high}, the stratospheric temperatures should vary across seasons, reaching a maximum in the summer hemisphere shortly after the peak insolation at summer solstice \citep{conrath1990temperature, greathouse2008general}. The imaging data bracket Neptune's 2005 southern summer solstice (\textit{L$_s$}$\sim$266--303), but rather than warming, we observe a decline in the disk-integrated stratospheric radiances and implied temperatures. Only between 2018 and 2020 do the polar regions appear to increase in radiance.

Likewise, the photochemical modeling of \citet{moses2018seasonal} predicts a peak in the southern hemisphere ethane abundance following the maximum solar flux at solstice. As methane is photolyzed by high energy photons, its photolysis leads to the production of ethane and other hydrocarbons.  Changes in photochemistry are expected to be greatest at lower pressures---where CH$_4$ photolysis rates are largest and chemical time constants are shortest---and higher latitudes---where insolation varies most greatly with seasons. As with the temperatures, the variation is muted and delayed with increasing pressure---in this case owing to longer advective timescales as the ethane slowly descends.  The \citet{moses2018seasonal} model predicts that ethane should vary significantly at pressures less than 1 mbar, varying annually at by a factor of nearly $\sim$2.5 near the poles at 0.1 mbar.  Across the southern hemisphere, the 0.1-mbar ethane volume mixing ratio (VMR) and column abundance is expected to peak around \textit{L$_s$}$\sim$300 (roughly 2019) and fall gradually to a minimum following winter solstice.  In contrast, the seasonal variation of methane should appear insignificant in comparison to its overall abundance at the pressures probed by the 7--8 $\mu$m observations. Therefore, variation in seasonal photochemistry should cause ethane radiances (11--12 $\mu$m) to increase near solstice, while the methane radiances (7--9 $\mu$m) remain constant. Yet, this appears inconsistent with our observations, which decline in radiance across all stratospheric-sensing wavelengths between 2003 and 2010. 

The explanation for the apparent inconsistency between seasonal models and observations may be resolved by the feedback between photochemical and radiative processes. While methane absorbs sunlight and warms the atmosphere, photochemically produced hydrocarbons---primarily ethane and acetylene---are powerful infrared emitters that serve to cool the stratosphere. The balance between this radiative heating and cooling changes as the amount of photochemical hydrocarbons changes.  Applying a coupled radiative-chemical model to Saturn's atmosphere,  \citet{hue20162d} showed that this interplay can produce a peak in summer temperatures \emph{prior} to maximum insolation at summer solstice, as the late spring production of infrared emitters overwhelmingly cools the stratosphere, counteracting the growing solar heating. Although the effect was limited to pressures less than 0.1 mbar in the Saturn study, an equivalent process in Neptune's atmosphere may help explain the observed decline in stratospheric radiances beginning in 2003---prior to the solstice.  Since the observed radiances are sensitive to both the temperature and abundance, the indirect effect of cooling the atmosphere would need to dominate over the direct effect of increasing the ethane to explain the decreasing trend at 11-13 $\mu$m radiances. Indeed, our modeling in Figure \ref{fig:stratvstime} shows that expected seasonal variation in ethane would produce only a very slight increase in the disk radiances ($\lesssim$1$\times10^{-8}$ W/cm$^2$/sr/$\mu$m) over the observed period.  The likely important effect of photochemistry on the seasonal temperatures and radiances suggested by our results should be further investigated with coupled radiative-chemical modeling in future work.

Nonetheless, given Neptune's 165-year orbital period, any seasonal changes are expected to occur gradually over decades. The rapid changes observed between 2018 and 2020 appear surprisingly swift for seasonal response, particularly considering that the south pole has been constantly illuminated since 1963. Nor can the slowly changing sub-solar latitude explain the sudden change in meridional gradients between 2006 and 2008. Additional processes appear to be operating in Neptune's atmosphere on sub-seasonal timescales, and on both regional and global scales.


\subsubsection{Weather and Variation in Visible and Near-IR Observations}

\emph{Cloud and haze activity---} As evidenced by sporadic clouds and occasional vortices, Neptune clearly exhibits variable weather in the upper troposphere and lower stratosphere on sub-seasonal timescales; therefore, it is worth considering if this weather can be related to the stratospheric variability. 

In an analysis of Hubble Space Telescope (HST) visible and NIR imaging, \citet{karkoschka2011neptune} found that trends in discrete cloud activity changed on a timescale of $\sim$5 years.  These discrete clouds were inferred to form near the tropopause. It was speculated that discrete cloud variation may be related to variation in large-scale dynamics at these pressures. \citet{karkoschka2011neptune} also reported variation in the albedo of hazes and dark bands on similar timescales, but these were interpreted as changes in the aerosol abundances located much deeper in the atmosphere ($>$ 1 bar). 

The observed maximum mid-infrared radiances in 2003 roughly coincided with the peak in the discrete cloud coverage in 2002 as determined by \citet{karkoschka2011neptune}.  These peaks also roughly corresponded with the apparent seasonal peak in Neptune's disk-integrated visual albedos (472 nm and 551 nm), which had steadily brightened since the 1950s before levelling off in the early 2000s \citep{lockwood2002photometric, sromovsky2003seasonal_response, lockwood2006photometric, lockwood2019final}. The subsequent decline in the stratospheric emission across multiple filters from 2003 to 2009 then coincided with a period of declining discrete cloud and plateauing visual magnitudes.

These coinciding trends suggest an intriguing connection between the mid-IR, discrete cloud cover, and the visual albedo, but the physical mechanism linking these variables is not obvious.  Most easily observed at near-IR wavelengths, the high discrete clouds appear to contribute relatively little to overall visual ($\sim$500 nm) albedo, which is thought to be most sensitive to deeper aerosols \citep{karkoschka2011neptune}. Correlated variability would therefore suggest a link over many pressure scale heights, linking the upper troposphere and stratosphere.  Implied variation near the tropopause is important because it marks the broad, cold boundary across which tropospheric methane must be transported to the stratosphere to produce ethane and other photochemical derivatives.  Precisely how methane is transported from the troposphere to the stratosphere is unresolved given the tropopause saturation cold trap \citep{smith1989voyager, baines_UV_neptune_1990,orton2007evidencehotspot, dePater2014neptune}, but moist convection has been suggested to play a role \citep{stoker1986moist,lunine1989abundance, sinclair2020spatial}. If abundant clouds and hazes are indicative of convection or higher methane humidity, then perhaps some meteorological upwelling mechanism temporarily enriched the tropopause and lower stratosphere in hydrocarbons prior to 2003.  As a consequence, the radiative heating rates at these levels could have temporarily increased, until the upwelling ceased, leaving the methane, clouds, and temperatures to decline over the following years.  However, this is purely speculative, as no variation in the stratospheric methane abundances has been identified.  

Alternatively, inertia-gravity waves originating from intermittent convective plumes, vortices, or other long-lived, cloud-generating disturbances may cause variable heating as waves break in the stratosphere. Indeed, \citet{roques1994neptune} showed variable gravity wave heating could explain anomalous, variable structures in the upper stratospheric temperature profiles inferred from stellar occultations between 1983 and 1990.  Periodic, sub-seasonal variation in this heating may be possible if it is associated with planetary oscillations, such as those seen on Jupiter and Saturn, which perturb stratospheric temperatures at low latitudes \citep{leovy1991quasiquadrennial,friedson1999_QBO,orton2008saturnoscillations,fletcher2017saturndisruption,guerlet2018equatorial}. If analogous oscillations are operating on Neptune, they may help explain the mid-IR variation at lower latitudes, but many decades of consistent observations would be needed to establish a periodicity with statistical significance. The implied possible correlation between clouds and temperature so far is still based on a limited number of observations and not without exception, as an unusually cloudy outburst at the equator in 2017 \citep{molter2019neptune} has shown.  

Other recent visible changes in clouds and hazes at greater pressures include a dark vortex in the north tropics, first observed by the Hubble Space Telescope (HST) beginning in 2018 \citep{simon2019darkspot}. Dark vortices are examples of occasional disturbances that can potentially generate waves and affect weather over a large vertical range. They appear to be localized regions of reduced scattering or upwelling of low-albedo aerosols from below, often triggering orographic-like companion clouds at higher altitudes \citep[\textit{e.g.},][]{smith1989voyager, sromovsky1993dynamics,dePater2014neptune,huesoneptunelonglived2017}. The precise height and vertical extent of Neptune's dark vortices are unknown, but retrievals suggest they are centered well below the stratosphere, nearer a pressure of 7 bars or more \citep{irwin2022hazy}. 
\begin{figure}[h]
    \centering
    \includegraphics[clip, trim=.0in 0.0in 0.in 0.0in,scale=0.85]{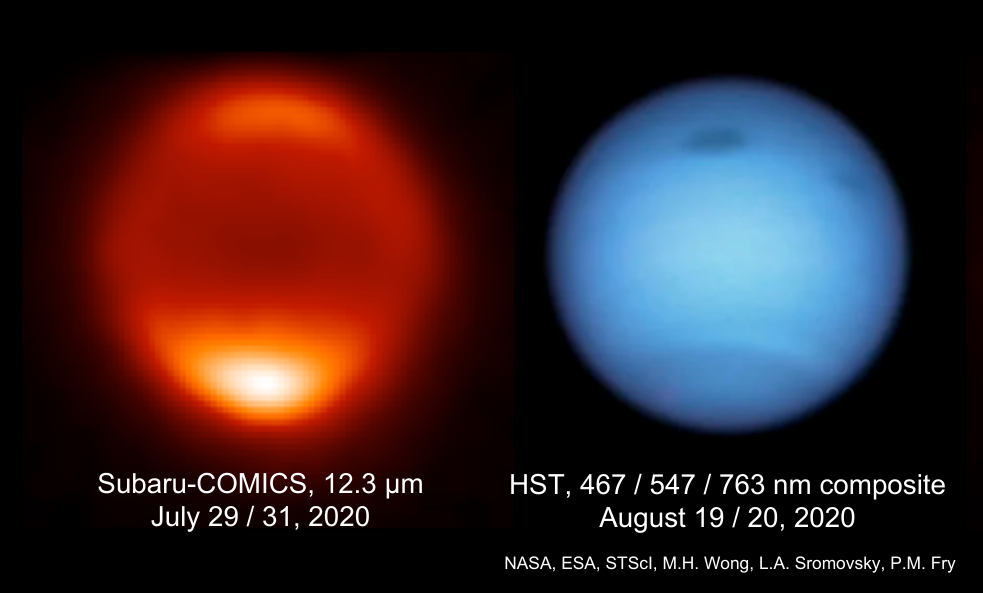}
    \caption{Comparison between a mid-infrared and a visible images acquired only weeks apart in 2020. (Left) Subaru-COMICS 12.3-$\mu$m image averaged from observations on July 29 and 31, 2020, with prominent south polar emission and dim low- to mid-latitude and equatorial regions. (Right) Hubble Space Telescope composite visible image (combining F467, F547, and F763 filtered images) from August 19 and 20, 2020, with albedo gradients demarcating the south polar region ($\sim$ 60$^{\circ}$ S) and two dark vortices at northern low-latitudes  ($\sim$ 15$^{\circ}$ N) (HST Image credit: NASA, ESA, STScI, M.H. Wong (University of California, Berkeley), and L.A. Sromovsky and P.M. Fry (University of Wisconsin-Madison). }
    \label{fig:spot} 
\end{figure}

Dynamical modeling of the dark vortex predicts that these features are associated with a decrease in the methane mixing ratio within the vortex, along with a slight drop in temperature above \citep{stratman2001epic,hadland2020epic}, although Voyager detected very little if any temperature anomaly over the Great Dark Spot \citep{conrath1989neptune} in 1989. The recent prominent dark vortices were imaged by HST at visible wavelengths multiple times, including in August 2020, just three weeks after our mid-infrared Subaru-COMICS observations (Figure \ref{fig:spot}). Our imaging data show no obvious localized anomaly in the upper troposphere or lower stratosphere in 2020, but the low SNR and blurring of long integration times render such detection unlikely.  However, we do note that the equatorial stratosphere appears anomalously cold over most of the disk in recent years, with exception to the south pole. Aside from the Voyager IRIS spectra in 1989, no spatially-resolved mid-IR data have been available during any past vortex appearances \citep[\textit{e.g.}, see][]{wong2018new}, and so it is unclear whether the relatively large stratospheric temperature gradient between the cool equator and warmer, northern mid-latitudes is in any way related to the dynamical environment in which these vortices form. We also note that the discrete cloud features---so vividly associated with the dark vortex in Voyager images---appear largely absent from this present spot in 2020 (Figure \ref{fig:spot}) and 2021, despite earlier companion clouds seen in 2018 and 2019. Although a number of dynamical factors may explain the current lack of clouds (\textit{e.g.}, the vertical extent of the vortex), the limited condensation, combined with our temperature findings, may suggest that the overlying atmosphere has grown not only colder but drier in recent years. 

Establishing whether any of the meteorological changes discussed above are related to the trends in the stratosphere will require additional observations at mid-IR, Near-IR and visible wavelengths. Potential processes coupling the troposphere and stratosphere should continue to be investigated in future work.

\subsubsection{Solar Effects}
Aside from seasonal modulation, the solar flux varies as part of the roughly 11-year solar cycle. While the total solar irradiance differs by little more than a 0.1\% over a typical solar cycle \citep{Kopp2019solar}\footnote{SORCE Level 3 Total Solar Irradiance Daily Means, version 018, 
Accessed 2021 March 11 at doi:10.5067/D959YZ53XQ4C}, variation in regions of the ultraviolet---as represented by a time series of 121.57-nm Lyman-$\alpha$ irradiance measured at Earth in Figure \ref{fig:time_trends}---exceeds 40\%. High energy photons (of wavelengths less than $\sim$ 145 nm) are the main drivers behind the methane photochemistry \citep{moses2020icegiantchem}, and so modulation in the ultraviolet flux can potentially produce observable variation in photochemistry if chemical timescales are sufficiently short \citep{moses2005latitudinal}.  

A strong solar maximum occurred in $\sim$2001 (cycle 23), followed by a drop in the Lyman-$\alpha$ over the subsequent 10 years. The deep minimum in 2009 (cycle 24) gave way to a weak solar maximum in $\sim$2014. Currently, solar cycle 25 is rebounding from another deep minimum in 2019. Plotted with the time series of Lyman-$\alpha$, Neptune's mid-IR radiances followed a roughly similar trend. 

However, as with the seasonal cycle, most of the photochemical response to changing far-ultraviolet flux of the solar cycle is expected to occur at relatively lower stratospheric pressures than the mid-IR images sense---nearer to 1 $\mu$bar, where H-Lyman-alpha and other short-wavelength radiation is absorbed and the radiative, chemical, and transport time-scales are shortest. At the 0.05--1-mbar pressures sensed by the ethane filters, the photochemical variation will be dampened \citep{moses2018seasonal}. And although methane filters are sensitive to somewhat lower pressures (0.005--0.5 mbar), the methane mole fraction is not expected to show appreciable change given that its photochemical loss is easily replenished from the abundant source below. Combined with the quadrupole observations, the observed mutual variation in the radiances would therefore again suggest a change in the temperatures, rather than chemistry, with a mechanism possibly linked to solar cycle.   

A possible correlation between the solar cycles and upper stratospheric temperatures at even lower pressures had been reported previously by \citet{roques1994neptune}. Analyzing ground-based stellar occultations dating between 1983 and 1990, \citet{roques1994neptune} retrieved temperatures at pressures of 0.01--0.03 mbar. They found temporal variations of nearly 60 $\pm$ 20 K ($\sim$150--210 K) that appeared roughly correlated in time with the sunspot numbers and Lyman-$\alpha$ flux of solar cycles 21 and 22, but lagging by roughly one year. 

To explain these possibly correlated temperature changes, \citet{roques1994neptune} proposed a mechanism in which the enhanced ultraviolet flux leads to the formation of aerosols, which in turn absorb solar radiation and heat the atmosphere.  

Chemical and microphysical models have predicted the presence of hydrocarbon hazes \citep[\textit{e.g.},][]{romani1988methane, romani1989stratospheric,moses1992hydrocarbon,romani1993methane,toledo2020neptune}, and the presence of ultraviolet-absorbing hazes appears common in stratospheres and tropospheres of all the giant planets \citep[\textit{e.g.},][]{west1986clouds,karkoschka2009haze,karkoschka2011neptune,roman2013saturn,zhang2013stratospheric}. On Neptune, stratospheric aerosol layers were initially inferred by \citet{smith1989voyager} at heights of $\sim$150 km above the tropopause using Voyager high-phase angle limb scans, placing them at the necessary sub-millibar pressure levels. Subsequent analysis of Voyager high-phase angle data by \citet{moses1995neptunehazes} indicated the presence of excess extinction consistent with sub-micron haze particles at pressures less than 15 mbar, including particles near 0.5 mbar.  Although ethane, acetylene, and other relevant hydrocarbons do not condense at sub-millibar pressures, refractory hydrocarbons (such as PAHs) that are theoretically produced from higher-altitude ion chemistry can potentially condense at these levels \citep{dobrijevic20161dphotochem}.  Stellar occultation data analyzed by \citet{roques1994neptune} also suggest the possibility of an aerosol layer even higher, extending from 50--100 $\mu$bar to 0.001--0.01 $\mu$bars.  However, it remains to be seen whether photochemically produced aerosols at these pressures actually vary in time, and if so, whether solar heating of such aerosols can even provide heating rates large enough to account for the observed thermal variation. \citet{moses1995neptunehazes} inferred a cumulative haze extinction optical depth of only $\sim$3$\times10^{-3}$ in Voyager's clear filter (280--640 nm) at pressures less than 15 mbar using Voyager imaging. Given these low optical depths, they concluded that, if typical, these hazes may not contribute significantly to the stratospheric heating rates.


Deeper in the atmosphere ($\ge\sim$100 mbar), solar cycle-induced variation in the abundance of absorbing tropospheric aerosols would also be consistent with observed correlations in visual (472 nm and 551 nm) magnitudes. Detrended to remove the longer term variability associated with seasonal changes (see Figure \ref{fig:time_trends}), Neptune's magnitude appeared dimmer near solar maxima and brighter near solar minima from the 1970s to the mid-1990s (cycles 20-22), with changes lagging about three years behind the Sun \citep{lockwood2002photometric,lockwood2006photometric,aplin2016determining,lockwood2019final}. \citet{baines_UV_neptune_1990} proposed that these apparent changes were the result of solid-state chemical "tanning" of stratospheric aerosols by enhanced far-ultraviolet flux.  The albedo change would naturally result in greater heating of the aerosol layers, and this heating would increase with height, presumably following a similar or longer lag relative to the aerosol production rates. 

Interestingly, despite decades of seemingly rough agreement, the purported correlation between Neptune's albedo and the solar cycle seemed to break in solar cycle 23, when Neptune's visible albedo appeared \emph{brighter} near the 2001 solar maximum---opposite to the previously inferred trend. The cause of this apparent break---if the solar correlation had indeed ever existed---has not yet been explained. 

We note that the fluctuations in Neptune's magnitude determined by \citet{aplin2016determining} (reproduced in Figure \ref{fig:time_trends}, top panel) appear very similar to the trend in mid-infrared brightness seen in the imaging. The early 2000s were the warmest in our limited record of the stratospheric temperatures, despite the planet's higher albedo at the time.  Additionally, as we noted above, that the early 2000s were also brighter at near-IR wavelengths owing to increased coverage of discrete high clouds, seen most clearly in methane band images \citep{karkoschka2011neptune, Roman2013DPS}. While the aforementioned albedo studies \citep{lockwood2006photometric,aplin2017solar,lockwood2019final} mostly considered the modulation of clouds and hazes in terms of disk-integrated visual magnitudes, the correlation of discrete cloud coverage was only previously hinted at from a few contemporary near-IR observations \citep{lockwood2002photometric}. 

In a study of HST images, \citet{karkoschka2011neptune} described the temporal variation in the discrete cloud abundance (represented by an albedo enhancing filling factor) as varying on roughly five-year timescales; however, as shown in Figure \ref{fig:time_trends} (middle panel), the discrete cloud filling factor appears remarkably well correlated with the solar cycle Lyman-$\alpha$, albeit these published cloud observations are currently limited to a single solar cycle. As noted, HST observations in recent years also show a dearth of discrete cloud cover, roughly at a time when the solar cycle reached its most recent minimum (between Solar Cycle 24 and 25 in December 2019). The possible correlation between the two is interesting because it appears contrary to theoretical expectations.   

While statistical analysis by \citet{aplin2016determining} claimed to show that the Lyman-$\alpha$ flux was generally anti-correlated with Neptune's visual brightness, variations in the count of galactic cosmic rays were also claimed to have a statistically significant contribution to the magnitude. Galactic cosmic rays fluxes are largely inversely correlated with the solar cycle, reaching maximum fluxes near solar minimum (see Figure \ref{fig:time_trends}, bottom panel).  \citet{moses1989neptune} previously proposed that Neptune's brightness variation could be explained by ion-induced cloud nucleation during solar minima by increasing the cloud coverage---similar to what is seen on Earth \citep{svensmark1997variation}. However, the data appear to suggest the opposite in this case. Discrete cloud coverage peaked during the 2001 solar \emph{maximum}, and appears anti-correlated with the cosmic ray rate\footnote{Oulu corrected cosmic ray counts at Earth, corrected for barometric pressure and efficiency, from the Sodankyla Geophysical Observatory at \url{http://cosmicrays.oulu.fi.}} for the years in which spatially resolved imaging is available.  Regular measurements of cloud coverage for the prior maximum are unavailable, but \citet{lockwood2002photometric} notes prominent "outbursts" in near-infrared observations in 1977 and 1986--1989---corresponding to times of solar \emph{minima} and \emph{greater} cosmic ray counts. The most recent solar maximum was weak, but the discrete cloud coverage at the time still appeared greater than what was seen during the following solar minimum of 2019 (with the notable exception of 2017's cloudy outburst \citep{molter2019neptune}).  

Altogether, this suggests that the stratospheric temperatures and albedo at visible and near-IR wavelengths may potentially be correlated, and the solar cycle may provide the physical link.  But observations are too limited to draw conclusions yet, and any solar effect on clouds and hazes have seemingly changed in the late 1990s for reasons unknown. It is possible that the expected ultraviolet "tanning" effect on the albedo acts independently of the discrete cloud coverage, but unless we are seeing a purely meteorological coincidence, the potential physical mechanisms possibly linking solar cycle variation, stratospheric temperatures, and cloud activity should be further examined. Given that possible changes occurred in the transition from southern spring to summer (2005 solstice, Figure \ref{fig:time_trends}, bottom panel), we might find that the aerosol response is somehow seasonally dependent or altered by sufficiently disruptive weather events. More observations over the next solar cycle will be key to assessing the consistency and physical mechanism behind this intriguing but uncertain potential relationship.

\subsection{Implied Wind Changes}
Regardless of their cause, the large temperature changes in Neptune's stratosphere should be diagnostic of the atmospheric dynamics.  For an atmosphere in geostrophic balance, the meridional temperature gradient is related to the vertical shear of the zonal winds through the thermal wind relationship \citep[\textit{e.g.},][]{forsythe1945thermalwind,holton1973introduction,conrath1983thermal}. The observed changes in the meridional temperature gradients over time should therefore be associated with changes in the vertical wind shear and consequent strength of zonal jets. 

The lower stratospheric temperatures reveal an averaged equator-to-south-pole contrast of $\Delta T\approx $ 9 K at 0.5 mbar, increasing in temperature to the south. By 2018, the low latitude temperatures decreased further, and the magnitude of this contrast increased to $\Delta T\approx$ 17 K, with a maximum gradient at mid-latitudes. Within two years, the warming south pole increased the contrast yet further to $\Delta T\approx$ 30 K at 0.5 mbar.  This tripling of the meridional gradient ($\Delta T/\Delta y$), rising in temperature towards the south pole, would imply a similar increase in the magnitude of the vertical wind shear at mid-latitudes, all else being equal. According to the thermal wind relationship, \(\partial u/\partial ln(p) = \frac{R}{f} (\partial T/\partial y)_p\), where $u$ is the geostrophic wind in the zonal direction, $p$ is pressure, $R$ is the specific gas constant, $f$ is the Coriolis parameters,  and $(\partial T/\partial y)_p$ is the meridional temperature gradient at pressure $p$.  Given that temperatures reach a maximum at the south pole (\textit{i.e.,} \(\partial T/\partial y <0\)), and $f$ is negative in the southern hemisphere, this implies a shear of increasing westward velocity with height (\textit{i.e.,} \(\partial u/\partial ln(p)>0\rightarrow\partial u/\partial z<0\)).  Based on cloud tracking in the upper troposphere and lower stratosphere \citep[\textit{e.g.},][]{limaye1991winds,sromovsky2001coordinated,sromovsky2001neptune, tollefson2018thermalwind}, the zonal winds are expected to blow eastward at high latitudes, westward at low latitudes, and change in sign at mid-latitudes.  The stratospheric temperature gradients near 45-70$^{\circ}$S are broadly consistent with that observed in the troposphere, suggesting a weakening of the prograde jet with altitude (\textit{i.e.}, westward du/dz).  At the equator $\pm$20$^{\circ}$, tropospheric gradients suggest a weakening of the retrograde jet with height (\textit{i.e.,} eastward \(\partial u/\partial z\)), but stratospheric gradients are considerably weaker and highly variable.


\citet{tollefson2018thermalwind} examined the vertical shear in H and K'--band Keck images from 2013 and 2014 by tracking clouds at the different heights sensed. They determined that equatorial winds increased with height (\textit{i.e.}, became increasingly westward).  However, their observations only showed shear at the equator, as opposed to the maximum in mid-latitudinal shear implied by our observed temperatures.  Furthermore, the near-IR measurements were only deemed sensitive to shear between the 1--2-bar and 10--100-mbar levels---far deeper than the shear implied by our mid-IR stratospheric measurements.  And as \citet{tollefson2018thermalwind} note, compositional gradients---particularly the latitudinal variation of the methane abundance---will affect conclusions drawn from the thermal wind relationship. Finally, it is worth stressing that the limited vertical resolution of the broadly-sensing imaging data still leaves significant uncertainty in the temperature field, which could alter conclusions regarding the vertical shear.  A more rigorous analysis will require greater constraints on the temperature and chemistry, independently, over a wide range of pressures, but this is beyond the scope of this work.  

\subsection{Comparison with Uranus}
Finally, given the observed variability in Neptune's mid-infrared emission, it is worth briefly considering the case for similar variation in Uranus' atmosphere. 

Uranus displays well documented seasonal variability in reflected light \citep{lockwood2019final}, with pronounced oscillation in the albedo near its poles \citep{rages2004evidence,hammel2007long,irwin2012uranus_seasons,roman2018aerosols,toledo2018uranus}. Like Neptune, Uranus' magnitude also varies in response to the solar cycle, but to a lesser extent \citep{aplin2017solar}. Stellar occultation data show Uranus' upper stratospheric temperatures also vary with time \citep{sicardy1985_uranus_variations,baron1989oblateness,roques1994neptune,young2001uranus}, but with the limited published data (mostly from the 1970s and 1980s), it is difficult to disentangle coinciding solar and seasonal response. 

Compared to Neptune, relatively few mid-infrared observations of Uranus exist, and most of those are images in the Q-band, sensing upper-tropospheric temperatures.  Similar to what we find for Neptune, comparisons between these ground-based images and Voyager have shown little if any changes in the upper tropospheric temperatures \citep{orton2015thermal,roman2020uranus}. But unlike Neptune, a comparison of images sensing Uranus' stratosphere (at 0.1 mbar via 13-$\mu$m acetylene emission) in 2009 and 2018 has shown no significant change in radiance \citep{orton2018neii,roman2020uranus}. This might indicate that Uranus' lower stratosphere is less variable than Neptune's owing to the inferred weaker vertical mixing, limited hydrocarbon abundances, and longer radiative time constants compared to Neptune \citep{conrath1990temperature,moses2018seasonal,li2018high,moses2020icegiantchem}, and it is consistent with typically less discrete cloud activity \citep[\textit{e.g.,}][]{hammel2005new,sromovsky2005dynamics, sromovsky2009uranus,hammel2007long,depater2011Uranus, depater2015record,roman2018aerosols}. However, \citet{rowe2021longitudinal} reported slight variation ($<$15$\%$) in 2007 N-band Spitzer-IRS spectra covering four different longitudes, possibly indicating the effect of significant meteorological activity.   Furthermore, given Uranus' relatively lower 12-$\mu$m radiance, no ethane-sensing images currently exist for Uranus; likewise, no acetylene-sensing images exist for Neptune, so a direct comparison of their variability is not currently possible.  Finally, we note that the limited observations of Uranus' stratospheric emission---2007 Spitzer observations \citep{orton2014a,orton2014mid, rowe2021longitudinal}; the 2009 and 2018 VISIR ground-based observations \citep{roman2020uranus}; and even earlier mid-IR spectra from 1987 \citep{orton1987spectra}---all happen to coincide with a similar phase of the solar cycle---a phase just approaching solar minimum.  So until more observations are made coinciding with other phases of the solar cycle, the question of variability in Uranus' lower stratosphere remains open. 

\section{Conclusions}\label{sec:conclude}

The collective mid-infrared imaging data reveal significant temporal variability in Neptune's stratosphere between 2003 and 2020. These observations provide the strongest evidence to date that processes produce sub-seasonal variation on both global and regional scales. We highlight the following conclusions:
\begin{itemize}

  \item Q-band images (17.5--25 $\mu$m), sensitive to temperatures at pressures of roughly 50--300 mbar, continue to show a pattern of cooler mid-latitudes and warmer equator, consistent with Voyager-era measurements.  Only at the pole do we see significant variation at these wavelengths, with temperatures 6 $\pm$ 3 K warmer in 2003 and 2006 compared to 2018 and 2020. 

  \item Images sensitive to stratospheric ethane (12--13 $\mu$m), methane (7--8 $\mu$m), and CH$_3$D (8--9 $\mu$m) show significant variation in time.  Disk-integrated radiances generally decreased from a maximum in 2003 to a near minimum in 2010. Limited data over the next decade show a possible but uncertain radiance increase in 2011 and 2012, dropping to a minimum in 2018 and possibly rising again in 2020.   

  \item Stratospheric temperatures inferred from 17--$\mu$m H$_2$ S(1) hydrogen quadrupole spectra show a similar decline in radiances over time, and appear remarkably similar to contemporaneous ethane (12--13 $\mu$m) images.  This suggests that observed variation in stratospheric emission is primarily due to temperature variation. 
  

  \item The south pole dramatically brightened in stratospheric images between 2018 and 2020, while the mid- and low-latitudes remained dimmer than in previous years. Similar changes were also observed in the hydrogen H$_2$ S(1) quadrupole spectra between 2007 and 2019.   

  \item If radiance changes are attributed to temperatures, the meridional temperature contrasts between the warmer south pole and cooler equator increased from roughly 8 $\pm$ 2 K in 2003 to 28 $\pm$ 2 K in 2020; polar temperatures rose from 152 $\pm$ 2  K to 163 $\pm$ 2  K between 2018 and 2020 alone.  This increase in gradient implies a corresponding increasing westward vertical shear with height at southern mid-latitudes.

  \item The temporal changes in relative radiances from stratospheric data appear to parallel changes in Neptune's visible albedo anomalies and discrete cloud coverage. The physical mechanism linking the stratospheric temperatures and tropospheric clouds and hazes over many scale heights is unknown, but we speculate that it may be related to seasonal forcing, meteorological phenomenon, or solar cycle variations in Lyman-$\alpha$ flux. Finally, although observations are limited in time, we also note an intriguing potential correlation between Neptune's discrete cloud cover \citep{karkoschka2011neptune} and the recent solar cycles.
\end{itemize}

Neptune's stratospheric temperatures have changed, and what we can expect to see in the years ahead is unknown. Although revealing, the observations examined in this work ultimately comprise less than half a Neptunian season.  The greater seasonal context and fundamental cause of the observed variation remains unknowable without additional and repeated observations extending well into the future. We were fortunate to capture the surprising recent changes given the sporadic history of observations, but we now have the opportunity to observe how the changing stratosphere evolves in time. In particular, with the new solar cycle beginning to ramp up, regular observations over the next decade will be crucial for understanding the nature and trends shaping the stratospheric variability of Neptune.
\clearpage

\begin{acknowledgements}
MTR, LNF and NRG were supported by a European Research Council Consolidator Grant, under the European Union’s Horizons 2020 research and innovation program, grant number 723890. LNF was also supported by a Royal Society Research fellowship. NRG was also supported by NASA under award number 80GSFC21M0002. GSO and JS were supported by NASA through funds distributed to the Jet Propulsion Laboratory, California Institute of Technology, under contract 80NM0018D0004. JIM acknowledges support from NASA grant 80NSSC19K0536. IdP was partially supported by the National Science Foundation, NSF Grant AST-1615004 to the University of California, Berkeley. AA is supported by the Spanish project PID2019-109467GB-I00924 (MINECO/FEDER, UE) and Grupos Gobierno Vasco IT1366-19. 
This work used the ALICE supercomputing facilities provided by the University of Leicester. This analysis was based on observations collected at the European Organisation for Astronomical Research in the Southern Hemisphere under ESO programs 077.C-0571(A), 081C-0496(A), 083.C-0163(A,B), and 0101.C-0044(A).

This research has made use of the Keck Observatory Archive (KOA), which is operated by the W. M. Keck Observatory and the NASA Exoplanet Science Institute (NExScI), under contract with the National Aeronautics and Space Administration. 2003 Datasets were acquired through the KOA, originating with PIs M. Brown (C14LSN) and I. de Pater (U38LS).

This research is based in part on data collected at Subaru Telescope (P.I. Y. Kasaba; programs: o07150, o08161, o12143), which is operated by the National Astronomical Observatory of Japan. 

This research is based on observations obtained at the international Gemini Observatory, a program of NSF’s NOIRLab, which is managed by the Association of Universities for Research in Astronomy (AURA) under a cooperative agreement with the National Science Foundation on behalf of the Gemini Observatory partnership: the National Science Foundation (United States), National Research Council (Canada), Agencia Nacional de Investigaci\'{o}n y Desarrollo (Chile), Ministerio de Ciencia, Tecnolog\'{i}a e Innovaci\'{o}n (Argentina), Minist\'{e}rio da Ci\^{e}ncia, Tecnologia, Inova\c{c}\~{o}es e Comunica\c{c}\~{o}es (Brazil), and Korea Astronomy and Space Science Institute (Republic of Korea).

The authors wish to recognize and acknowledge the very significant cultural role and reverence that the summit of Maunakea has always had within the indigenous Hawaiian community.  We are most fortunate to have the opportunity to conduct observations from this mountain.

Finally, we thank the many observers and telescope operators who work long nights to provide the community with the data needed to advance planetary science.

\end{acknowledgements}

\begin{figure*}
    \centering
    \includegraphics[clip, trim=.0in 0.0in 0.in 0.0in,scale=1]{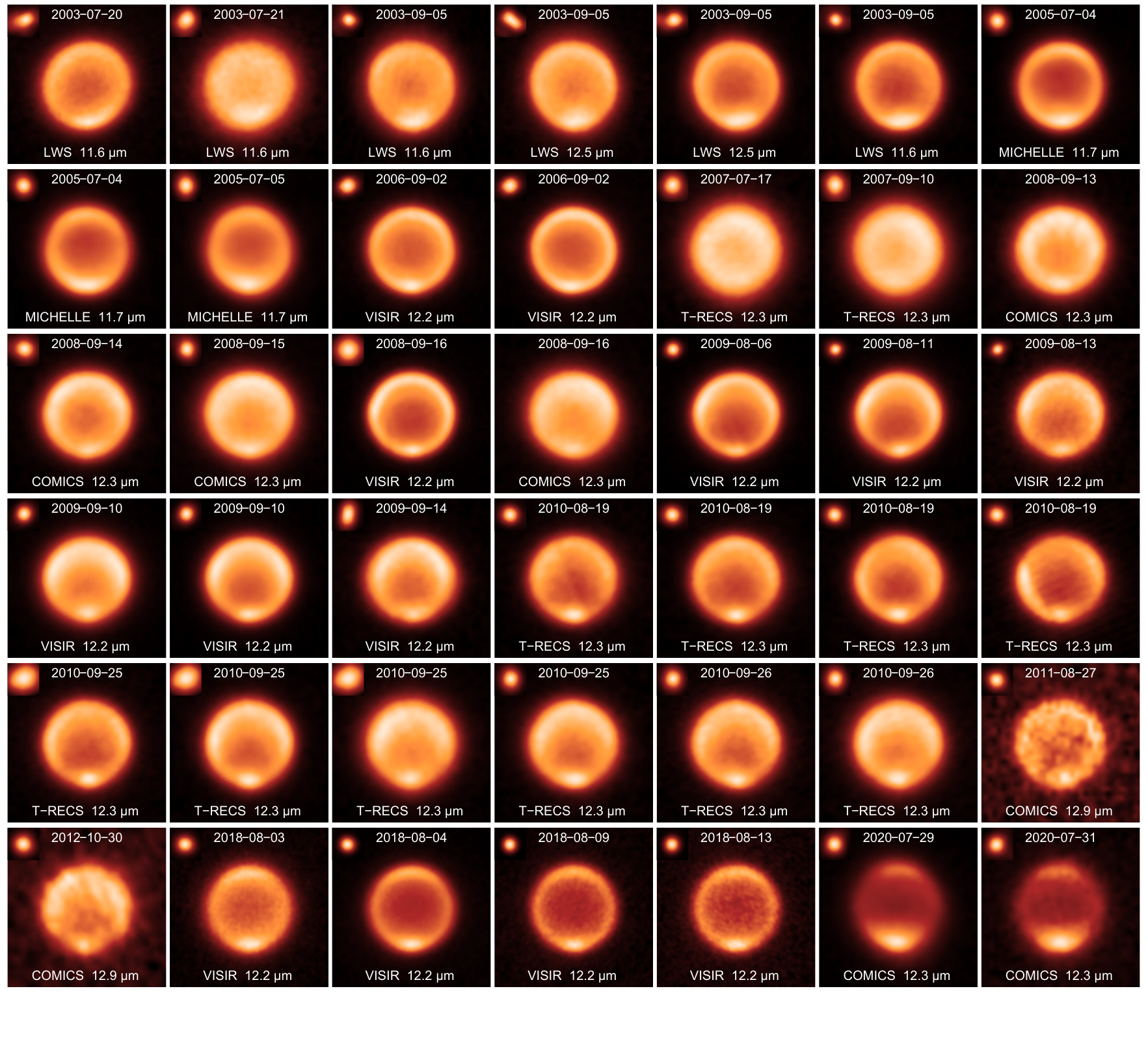}
    \caption{N-band images of Neptune sensitive to stratospheric ethane emission with effective filtered wavelengths of 11.6\textendash12.5 $\mu$m, dating from 2003 to 2020. The date (yyyy-mm-dd), imaging instrument, and effective filter wavelength for each image are stated. Insets of the accompanying calibration stars, when available, are shown to indicate the respective spatial resolutions of the seeing disks near the time of the observations. Image details are provided in Table \ref{tab:ethane}}.
    \label{fig:ethane_imgs}
\end{figure*}
\begin{figure*}
    \centering
    \includegraphics[clip, trim=0in .00in 0.in 0.0in,scale=1.00]{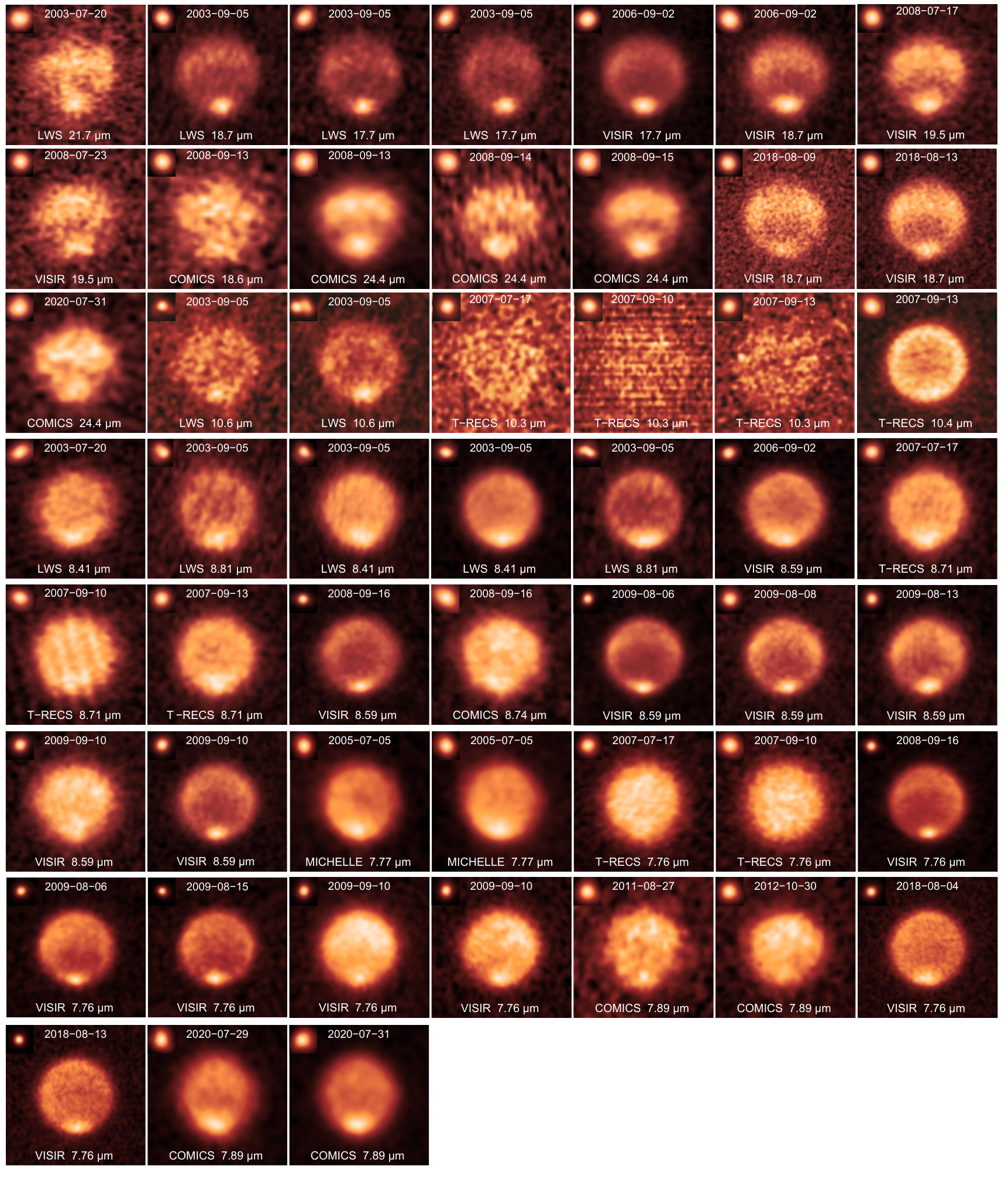}
    \caption{As in Figure \ref{fig:ethane_imgs}, but for the remaining spectral groups, including images sensitive to upper-tropospheric and lower-stratospheric temperatures via emission from hydrogen at effective wavelengths of 17.7\textendash24.4 $\mu$m; images sensitive to hydrogen and hydrocarbon emission from a broad range of pressures at effective filtered wavelengths of 10--11 $\mu$m; images sensitive to stratospheric monodeuterated methane (CH$_3$D) emission at wavelengths of 8\textendash9 $\mu$m; and images sensitive to stratospheric methane emission at wavelengths of 7\textendash8 $\mu$m. The date (yyyy-mm-dd), imaging instrument, and effective filter wavelength for each image are stated. Insets of the accompanying calibration stars, when available, are shown to indicate the respective spatial resolutions of the seeing disks near the time of the observations. Image details are provided in Tables \ref{tab:images_h2_ch4_ch3d_imgs} and \ref{tab:methane}}
    \label{fig:images_h2_ch4_ch3d_imgs}
\end{figure*}

\appendix \label{appendix}
All the ground-based mid-infrared images of Neptune used in this study are presented in Figures \ref{fig:ethane_imgs} and \ref{fig:images_h2_ch4_ch3d_imgs}, with corresponding details provided in Tables \ref{table:ethanetab}, \ref{tab:images_h2_ch4_ch3d_imgs}, and \ref{table:methanetab}.  As far as we are aware, this comprises all available mid-infrared, spatially-resolved imaging of Neptune to date. 

\begin{figure*}
   \centering
    \includegraphics[clip, trim=.0in 0.0in 0.in 0.0in,scale=0.84]{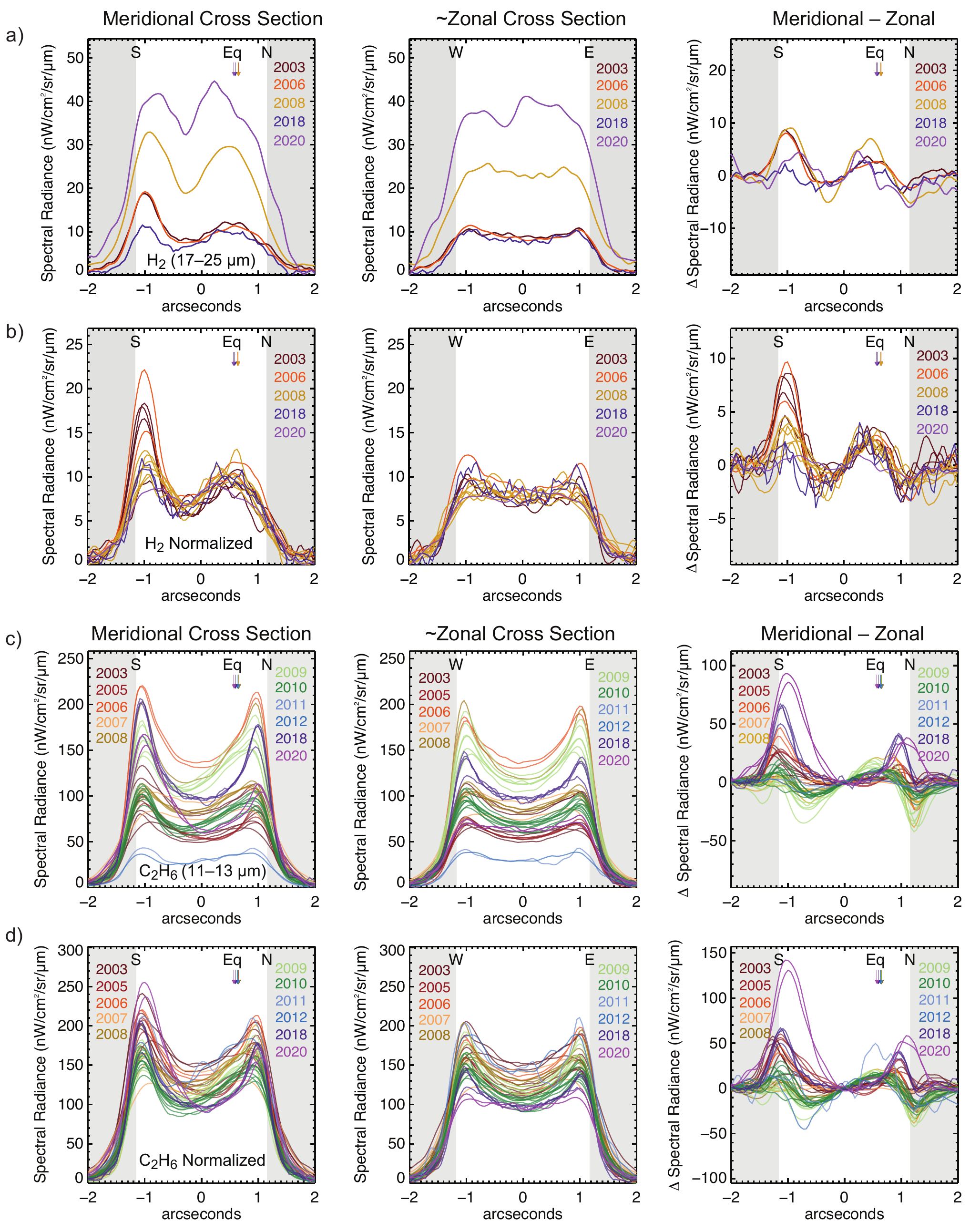}
    \caption{Profiles of radiances across the disk showing the meridional, roughly zonal, and difference cross-sections (meridional--zonal) for individual images, color coded by year. a) Profiles for all Q-band hydrogen imaging (17--25 $\mu$m), as calibrated, and b) the same images, now normalized by their passband-integrated Spitzer radiances and scaled by the corresponding VISIR Q2 filter (18.8 $\mu$m) radiance. For clarity, error bars are omitted but taken to be 30$\%$ or less. c) Similarly, profiles for all ethane images (11--13 $\mu$m) at their observed radiances, and d) Spitzer-normalized and scaled by for the VISIR NeII$\_$1 filter (12.2 $\mu$m), with assumed uncertainty of 10$\%$ or less.} 
    \label{fig:diskh2_c2h6all}
\end{figure*}

\begin{figure*}
   \centering
    \includegraphics[clip, trim=.0in 0.0in 0.in 0.0in,scale=0.84]{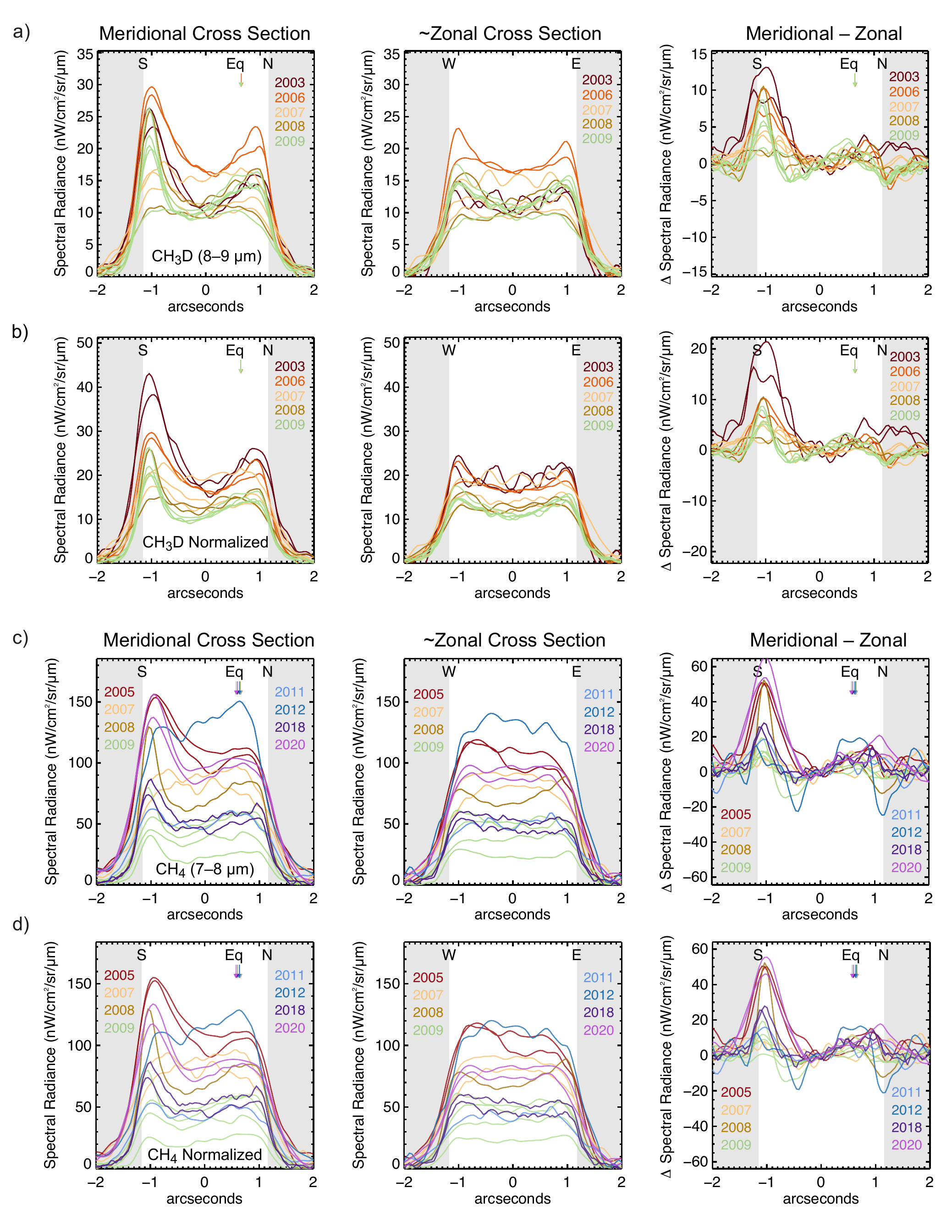}
    \caption{As in Figure \ref{fig:diskh2_c2h6all}, but for CH$_3$D and CH$_4$ images. a) Profiles for all CH$_3$D imaging (8--9 $\mu$m), as calibrated, and b) the same images, now normalized by their passband-integrated Spitzer radiances and scaled by the corresponding VISIR PAH1 filter (8.9 $\mu$m) radiance. For clarity, error bars are omitted but taken to be 20$\%$ or less. c) Similarly, profiles for all methane images (7--8 $\mu$m) at their observed radiances, and d) Spitzer-normalized and scaled by for the VISIR J7.9 filter (7.8 $\mu$m).}
    \label{fig:diskch3dplusch4all}
\end{figure*}

Disk profiles for all images are provided in Figures \ref{fig:diskh2_c2h6all} and \ref{fig:diskch3dplusch4all}, sorted by spectral group, for both native and normalized radiances. In these figures, profiles of observed radiances across the disk show the meridional cross-section (vertically bisecting the image, north-south), the perpendicular cross-sections ($\sim$zonal, horizontally bisecting the image, roughly west--east), and the difference between the profiles (meridional--zonal) to help separate latitudinal variation from center-to-limb behavior. Plots average 9 central lines for each profile, amounting to 0.4'', or roughly just over 17$\%$ of the disk diameter in our normalized resolution images (see Figure \ref{fig:ave_imgs}).  Observation years are defined by color, and the location of the disk edges and changing equator are also indicated.

\begin{deluxetable*}{ccccccccc}
\centering
\tabletypesize{\footnotesize}
\tablecolumns{9} 
\tablewidth{8.0in}
\tablecaption{ N-band Images: C$_2$H$_6$ (11--13 $\mu$m)}
 \label{table:ethanetab}
\tablehead{
\head{2.9cm}{\vspace{-0.1cm} Date, Time (yyyy-mm-dd, hr:mn)} & \head{1.4cm}{\vspace{-0.0cm}Instrument} & \head{1.6cm}{\vspace{-0.0cm}   Filter} & \head{1.6cm}{\vspace{-0.15cm}Effective Wavelength ($\mu$m)} & \head{2.6cm}{\vspace{-0.1cm} Disk Radiance \scriptsize($10^{-7}$W/$cm^2$/sr/$\mu$m)} & \head{1.2cm}{\vspace{-0.15cm}Radiance Ratio
 to Spitzer} & \head{1.cm}{\vspace{-0.0cm}Airmass} & \head{1.2cm}{\vspace{-0.15cm}Seeing Disk (arcsec)} & \head{0.8cm}{\vspace{-0.15cm}Calibration Star}}
 \startdata 
  2003-07-20  ,       11:00 &        LWS &              SiC & 11.66 &  1.00 $\pm$ 0.10 &  0.98 $\pm$ 0.11 & 1.26 & 0.59 &     HD199345
          \\
  2003-07-21  ,       10:24 &        LWS &              SiC & 11.66 &  1.23 $\pm$ 0.12 &  1.21 $\pm$ 0.14 & 1.30 & 0.58 &     HD139663
          \\
  2003-09-05  ,       05:49 &        LWS &             11.7 & 11.67 &  1.11 $\pm$ 0.11 &  0.89 $\pm$ 0.10 & 1.62 & 0.37 &    HD186791
          \\
  2003-09-05  ,       05:59 &        LWS &             12.5 & 12.57 &  1.60 $\pm$ 0.16 &  1.12 $\pm$ 0.13 & 1.57 & 0.46 &    HD186791
          \\
  2003-09-05  ,       07:25 &        LWS &             12.5 & 12.57 &  1.57 $\pm$ 0.16 &  1.11 $\pm$ 0.12 & 1.30 & 0.57 &     HD199345
          \\
  2003-09-05  ,       07:32 &        LWS &             11.7 & 11.67 &  1.05 $\pm$ 0.11 &  0.85 $\pm$ 0.09 & 1.28 & 0.37 &     HD199345
          \\
  2005-07-04  ,       11:10 &   MICHELLE &       Si5-11.6 & 11.66 &  1.02 $\pm$ 0.10 &  0.84 $\pm$ 0.09 & 1.34 & 0.47 &     HD199345
          \\
  2005-07-04  ,       14:56 &   MICHELLE &       Si5-11.6 & 11.66 &  1.02 $\pm$ 0.10 &  0.83 $\pm$ 0.09 & 1.49 & 0.47 &     HD199345
          \\
  2005-07-05  ,       10:43 &   MICHELLE &       Si5-11.6 & 11.66 &  1.03 $\pm$ 0.10 &  0.84 $\pm$ 0.09 & 1.44 & 0.50 &     HD199345
          \\
  2006-09-02  ,       01:14 &      VISIR &           NEII$\_$1  & 12.22 &  2.55 $\pm$ 0.25 &  0.99 $\pm$ 0.11 & 1.15 & 0.46 &     HD200914
          \\
  2006-09-02  ,       03:27 &      VISIR &           NEII$\_$1  & 12.22 &  2.45 $\pm$ 0.24 &  0.95 $\pm$ 0.11 & 1.01 & 0.46 &     HD200914
          \\
  2007-07-17  ,       05:45 &     T-RECS &       Si6-12.3 & 12.33 &  1.81 $\pm$ 0.20 &  1.08 $\pm$ 0.12 & 1.05 & 0.62 & HD199345
          \\
  2007-09-10  ,       23:47 &     T-RECS &       Si6-12.3 & 12.33 &  1.44 $\pm$ 0.16 &  0.91 $\pm$ 0.10 & 1.32 & 0.60 & HD199345
          \\
  2008-09-13  ,       08:07 &     COMICS &   F09C12.50 & 12.37 &  1.59 $\pm$ 0.16 &  0.94 $\pm$ 0.10 & 1.22 & ... &         ...
          \\
  2008-09-14  ,       09:40 &     COMICS &   F09C12.50 & 12.37 &  1.61 $\pm$ 0.16 &  0.96 $\pm$ 0.11 & 1.28 & 0.50 &     HD216032
          \\
  2008-09-15  ,       07:48 &     COMICS &   F09C12.50 & 12.37 &  1.66 $\pm$ 0.17 &  0.99 $\pm$ 0.11 & 1.23 & 0.44 &     HD216032
          \\
  2008-09-16  ,       04:04 &      VISIR &           NEII$\_$1  & 12.22 &  2.23 $\pm$ 0.22 &  0.87 $\pm$ 0.10 & 1.09 & 0.38 &     HD178345
          \\
  2008-09-16  ,       06:20 &     COMICS &   F09C12.50 & 12.37 &  1.67 $\pm$ 0.17 &  1.00 $\pm$ 0.11 & 1.42 & ... &           ...
          \\
  2009-08-06  ,       04:41 &      VISIR &           NEII$\_$1  & 12.22 &  2.23 $\pm$ 0.22 &  0.87 $\pm$ 0.10 & 1.04 & 0.37 &     HD216149
          \\
  2009-08-11  ,       08:14 &      VISIR &           NEII$\_$1  & 12.22 &  2.15 $\pm$ 0.22 &  0.84 $\pm$ 0.09 & 1.40 & 0.37 &     HD196321
          \\
  2009-08-13  ,       07:06 &      VISIR &           NEII$\_$1  & 12.22 &  2.04 $\pm$ 0.20 &  0.80 $\pm$ 0.09 & 1.16 & 0.31 &        HD787
          \\
  2009-09-10  ,       01:19 &      VISIR &           NEII$\_$1  & 12.22 &  2.03 $\pm$ 0.20 &  0.79 $\pm$ 0.09 & 1.14 & 0.41 &     HD178345
          \\
  2009-09-10  ,       06:33 &      VISIR &           NEII$\_$1  & 12.22 &  2.05 $\pm$ 0.20 &  0.80 $\pm$ 0.09 & 1.52 & 0.41 &     HD178345
          \\
  2009-09-14  ,       01:41 &      VISIR &           NEII$\_$1  & 12.22 &  1.86 $\pm$ 0.19 &  0.73 $\pm$ 0.08 & 1.07 & 0.50 &     HD177716
          \\
   2010-08-19  ,       01:00 &     T-RECS &       Si6-12.3 & 12.33 &  1.21 $\pm$ 0.13 &  0.76 $\pm$ 0.09 & 1.72 & 0.45 & HD199345
          \\
  2010-08-19  ,       03:45 &     T-RECS &       Si6-12.3 & 12.33 &  1.34 $\pm$ 0.15 &  0.85 $\pm$ 0.09 & 1.08 & 0.45 &     HD199345
          \\
  2010-08-19  ,       06:02 &     T-RECS &       Si6-12.3 & 12.33 &  1.32 $\pm$ 0.14 &  0.83 $\pm$ 0.09 & 1.10 & 0.45 &     HD199345
          \\
  2010-08-19  ,       07:43 &     T-RECS &       Si6-12.3 & 12.33 &  1.23 $\pm$ 0.14 &  0.78 $\pm$ 0.09 & 1.41 & 0.45 &     HD199345
          \\
  2010-09-25  ,       00:15 &     T-RECS &       Si6-12.3 & 12.33 &  1.26 $\pm$ 0.14 &  0.80 $\pm$ 0.09 & 1.19 & 1.03 &     HD216032
          \\
  2010-09-25  ,       03:42 &     T-RECS &       Si6-12.3 & 12.33 &  1.39 $\pm$ 0.15 &  0.88 $\pm$ 0.10 & 1.12 & 1.03 &     HD216032
          \\
  2010-09-25  ,       05:08 &     T-RECS &       Si6-12.3 & 12.33 &  1.32 $\pm$ 0.14 &  0.83 $\pm$ 0.09 & 1.38 & 1.03 &     HD216032
          \\
  2010-09-25  ,       23:53 &     T-RECS &       Si6-12.3 & 12.33 &  1.31 $\pm$ 0.14 &  0.83 $\pm$ 0.09 & 1.24 & 0.63 &     HD199345
          \\
  2010-09-26  ,       01:56 &     T-RECS &       Si6-12.3 & 12.33 &  1.29 $\pm$ 0.14 &  0.82 $\pm$ 0.09 & 1.05 & 0.63 &     HD199345
          \\
  2010-09-26  ,       04:23 &     T-RECS &       Si6-12.3 & 12.33 &  1.32 $\pm$ 0.15 &  0.84 $\pm$ 0.09 & 1.22 & 0.63 &     HD199345
          \\
  2011-08-27  ,       10:57 &     COMICS &   F30C12.81 & 12.92 &  0.57 $\pm$ 0.06 &  1.08 $\pm$ 0.12 & 1.21 & 0.45 &     HD206445
          \\
  2012-10-30  ,       05:09 &     COMICS &   F30C12.81 & 12.92 &  0.59 $\pm$ 0.06 &  1.13 $\pm$ 0.13 & 1.20 & 0.45 &     HD206445
          \\
  2018-08-03  ,       06:03 &      VISIR &           NEII$\_$1  & 12.22 &  1.85 $\pm$ 0.19 &  0.72 $\pm$ 0.08 & 1.09 & 0.42 &     HD198048
          \\
  2018-08-04  ,       05:49 &      VISIR &           NEII$\_$1  & 12.22 &  1.96 $\pm$ 0.20 &  0.76 $\pm$ 0.09 & 1.10 & 0.36 &     HD198048
          \\
  2018-08-09  ,       09:15 &      VISIR &           NEII$\_$1  & 12.22 &  1.89 $\pm$ 0.19 &  0.73 $\pm$ 0.08 & 1.34 & 0.35 &        HD787
          \\
  2018-08-13  ,       06:42 &      VISIR &           NEII$\_$1  & 12.22 &  1.90 $\pm$ 0.19 &  0.74 $\pm$ 0.08 & 1.06 & 0.37 &     HD217902
          \\
  2020-07-29  ,       12:15 &     COMICS &   F09C12.50 & 12.37 &  1.54 $\pm$ 0.15 &  0.92 $\pm$ 0.10 & 1.14 & 0.46 &     HD216032
          \\
  2020-07-31  ,       12:12 &     COMICS &   F09C12.50 & 12.37 &  1.57 $\pm$ 0.16 &  0.93 $\pm$ 0.10 & 1.14 & 0.42 &     HD216032
          \\
\enddata
\label{tab:ethane}
\tablecomments{Corresponding images are shown in Figure \ref{fig:ethane_imgs}.}
\end{deluxetable*}

\begin{deluxetable*}{ccccccccc}
\centering
\tabletypesize{\footnotesize}
\tablecolumns{9} 
\tablewidth{8.0in}
\tablecaption{ Q-band and N-band Images: H$_2$ (17--25 $\mu$m); H$_2$, C$_2$H$_6$, C$_2$H$_4$ (10--11 $\mu$m); CH$_3$D (8--9 $\mu$m)}
 \label{table:q_and_n_1_tab}
\tablehead{
\head{2.9cm}{\vspace{-0.1cm} Date, Time (yyyy-mm-dd, hr:mn)} & \head{1.4cm}{\vspace{-0.0cm}Instrument} & \head{1.6cm}{\vspace{-0.0cm}   Filter} & \head{1.6cm}{\vspace{-0.15cm}Effective Wavelength ($\mu$m)} & \head{2.6cm}{\vspace{-0.1cm} Disk Radiance \scriptsize($10^{-7}$W/$cm^2$/sr/$\mu$m)} & \head{1.2cm}{\vspace{-0.15cm}Radiance Ratio
 to Spitzer} & \head{1.cm}{\vspace{-0.0cm}Airmass} & \head{1.2cm}{\vspace{-0.15cm}Seeing Disk (arcsec)} & \head{0.8cm}{\vspace{-0.15cm}Calibration Star}}
 \startdata 
  2003-07-20  ,       11:45 &        LWS &             22.0 & 21.73 &  0.23 $\pm$ 0.07 &  0.72 $\pm$ 0.22 & 1.25 & 0.63 &    HD199345
        \\
  2003-09-05  ,       06:28 &        LWS &             10.7 & 10.68 &  0.07 $\pm$ 0.02 &  0.64 $\pm$ 0.20 & 1.43 & 0.36 &    HD186791
          \\
  2003-09-05  ,       06:42 &        LWS &            18.75 & 18.72 &  0.12 $\pm$ 0.04 &  0.87 $\pm$ 0.27 & 1.37 & 0.48 &    HD186791
          \\
  2003-09-05  ,       07:11 &        LWS &            17.65 & 17.77 &  0.14 $\pm$ 0.04 &  0.91 $\pm$ 0.28 & 1.32 & 0.52 &     HD199345
          \\
  2003-09-05  ,       08:14 &        LWS &             10.7 & 10.68 &  0.06 $\pm$ 0.02 &  0.56 $\pm$ 0.17 & 1.26 & 0.54 &     HD199345
          \\
  2003-09-05  ,       08:28 &        LWS &            17.65 & 17.77 &  0.14 $\pm$ 0.04 &  0.92 $\pm$ 0.28 & 1.26 & 0.52 &     HD199345
          \\
  2006-09-02  ,       02:35 &      VISIR &               Q1 & 17.76 &  0.21 $\pm$ 0.06 &  1.41 $\pm$ 0.43 & 1.02 & 0.55 &     HD200914
          \\
  2006-09-02  ,       02:46 &      VISIR &               Q2 & 18.75 &  0.16 $\pm$ 0.05 &  1.11 $\pm$ 0.34 & 1.04 & 0.64 &     HD186791
          \\
  2007-07-17  ,       05:42 &     T-RECS &       Si4-10.4 & 10.39 &  0.02 $\pm$ 0.01 &  0.85 $\pm$ 0.26 & 1.14 & 0.59 & HD199345
          \\
  2007-09-10  ,       23:44 &     T-RECS &       Si4-10.4 & 10.41 &  0.03 $\pm$ 0.01 &  1.29 $\pm$ 0.39 & 1.04 & 0.63 &     HD216032
          \\
   2007-09-13  ,       05:06 &     T-RECS &       Si4-10.4 & 10.41 &  0.01 $\pm$ 0.01 &  0.24 $\pm$ 0.07 & 1.59 & 0.46 &     HD199345
          \\
  2008-07-17  ,       08:08 &      VISIR &               Q3 & 19.54 &  0.15 $\pm$ 0.05 &  0.90 $\pm$ 0.28 & 1.08 & 0.55 &     HD198048
          \\
  2008-07-23  ,       06:27 &      VISIR &               Q3 & 19.54 &  0.16 $\pm$ 0.05 &  0.97 $\pm$ 0.30 & 1.02 & 0.53 &     HD178345
          \\
  2008-09-13  ,       09:39 &     COMICS &   F37C18.75 & 18.68 &  0.15 $\pm$ 0.05 &  1.08 $\pm$ 0.33 & 1.27 & 0.60 &     HD217906
          \\
  2008-09-13  ,       08:20 &     COMICS &   F42C24.50 & 24.49 &  0.63 $\pm$ 0.19 &  0.93 $\pm$ 0.28 & 1.21 & 0.73 &     HD217906
      \\
  2008-09-14  ,       09:57 &     COMICS &   F42C24.50 & 24.49 &  0.63 $\pm$ 0.19 &  0.92 $\pm$ 0.28 & 1.32 & 0.67 &     HD186791
          \\
  2008-09-15  ,       08:16 &     COMICS &   F42C24.50 & 24.49 &  0.67 $\pm$ 0.20 &  0.99 $\pm$ 0.30 & 1.21 & 0.69 &     HD217906
          \\
  2018-08-09  ,  09:49 & VISIR &   Q2 & 18.75  &  0.13 $\pm$ 0.04 &  0.94 $\pm$ 0.29 & 1.52 & 0.50 & HD787 \\
 2018-08-13  , 07:10  & VISIR & Q2 & 18.75 &  0.12 $\pm$ 0.04 &  0.83 $\pm$ 0.25 & 1.07 & 0.51 &     HD217902
          \\
  2020-07-31  ,       13:17 &     COMICS &   F42C24.50 & 24.49 &  0.64 $\pm$ 0.19 &  0.95 $\pm$ 0.29 & 1.10 & 0.68 &      HD12929 
  \\
    2003-09-05  ,       06:28 &        LWS &             10.7 & 10.68 &  0.07 $\pm$ 0.02 &  0.64 $\pm$ 0.20 & 1.43 & 0.36 &    HD186791
          \\
  2003-09-05  ,       08:14 &        LWS &             10.7 & 10.68 &  0.06 $\pm$ 0.02 &  0.56 $\pm$ 0.17 & 1.26 & 0.54 &     HD199345
          \\
  2007-07-17  ,       05:42 &     T-RECS &       Si4-10.4 & 10.39 &  0.02 $\pm$ 0.01 &  0.85 $\pm$ 0.26 & 1.14 & 0.59 & HD199345
          \\
  2007-09-10  ,       23:44 &     T-RECS &       Si4-10.4 & 10.41 &  0.03 $\pm$ 0.01 &  1.29 $\pm$ 0.39 & 1.04 & 0.63 &     HD216032
          \\
   2007-09-13  ,       05:06 &     T-RECS &       Si4-10.4 & 10.41 &  0.01 $\pm$ 0.01 &  0.24 $\pm$ 0.07 & 1.59 & 0.46 &     HD199345
          \\
  2007-09-17  ,       01:32 &     T-RECS &                N & 10.04 &  9.29 $\pm$ 8.36 & 13.28 $\pm$ 11.95 & 1.08 & 0.65 &     HD216032
          \\
  2003-07-20  ,       11:04 &        LWS &              8.0 &  8.43 &  0.93 $\pm$ 0.19 &  1.31 $\pm$ 0.27 & 1.26 & 0.54 &     HD199345
          \\
  2003-09-05  ,       06:09 &        LWS &              8.9 &  8.82 &  0.22 $\pm$ 0.04 &  1.35 $\pm$ 0.28 & 1.52 & 0.41 &    HD186791
          \\
  2003-09-05  ,       06:18 &        LWS &              8.0 &  8.43 &  1.08 $\pm$ 0.22 &  1.52 $\pm$ 0.32 & 1.47 & 0.33 &    HD186791
          \\
  2003-09-05  ,       07:46 &        LWS &              8.0 &  8.43 &  1.10 $\pm$ 0.22 &  1.54 $\pm$ 0.32 & 1.27 & 0.32 &     HD199345
          \\
  2003-09-05  ,       08:00 &        LWS &              8.9 &  8.82 &  0.20 $\pm$ 0.04 &  1.27 $\pm$ 0.27 & 1.26 & 0.46 &     HD199345
          \\
  2006-09-02  ,       00:39 &      VISIR &             PAH1 &  8.60 &  0.29 $\pm$ 0.06 &  1.11 $\pm$ 0.23 & 1.26 & 0.37 &     HD200914
          \\
  2006-09-02  ,       07:20 &      VISIR &             PAH1 &  8.60 &  0.29 $\pm$ 0.06 &  1.11 $\pm$ 0.23 & 1.90 & 0.37 &     HD200914
          \\
  2007-07-17  ,       05:38 &     T-RECS &        Si2-8.8 &  8.72 &  0.21 $\pm$ 0.05 &  1.00 $\pm$ 0.21 & 1.07 & 0.66 & HD199345
          \\
   2007-09-10  ,       23:40 &     T-RECS &        Si2-8.8 &  8.73 &  0.27 $\pm$ 0.06 &  1.38 $\pm$ 0.29 & 1.08 & 0.68 &     HD216032
          \\
   2007-09-13  ,       04:14 &     T-RECS &        Si2-8.8 &  8.73 &  0.17 $\pm$ 0.04 &  0.86 $\pm$ 0.18 & 1.35 & 0.49 &     HD216032
          \\
  2008-09-14  ,       09:51 &     COMICS &   F05C08.70 &  8.75 &  0.16 $\pm$ 0.03 &  0.86 $\pm$ 0.18 & 1.31 & ... &         ...
          \\
  2008-09-16  ,       02:22 &      VISIR &             PAH1 &  8.60 &  0.20 $\pm$ 0.04 &  0.78 $\pm$ 0.16 & 1.02 & 0.32 &     HD178345
          \\
  2008-09-16  ,       07:32 &     COMICS &   F05C08.70 &  8.75 &  0.16 $\pm$ 0.03 &  0.83 $\pm$ 0.17 & 1.24 & 0.69 &     HD175775
          \\
  2009-08-06  ,       05:24 &      VISIR &             PAH1 &  8.60 &  0.21 $\pm$ 0.04 &  0.81 $\pm$ 0.17 & 1.02 & 0.28 &     HD216149
          \\
  2009-08-08  ,       03:48 &      VISIR &             PAH1 &  8.60 &  0.20 $\pm$ 0.04 &  0.76 $\pm$ 0.16 & 1.11 & 0.85 &     HD178345
          \\
  2009-08-13  ,       07:39 &      VISIR &             PAH1 &  8.60 &  0.19 $\pm$ 0.04 &  0.74 $\pm$ 0.16 & 1.27 & 0.33 &     HD220954
          \\
  2009-09-10  ,       00:38 &      VISIR &             PAH1 &  8.60 &  0.16 $\pm$ 0.03 &  0.62 $\pm$ 0.13 & 1.26 & 0.37 &      HD12524
          \\
  2009-09-10  ,       05:55 &      VISIR &             PAH1 &  8.60 &  0.18 $\pm$ 0.04 &  0.67 $\pm$ 0.14 & 1.31 & 0.37 &      HD12524
          \\
\enddata
\label{tab:images_h2_ch4_ch3d_imgs}
\tablecomments{Corresponding images are shown in Figure \ref{fig:images_h2_ch4_ch3d_imgs}. The N-filter image was used for spectroscopic acquisition; its flux calibration is highly uncertain owing to its short integration.}
\end{deluxetable*}

\begin{deluxetable*}{ccccccccc}
\centering
\tabletypesize{\footnotesize}
\tablecolumns{9} 
\tablewidth{7.in}
\tablecaption{ N-band Images: CH$_4$ (7--8 $\mu$m)}
 \label{table:methanetab}
\tablehead{
 \head{2.9cm}{\vspace{-0.1cm} Date, Time (yyyy-mm-dd, hr:mn)} & \head{1.4cm}{\vspace{-0.0cm}Instrument} & \head{1.6cm}{\vspace{-0.0cm}   Filter} & \head{1.6cm}{\vspace{-0.15cm}Effective Wavelength ($\mu$m)} & \head{2.6cm}{\vspace{-0.1cm} Disk Radiance \scriptsize($10^{-7}$W/$cm^2$/sr/$\mu$m)} & \head{1.2cm}{\vspace{-0.15cm}Radiance Ratio
 to Spitzer} & \head{1.cm}{\vspace{-0.0cm}Airmass} & \head{1.2cm}{\vspace{-0.15cm}Seeing Disk (arcsec)} & \head{0.9cm}{\vspace{-0.15cm}Calibration Star}}
 \startdata 
  2005-07-05  ,       11:04 &   MICHELLE &        Si1-7.9 &  7.76 &  1.66 $\pm$ 0.42 &  0.65 $\pm$ 0.17 & 1.36 & 0.50 &     HD199345
          \\
  2005-07-05  ,       14:07 &   MICHELLE &        Si1-7.9 &  7.76 &  1.65 $\pm$ 0.41 &  0.64 $\pm$ 0.17 & 1.34 & 0.54 &     HD199345
          \\
  2007-07-17  ,       05:34 &     T-RECS &        Si1-7.9 &  7.76 &  1.36 $\pm$ 0.37 &  0.54 $\pm$ 0.14 & 1.04 & 0.65 & HD199345
          \\
  2007-09-10  ,       23:36 &     T-RECS &        Si1-7.9 &  7.81 &  1.24 $\pm$ 0.34 &  0.51 $\pm$ 0.13 & 1.18 & 0.67 &     HD216032
          \\
  2008-09-16  ,       03:13 &      VISIR &             J7.9 &  7.76 &  0.99 $\pm$ 0.25 &  0.38 $\pm$ 0.10 & 1.03 & 0.32 &     HD178345
          \\
  2009-08-06  ,       06:13 &      VISIR &             J7.9 &  7.76 &  0.70 $\pm$ 0.18 &  0.27 $\pm$ 0.07 & 1.03 & 0.28 &     HD216149
          \\
  2009-08-15  ,       05:40 &      VISIR &             J7.9 &  7.76 &  0.95 $\pm$ 0.24 &  0.37 $\pm$ 0.09 & 1.04 & 0.24 &     HD196321
          \\
  2009-09-10  ,       01:58 &      VISIR &             J7.9 &  7.76 &  0.76 $\pm$ 0.19 &  0.29 $\pm$ 0.08 & 1.07 & 0.31 &     HD178345
          \\
  2009-09-10  ,       07:13 &      VISIR &             J7.9 &  7.76 &  0.52 $\pm$ 0.13 &  0.20 $\pm$ 0.05 & 1.87 & 0.31 &     HD178345
          \\
  2011-08-27  ,       10:44 &     COMICS &   F04C07.80 &  7.89 &  0.70 $\pm$ 0.17 &  0.32 $\pm$ 0.08 & 1.20 & 0.52 &     HD206445
          \\
  2012-10-30  ,       05:22 &     COMICS &   F04C07.80 &  7.89 &  1.29 $\pm$ 0.32 &  0.58 $\pm$ 0.15 & 1.19 & 0.49 &     HD206445
          \\
  2018-08-04  ,       06:47 &      VISIR &             J7.9 &  7.76 &  0.80 $\pm$ 0.20 &  0.31 $\pm$ 0.08 & 1.05 & 0.29 &     HD198048
          \\
  2018-08-13  ,       08:06 &      VISIR &             J7.9 &  7.76 &  0.68 $\pm$ 0.17 &  0.26 $\pm$ 0.07 & 1.16 & 0.28 &     HD217902
          \\
  2020-07-29  ,       12:41 &     COMICS &   F04C07.80 &  7.90 &  1.22 $\pm$ 0.31 &  0.55 $\pm$ 0.15 & 1.11 & 0.57 &     HD216032
          \\
  2020-07-31  ,       12:35 &     COMICS &   F04C07.80 &  7.90 &  1.28 $\pm$ 0.32 &  0.58 $\pm$ 0.15 & 1.11 & 0.51 &     HD216032
          \\
\enddata
\label{tab:methane}
\tablecomments{Corresponding images are shown in Figure \ref{fig:images_h2_ch4_ch3d_imgs}.}
\end{deluxetable*}

\bibliography{mybib}{}
\bibliographystyle{aasjournal}
 
\end{document}